\documentclass[]{aastex7}
\submitjournal{The Astrophysical Journal}
\accepted{February 2026}
\definecolor{ForestGreen}{rgb}{0,.4,.1}
\definecolor{Red}{rgb}{1,0,0}
\definecolor{Purple}{rgb}{1,0,1}
\definecolor{Orange}{rgb}{1,.5,0}
\definecolor{Teal}{rgb}{0,.5,.5}

\newcommand{\Bavg}{\ensuremath{B_{\mbox{\scriptsize avg}}}}

\newcommand{\Qavg}{\ensuremath{\langle Q\rangle}}
\newcommand{\GoF}{\ensuremath{\sum\delta^2}}
\newcommand{\Iobs}{\ensuremath{I_{\mbox{\scriptsize obs}}}}
\newcommand{\Imod}{\ensuremath{I_{\mbox{\scriptsize mod}}}}

\newcommand{\logT}{\ensuremath{\log T}\mbox{(K)}}

\newcommand{\bl}{\ensuremath{\gamma}}

\def\gx{{GX~Simulator}}
\shorttitle{Modeling of AR~12760 with GX~Simulator}
\shortauthors{Kucera et al.}


\begin{document}
\title{Modeling of AR~12760 with GX~Simulator and Evidence for the Extended Transition Region in Peripheral Active Region Loops}

\shorttitle{Modeling of AR~12760 with GX~Simulator}
\author[0000-0001-9632-447X]{Therese A.\ Kucera}
\affiliation{Heliophysics Science Division, NASA Goddard Space Flight Center, Greenbelt, MD 20771, USA}
\email[show]{therese.a.kucera@nasa.gov}  
\author[0000-0003-2846-2453]{Gelu M.\ Nita}
\affiliation{Center For Solar-Terrestrial Research, New Jersey Institute of Technology, Newark, NJ 07102, USA}
\email[show]{gnita@njit.edu}  
\author[0000-0003-2255-0305]{James A.\ Klimchuk}
\affiliation{Heliophysics Science Division, NASA Goddard Space Flight Center, Greenbelt, MD 20771, USA}
\email[show]{james.a.klimchuk@nasa.gov}  
\author[0000-0001-5557-2100]{Gregory D.\ Fleishman}
\affiliation{Center For Solar-Terrestrial Research, New Jersey Institute of Technology, Newark, NJ 07102, USA}
\email[show]{gfleishm@njit.edu}  
\affiliation{Institut f\"ur Sonnenphysik (KIS), Georges-Köhler-Allee 401 A, D-79110 Freiburg, Germany}

\begin{abstract}
In order to understand solar atmospheric heating it is important to test heating models against spatially resolved data from solar active regions. Here we model a small active region, AR~12760 observed on 2020 April 28, with the GX~Simulator package by fitting the extreme-ultraviolet (EUV) intensities in wave bands observed by Solar Dynamics Observatory's Atmospheric Imaging Assembly. We assume the temporally and spatially averaged heating rate along a loop has a power-law dependence on loop length, $L$ and average magnetic field strength along the loop, \Bavg.  We find that the best fit heating model for the 211~\AA\ band is $\Qavg\approx7\times 10^{-3} (\Bavg/{100\mbox{G}})^{1.5} (L/{10^9\mbox{cm}})^{-1}$~erg~cm$^{-3}$~s$^{-1}$ but that there is a range of parameters that give qualitatively reasonable fits, which we conclude is due to a correlation between \Bavg\ and $L$. In addition, we find that the models of the bands including cooler emission (131 and 171~\AA) greatly underestimate the extent of the emission in the legs of the longer loops at the peripheries of the active region that are the strongest contributors of the emission in those bands. We conclude that this is because the modeling assumes that all transition-region emission is confined to the loop foot points, but in reality the upper transition region of longer loops extends significantly farther into the loop. It is important to consider this aspect of the transition region in future efforts to model EUV emission. 
 \end{abstract}
 
\section{Introduction}
\label{s:intro}
 
Active regions are the chief source of solar emission in X-rays and much of the extreme ultraviolet (EUV), which are, in turn, critical for understanding how the Sun affects Earth's upper atmosphere.
Understanding the source of this emission and how to predict it is also closely connected to the long standing question of how the corona and transition region are heated.  There are many different aspects of this problem still to be addressed, but one key test of our understanding is the spatially resolved modeling of the emission from a specific active region in three dimensions. 

Active region modelers often take the approach of applying and constraining scaling laws describing heating to a set of model loops. Here, ``loop'' refers to a magnetic flux tube, not necessarily a bright feature in an image. There is a diffuse component of the corona comprised of observationally indistinguishable loops. There are many theoretically based relationships connecting the heating to quantities like loop geometry, magnetic field, and currents \citep{mandrini_00,cranmer_19} to which the results can be compared \citep{gudiksen_02,gudiksen_05a,schrijver_04,mok_05,mok_08,mok_16,warren_06,lundquist_08a,lundquist_08b,ugarte_17} for the purpose of bolstering or discarding particular heating models. Of special interest to us is the use of scaling laws based explicitly on episodic ``nanoflare'' heating models, employed, for instance, by \citet{barnes_19,mondal_25}.

Another technique is to heat the corona using self-consistent MHD simulations in which coronal heating is produced by numerical resistivity and viscosity \citep{gudiksen_02,gudiksen_05a,bourdin_13,rempel_17,warnecke_19}.  This has the advantage of creating a self-consistent model, sometimes called an \textit{ab initio} model, which does not rely on assumptions about coronal heating. However, the drawback of this approach is that the resistivity and viscosity in the models are orders of magnitude larger than solar values, and the mechanism of heating may be fundamentally different from the true solar mechanism, e.g., relatively passive Ohmic dissipation versus explosive magnetic reconnection.

In contrast to the works above that have their main (although not exclusive) focus on heating parameters, \citet{plowman_21,plowman_23} made fairly simple assumptions about the distribution of emission in loops so as to use the EUV emission to constrain the magnetic field.

In this paper we focus on testing the uses of the GX~Simulator model to simulate the emission of an active region observed in EUV in multiple wave bands by the Solar Dynamics Observatory's Atmospheric Imaging Assembly (SDO/AIA) in order to determine the dependence of emissions on loop length, magnetic field strength and average heating rate. 
GX~Simulator was originally developed for the purpose of modeling flare loops in microwaves and X-rays \citep{nita_15}, but has since been expanded to model active regions in radio and EUV \citep{nita_18,fleishman_21b,nita_23, Fleishman_2025,kuznetsov_25}. GX~Simulator uses  scaling-law based heating combined with magnetic field extrapolation calculated using a vector magnetogram (see Section~\ref{s:Process}). One advantage to GX~Simulator over related approaches is that it computes the emission from every voxel in the volume modeled rather than along particular loops extrapolated upward from a grid of footpoints on the photosphere. Models of the latter type either have large emission-free gaps between loops in the corona or non-physical loop overlap at low altitudes.

In Section~\ref{s:Process} we describe the process used  to model AR~12760 with GX~Simulator, including the magnetic field, plasma properties, and fitting of the model to the EUV intensity distribution of the active region. Section~\ref{s:FitResults} describes the fit results, with special focus on the 211 and 171~\AA\ bands. In Section~\ref{s:Discussion} we discuss the results including their indication of a relationship between the magnetic field and loop length, the importance of correctly modeling the upper transition region, and ways in which the modeling might be improved. In Section~\ref{s:Summary} we summarize our results.

\section{Modeling Process}
\label{s:Process}
\subsection{Overview}

GX~Simulator is described extensively by \citet{nita_18,nita_23} and is available as the gx\_simulator package in Solar Soft  \citep{freeland_98}. Here we discuss the points most relevant to this paper, starting with a very general overview and adding details in Sections~\ref{s:euvdata}-\ref{s:Fitting}.

Our goal is to calculate the dependence of the temporally and spatially averaged volumetric heating along a loop, \Qavg, as a function of the mean magnetic field, \Bavg, and loop length, $L$, on the assumption that it can be described by:
\begin{equation}
 \Qavg  = q_0 \bigg(\frac{\Bavg}{B_0}\bigg)^a \bigg(\frac{L}{L_0}\bigg)^{-b} \mbox{erg~cm$^{-3}$~s$^{-1}$}
 \label{e:Qavg}
\end{equation}
 where $B_0 = 100$~G,  $L_0 = 10^9$~cm and $q_0$ is the typical heating rate.  
  \begin{figure}
 \includegraphics[height=10cm, trim=100 20 100 50]{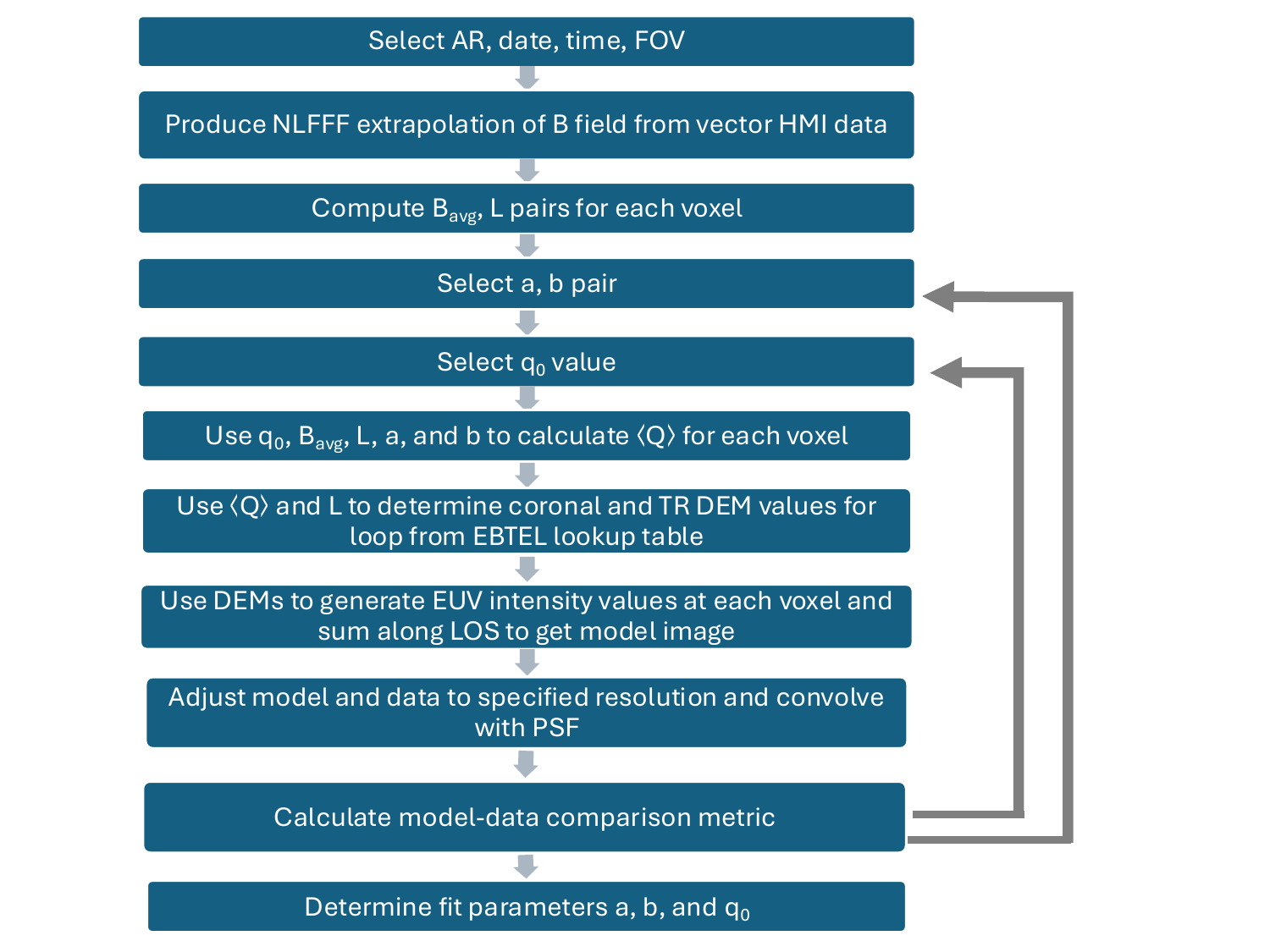}
 \caption{Flowchart showing the overall process used to fit EUV data with GX~Simulator}
 \label{f:flowchart}
 \end{figure}
 
The basic process is set out in Figure~\ref{f:flowchart}, and details are presented in the sections below. For a given active region, a 3D magnetic field model is developed via a nonlinear force-free field (NLFFF) extrapolation. The values for $L$ and \Bavg\ for each voxel are determined from the field line closest to the center of the voxel. Then, for a given $a$, $b$, and $q_0$ it is possible to calculate \Qavg\ using Equation~\ref{e:Qavg}. The code utilizes a lookup table, to determine the differential emission measures (DEMs) of the corona and transition region for the loop as a function of \Qavg\ and $L$. The DEM values are convolved with the instrument response function for a given EUV band to determine the amount of emission in that band for a particular voxel. The emission in each voxel along the line of sight is then integrated to produce a 2D model image. This image is compared to the EUV data to determine the best fit as a function of $a$, $b$, and $q_0$. 

Files needed to reproduce the modeling, including the input SDO/AIA data, fitting masks, magnetic field models, and GX~simulator result files are posted at doi:\href{https://doi.org/10.5281/zenodo.16896515}{10.5281/zenodo.16896516}.

\subsection{EUV Data}
\label{s:euvdata}
We selected for study Active Region (AR)~12760, a small compact active region that crossed the meridian  on 2020 April 28. It was chosen because of its compact structure, location near Sun center, lack of sunspots, and low level of activity. It was also the only significant active region on the Sun at the time, making it a good candidate for comparison with full-sun spectral irradiance measurements that, although not a part of this paper, pertains to a longer-term goal of a larger project. 

\begin{figure}
 \includegraphics[height=3.8cm, trim=40 10 40 10]{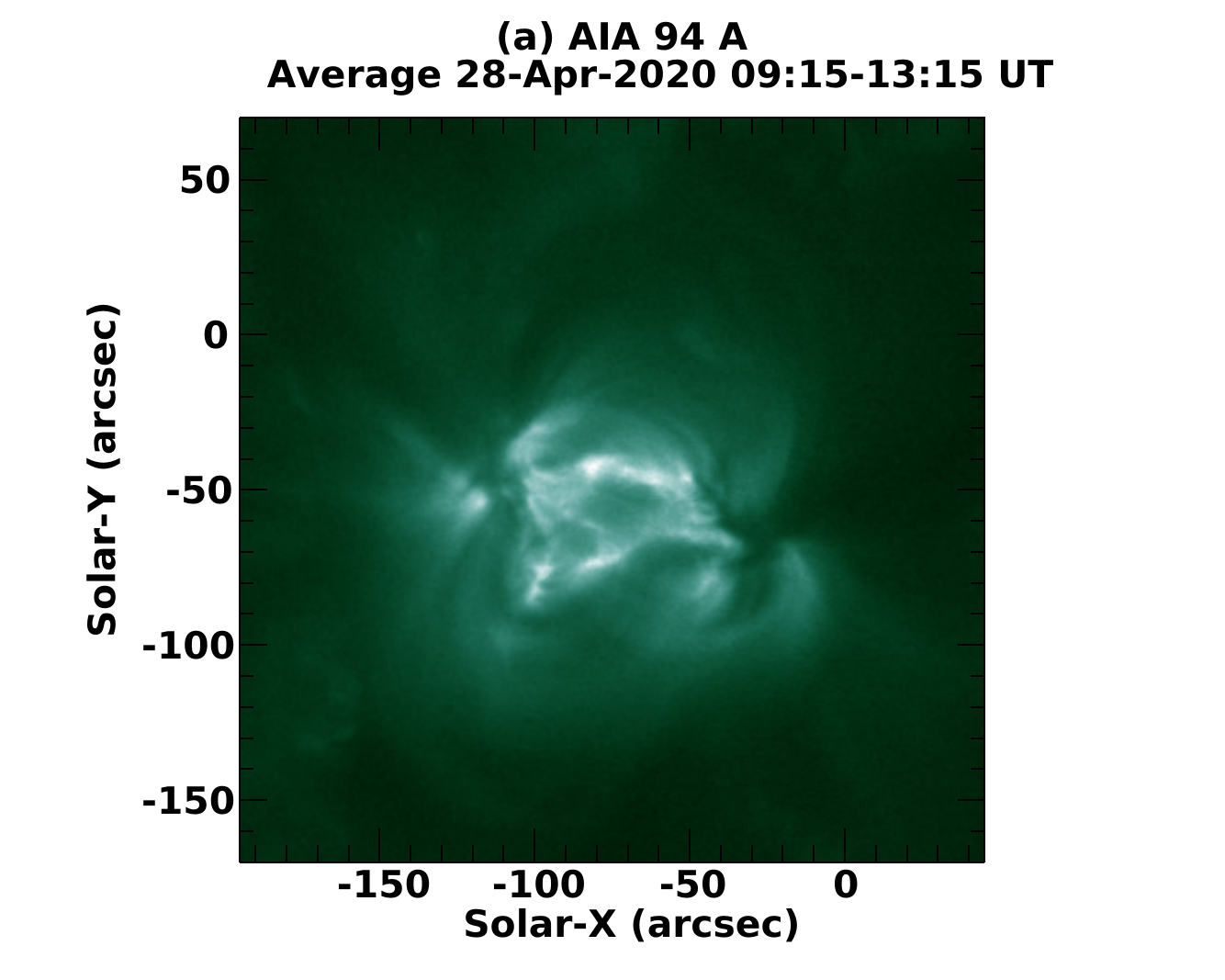}
 \includegraphics[height=3.8cm, trim=40 10 40 10]{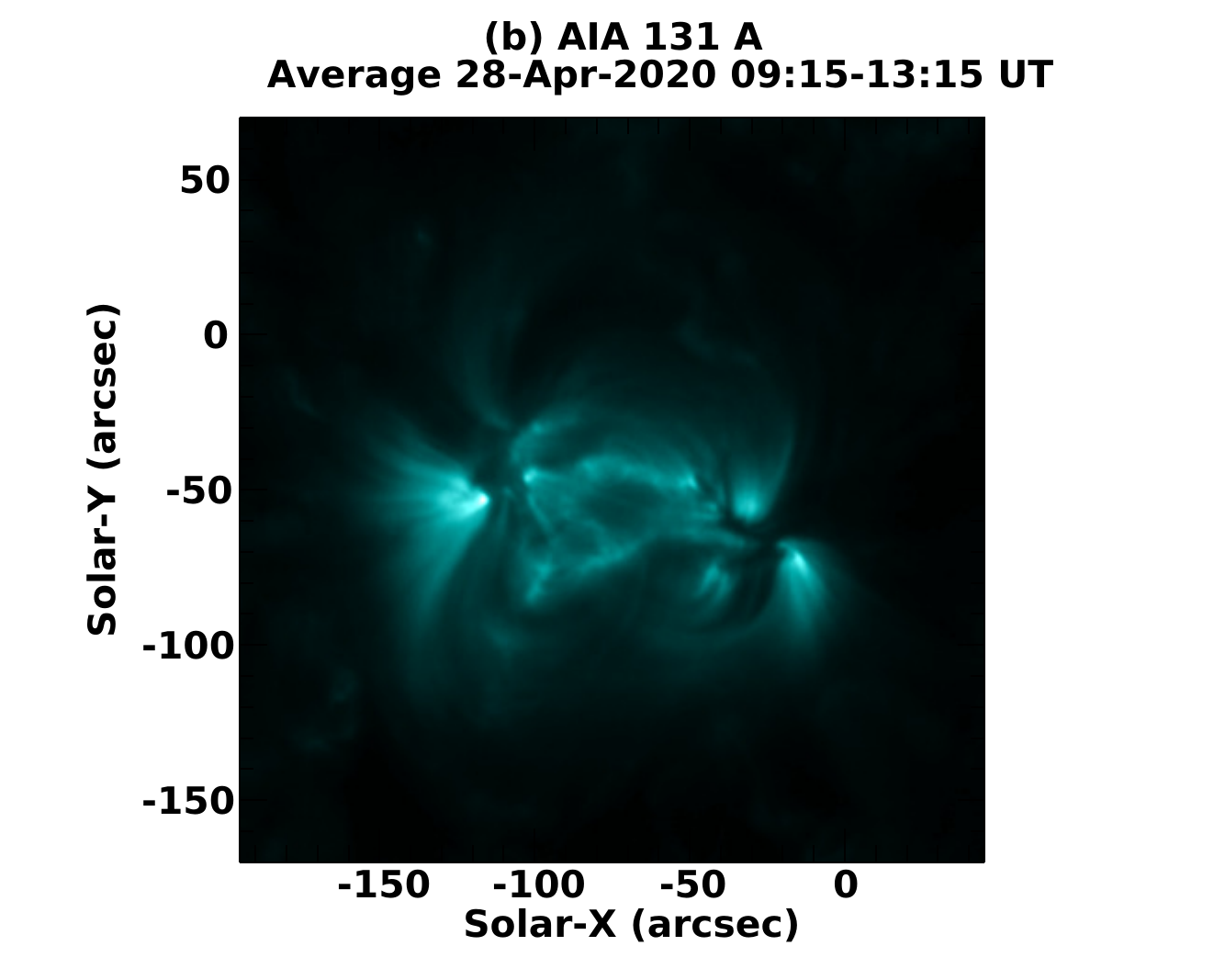}
 \includegraphics[height=3.8cm, trim=40 10 40 10]{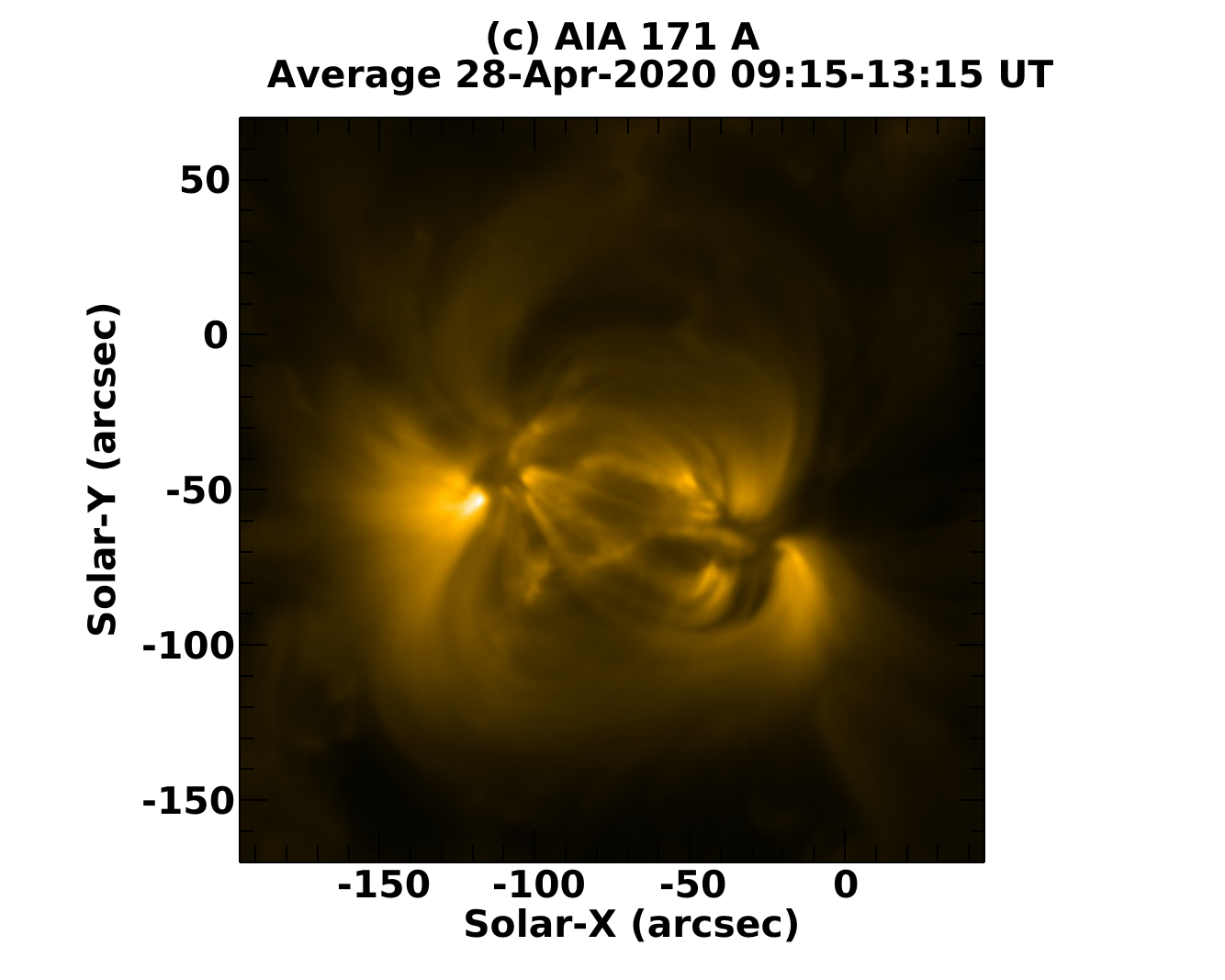}
 \includegraphics[height=3.8cm, trim=40 10 40 10]{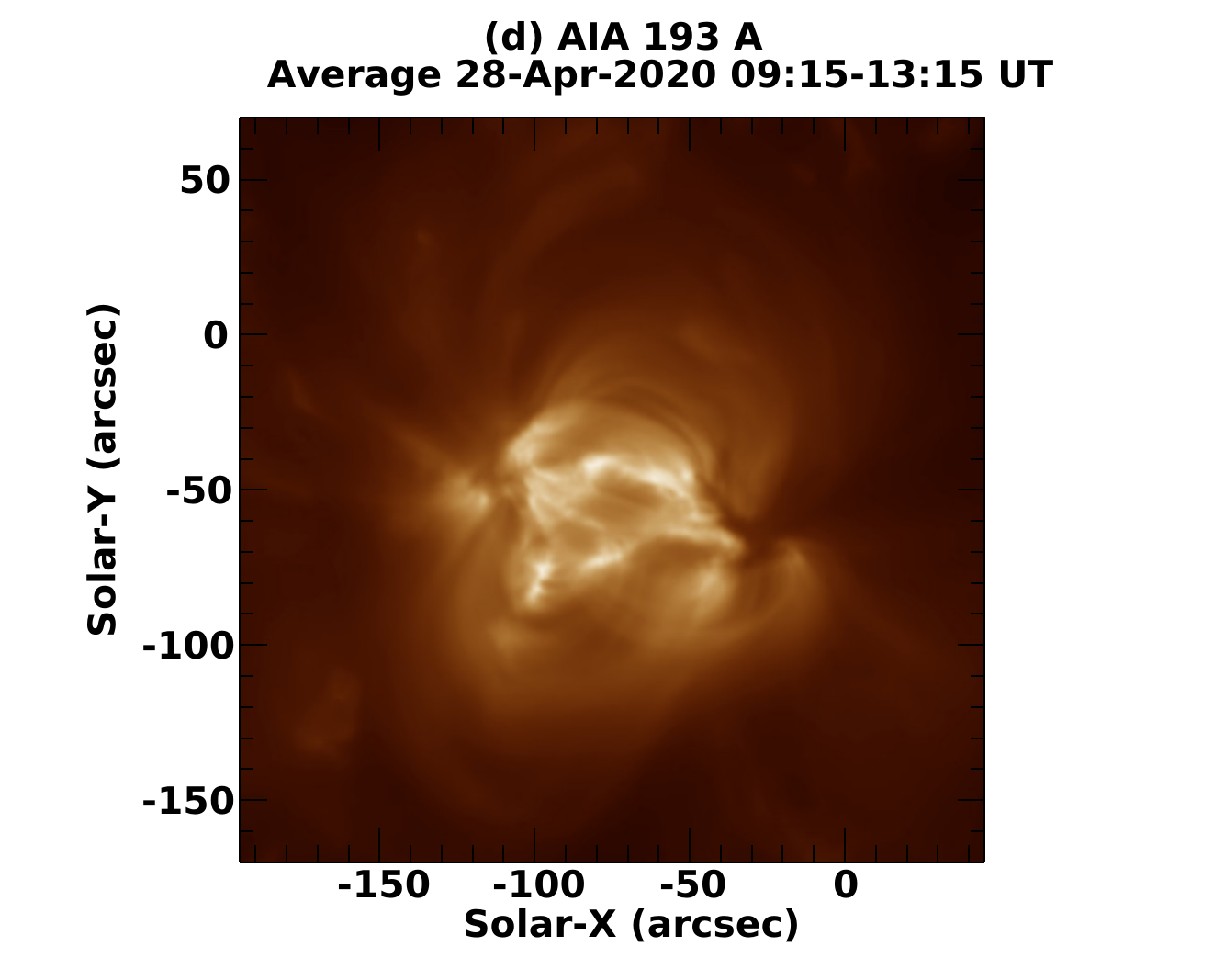}\\
 \includegraphics[height=3.8cm, trim=40 10 40 10]{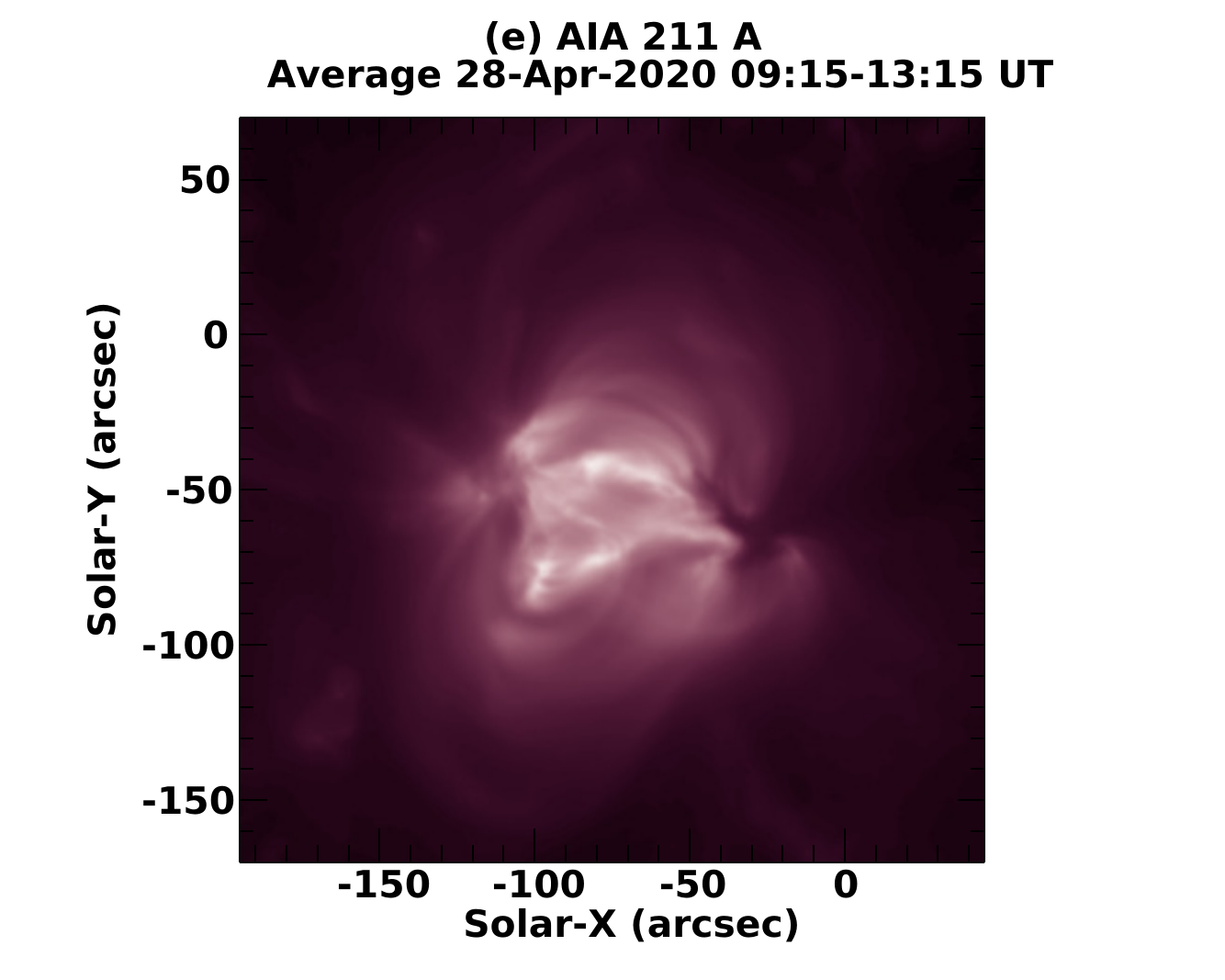}
 \includegraphics[height=3.8cm, trim=40 10 40 10]{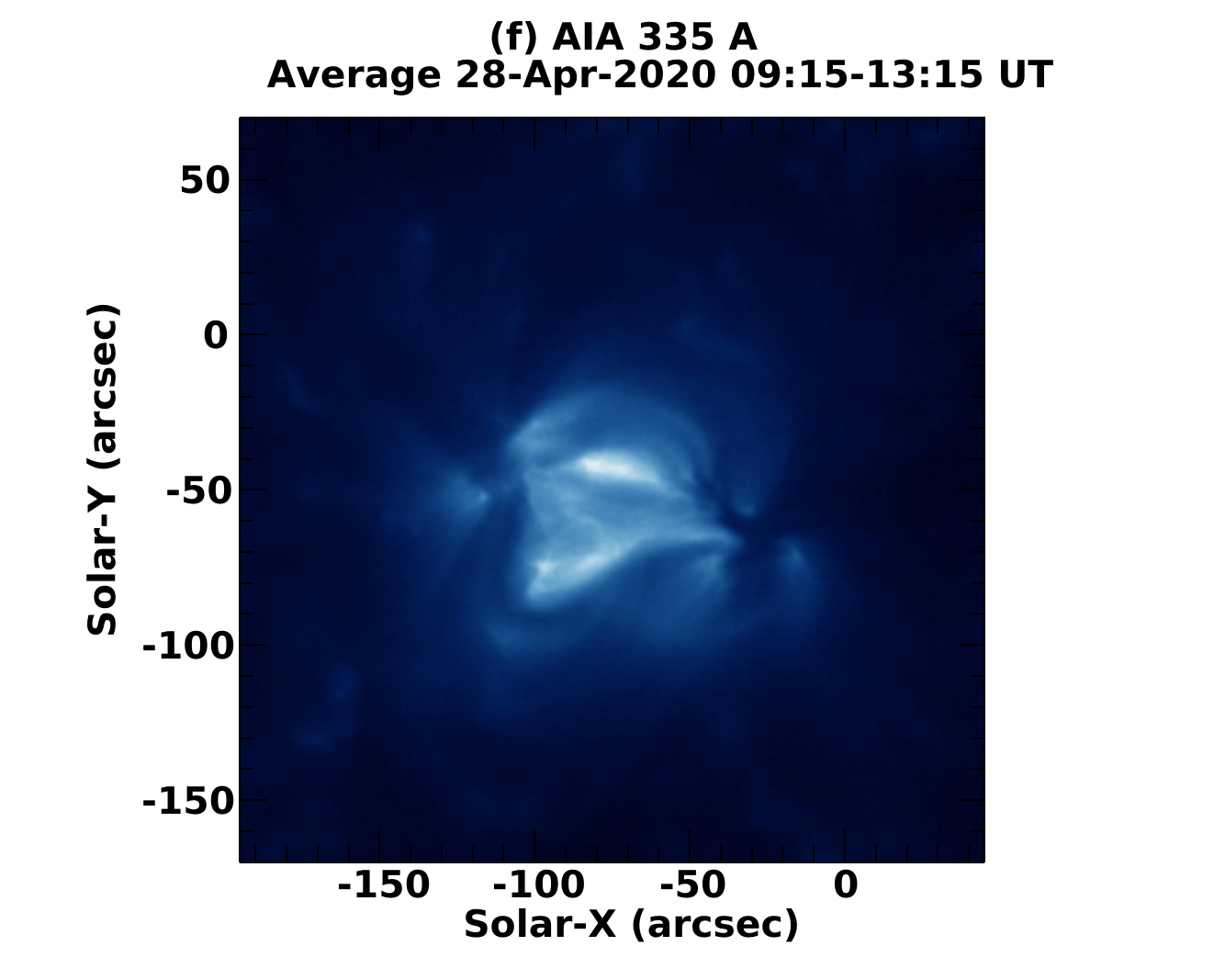}
 \includegraphics[height=3.8cm, trim=40 10 40 10]{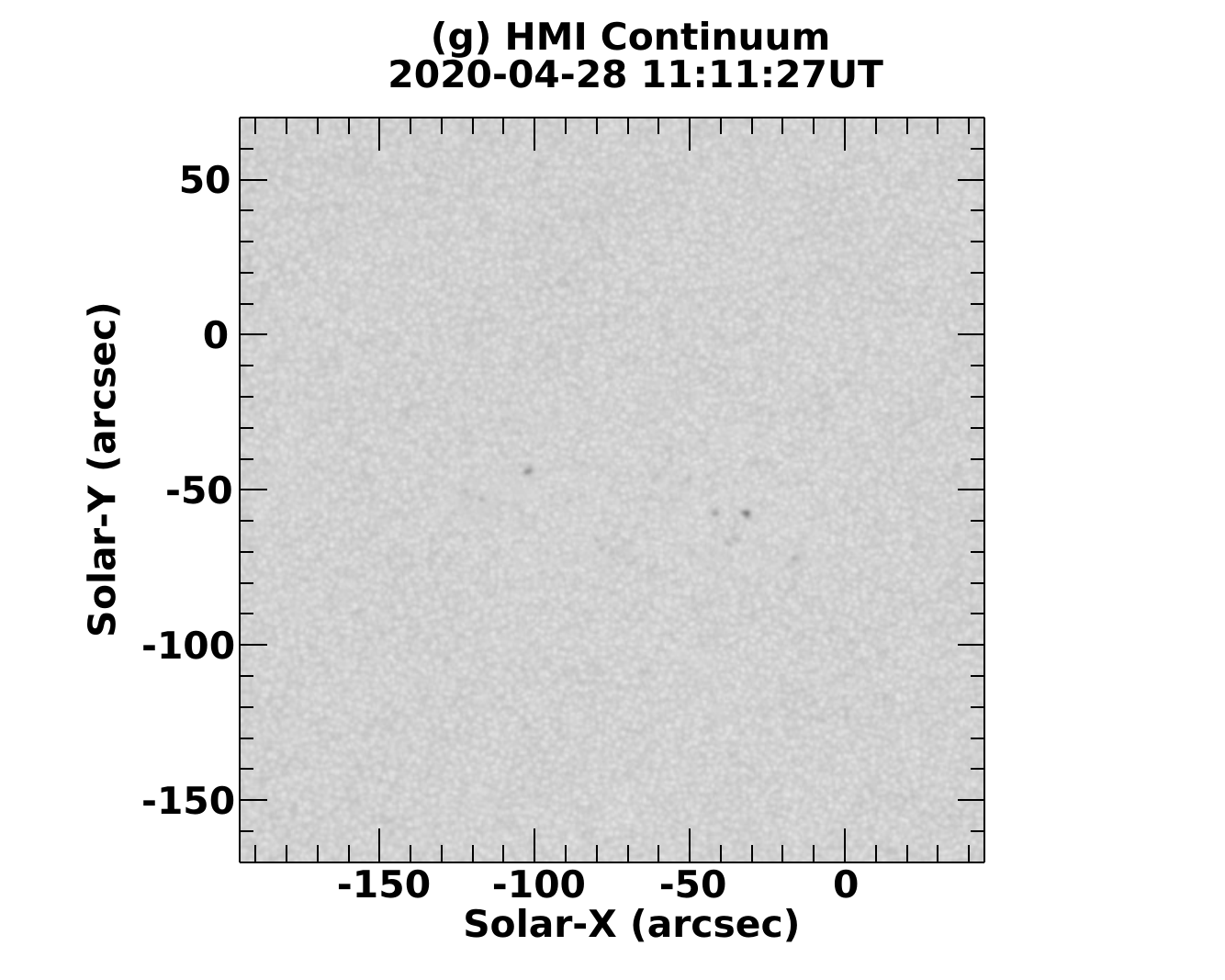}
 \includegraphics[height=3.8cm, trim=40 10 40 10]{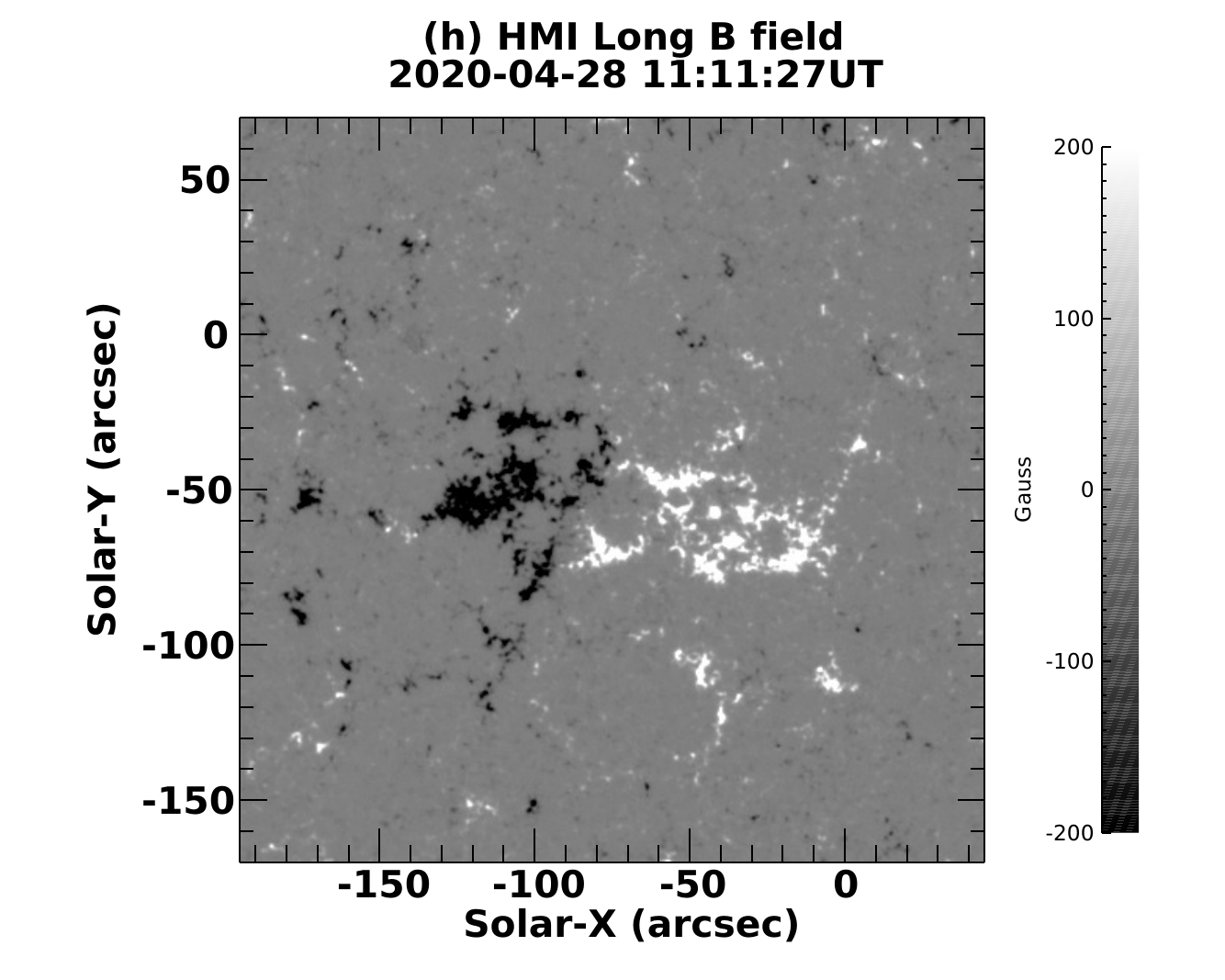}
\caption{(a)-(f) averages of SDO/AIA intensity for the 94, 131, 171, 193, 211, and 335~\AA\ bands from  9:15-13:15 UT shown with linear scaling. (g) HMI intensity and (h) longitudinal magnetic field data at 11:11 UT.}
\label{f:aia_img}
\end{figure}

In order to constrain the heating parameters we use data from the 94, 131, 171, 193, 211, and 335~\AA\ bands of Solar Dynamics Observatory's Atmospheric Imaging Assembly (SDO/AIA; \citet{pesnell_12,lemen_12}). 
Because we wish to simulate the general, time averaged behavior of the active region rather than its emissions at any one moment, we use an average of the AIA data at a 5 minute cadence from  9:15-13:15~UT (see Figure~\ref{f:aia_img}). This time period is slightly before the region crossed the meridian, but was selected to avoid small instances of emerging flux that occurred later in the day and could have affected the magnetic model. Each image was corrected for solar rotation before averaging, aligning it to the region's location at 11:15 UT. 

\begin{figure} 
\includegraphics[height=12cm, trim=40 10 40 10]{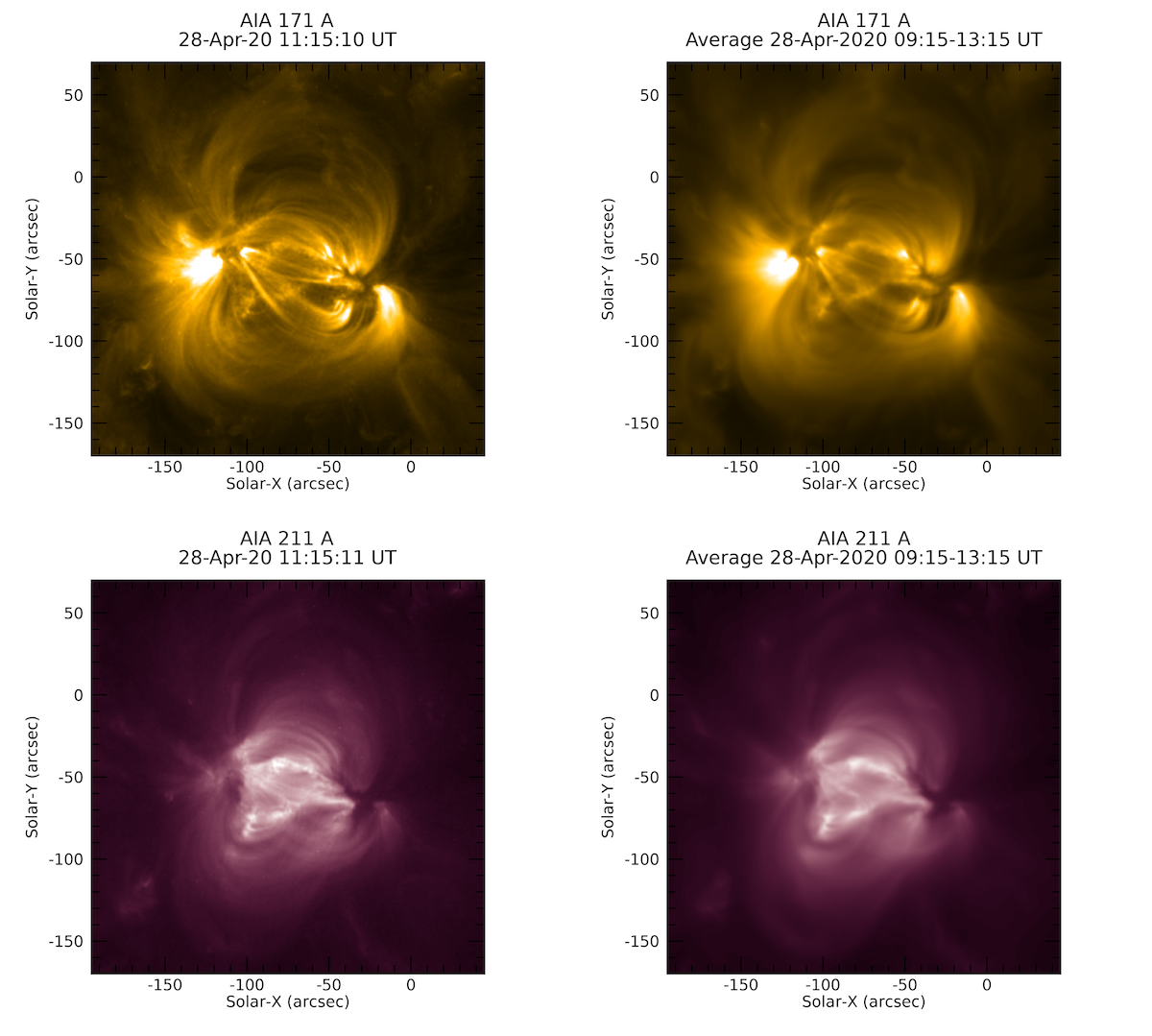}
\caption{The 171 and 211~\AA\ images at 11:15~UT (left) compared with the average images made using one image every five minutes from 09:15-13:15 UT (right). The on line-version of this paper includes an animation with the equivalent frame showing each image (one every 5 minutes from 09:15-13:15 UT) used to calculate the average. The video is 4~s long. }
\label{f:AIA1115vsAvg}
\end{figure}

At any given time, the observed active region corona is composed of a diffuse component and a component of bright loops (on the order of 1~Mm, or a few arcseconds in width in the AIA images). There is more total emission in the diffuse component than the loop component, as loops are typically only modest (10-40\%) enhancements over the background \citep{klimchuk_20}.  
These bright, thin loops appear and disappear at different locations with lifetimes of several thousand seconds. Over a 4-hour average they tend to combine to form larger structures (with widths on the order of $10\arcsec$), as can be seen in Figure~\ref{f:AIA1115vsAvg}, which compares individual images taken at 11:15~UT with the average over 4 hours. A successful GX~Simulator model would reproduce both the diffuse component and this medium-scale structure. Note that there is an abundance of theoretical evidence that both components are comprised of ultra-thin, spatially unresolved loops \citep[e.g.,][]{klimchuk_23a,johnston_25}.   

\subsection{Magnetic Field Model}
\label{s:BfieldModel}

To determine the coronal magnetic field we use the Automatic Model Production Pipeline \citep[AMPP,][]{nita_23} initiated with the SDO/Helioseismic and Magnetic Imager (HMI; \citet{scherrer_12}) vector magnetic field map from 11:12 UT.  

The magnetic field is calculated using a nonlinear force free field extrapolation as described in \citet{fleishman_17} and based on the methods of \citet{wiegelmann_04}. Following tests performed by \citet{fleishman_17} we did not apply any preprocessing to the photospheric boundary condition. The top and side boundary conditions are taken from the initial potential-field extrapolation that is initiated by the $B_z$ bottom boundary condition. To minimize the effect of these boundary conditions on the NLFFF solution, we follow \citet{wiegelmann_04} and employ buffer zones at the top and side boundaries that comprise 10\% of the corresponding size of the data cube as detailed in \citet{nita_23}. The C++ source code, along with set of compiled libraries for Windows, Linux, and Mac OS platforms, their calling IDL wrappers, and the GUI application, AMPP, are included in the \gx\ SSW distribution package. These source codes and platform specific compiled libraries are automatically updated from their independently maintained GitHub development repository\footnote{  \href{https://github.com/Alexey-Stupishin/Magnetic-Field_Library}{Magnetic-Field\_Library}}. A Python utility based on the same library is also available from GitHub \citep[pyAMPP,][]{yu_25}. In addition, this NLFFF package may be directly downloaded from a Zenodo digital repository
\citep{stupishin_20}\footnote{\href{https://doi.org/10.5281/zenodo.3896222}{Magnetic Field Library: NLFFF and magnetic lines}}.
The magnetic field strength in this active region is fairly low, making it somewhat difficult to model in detail, as will be discussed in Section~\ref{s:FitResults}.

\begin{figure}
 \includegraphics[height=8cm, trim=0 20 0 10]{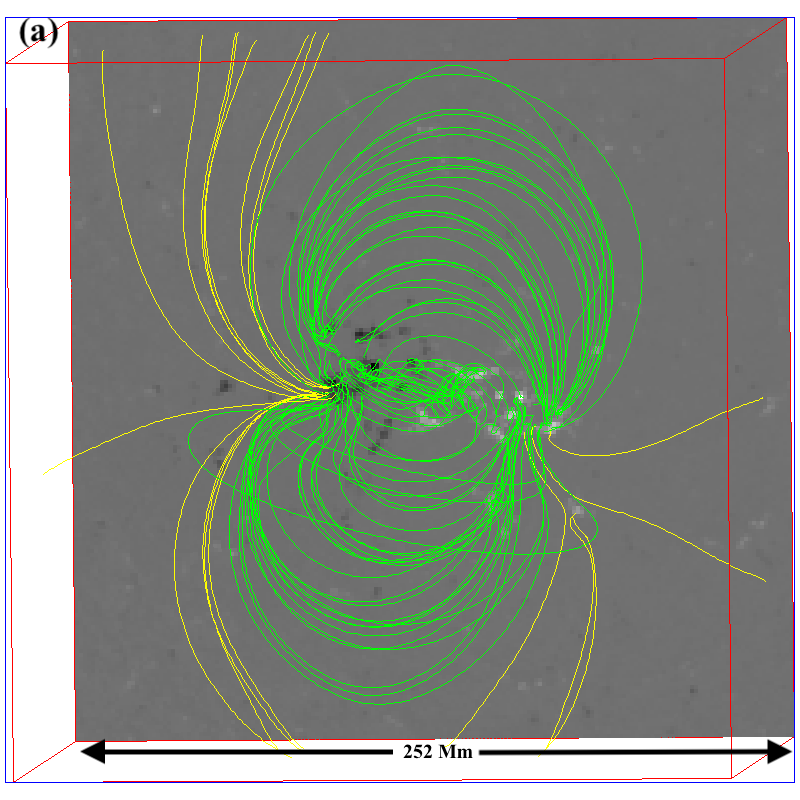}
 \includegraphics[height=8cm, trim=0 20 0 10]{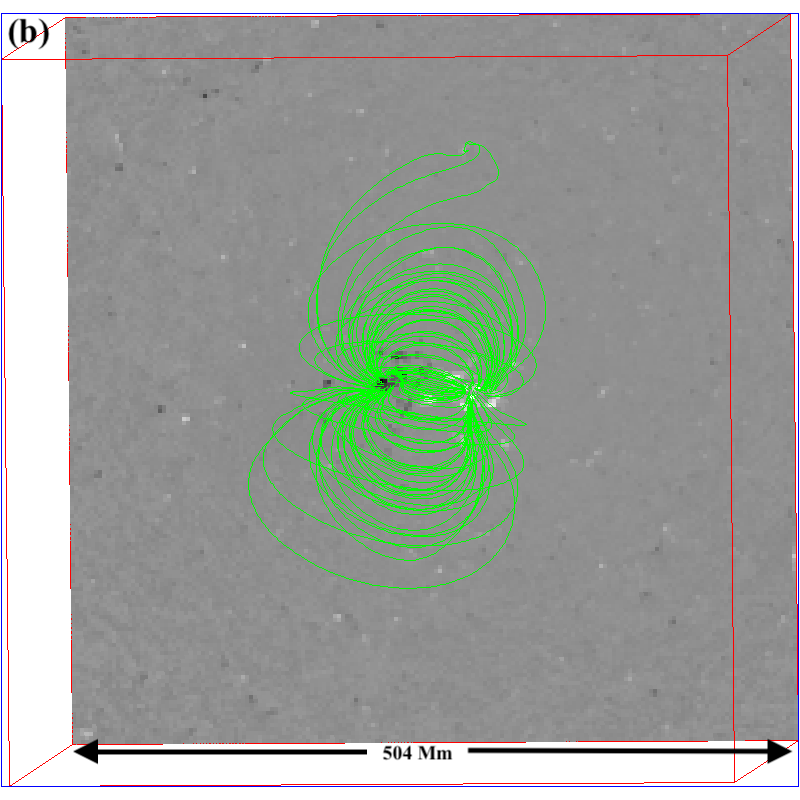}
 \caption{Magnetic field lines from (a) Magnetic Model~1  and (b) Magnetic Model~2 superimposed on a map of the radial photospheric field. Field lines closing within the model box are shown in green while ones that do not close are shown in yellow.}
 \label{f:Bfield_topdown}
\end{figure}

We developed two models, each of which is a compromise between resolution and field of view (FOV).  
Both models were centered at 2020 April 28 11:12 UT at $x=-75\arcsec$, $y=-50\arcsec$, and the point of view derives from those coordinates. The AIA data were adjusted to match that time. Thus, the view of the active region we are modeling is quite close to the actual view of AIA and not a simple top-down point of view. The models consisted of cubes with 180 cells in the $x$ and $y$ directions and 200 cells in the $z$ direction. 
Magnetic Model~1 (Figure~\ref{f:Bfield_topdown}a) has a cell size of 1400~km so as to better model the core field lines but some of the longer loops extending from the active region periphery are ``open'' and thus not included in our modeling of the emission (see Section~\ref{s:Plasma} below). 
Magnetic Model~2 (Figure~\ref{f:Bfield_topdown}b), conversely, covers an area twice as large while increasing the cell size to 2800~km to remain within the memory constraints of the computer being used. Model~2 includes the long loops, but produces a more simplified model of the core of the active region. 
Sample field lines from both models are shown in  Figure~\ref{f:Bfield_topdown} in which field lines that close within the model box are shown in green, while ones that extend outside the box are in yellow. 

\subsection{Loop Plasma}
\label{s:Plasma}

The active region volume is divided into voxels and a value for \Bavg\ and $L$ is assigned to each voxel using the NLFFF model. Each voxel is characterized using the field line passing closest to its center, providing a value for \Bavg\ and $L$. The plasma properties in the loop are determined using a look-up table as a function of $L$ and \Qavg, determined using the Enthalpy-Based Thermal Evolution of Loops (EBTEL, \citet{klimchuk_08,cargill_12a,cargill_12b}) model with a specific distribution of heating events. 
EBTEL is a ``0D'' model that computes the evolution of the coronal-averaged temperature, pressure, and density in a loop without spatial resolution. 
The instantaneous coronal DEM is assumed to be distributed uniformly over a temperature interval centered on the average temperature and having a width equal to twice the difference between the apex and average temperature. Because the time-averaged coronal DEM is determined by the behavior of many unresolved loops over time, it is reflective primarily of longer term loop evolution and not the instantaneous temperature distribution. Thus, we use the time average of long-duration EBTEL simulations that encompass many nanoflare energy releases. The time-averaged DEM is spread uniformly along the entire coronal portion of the loop. The EBTEL runs currently used to calculate GX~Simulator input tables assume no loop expansion.

For the transition region EBTEL computes a separate lookup table with the detailed DEM distribution at each time in the simulation based on energy balance and averages them to produce DEM curves for the \Qavg\ and $L$ grid.  EBTEL takes the temperature boundary between the corona and transition region to be the value where thermal conduction switches from a cooling term above to a heating term below. This temperature is roughly 60\% of the apex temperature. The lower and middle transition region by temperature are confined to low altitudes, but 1D loop modeling indicates that the upper transition region extends an appreciable distance from the loop footpoint. This plays an important role in the model images, as we discuss in Section~\ref{s:Discuss_TR}.
 
\begin{figure}
 \includegraphics[height=6cm, trim=40 10 40 10]{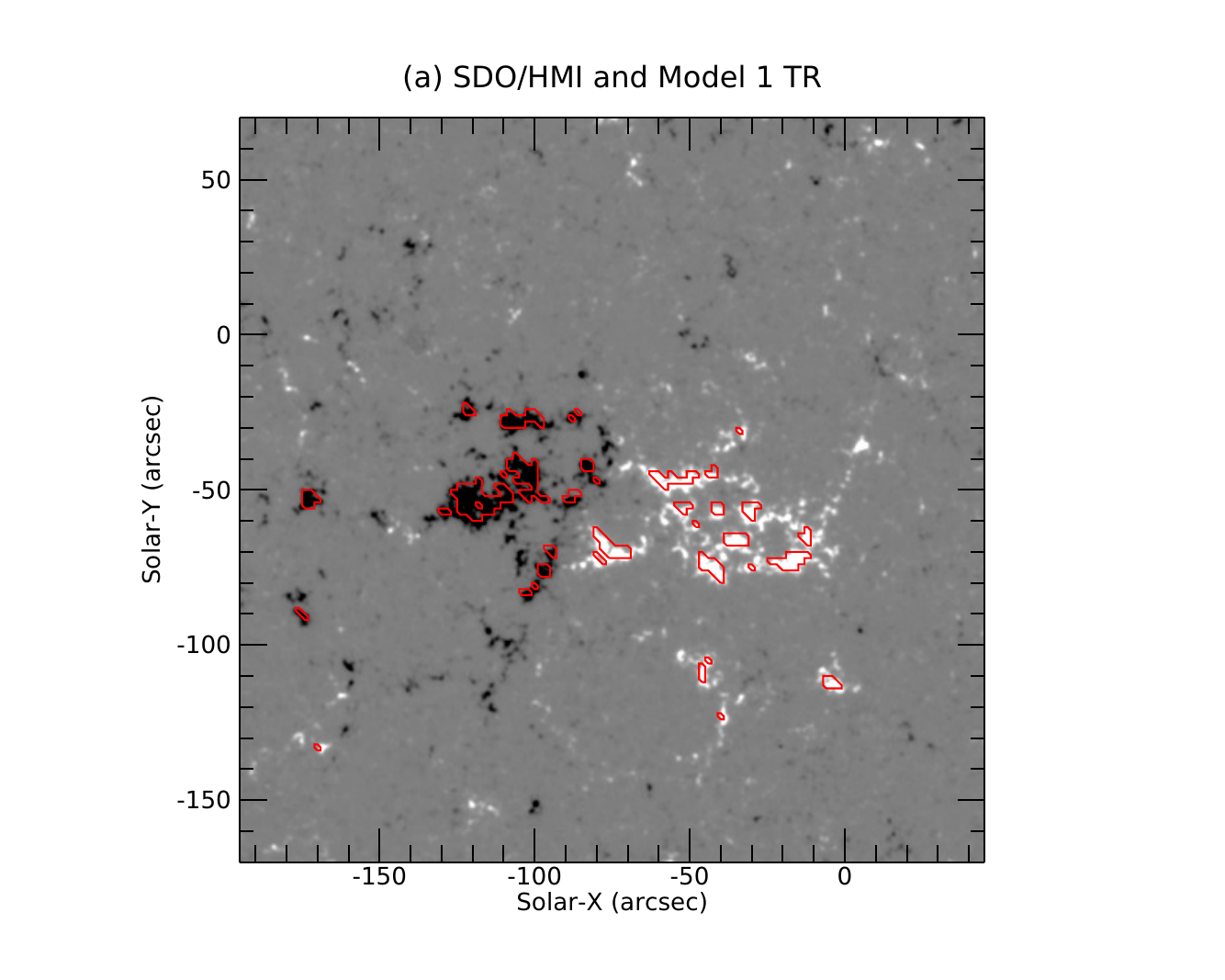}
 \includegraphics[height=6cm, trim=40 10 40 10]{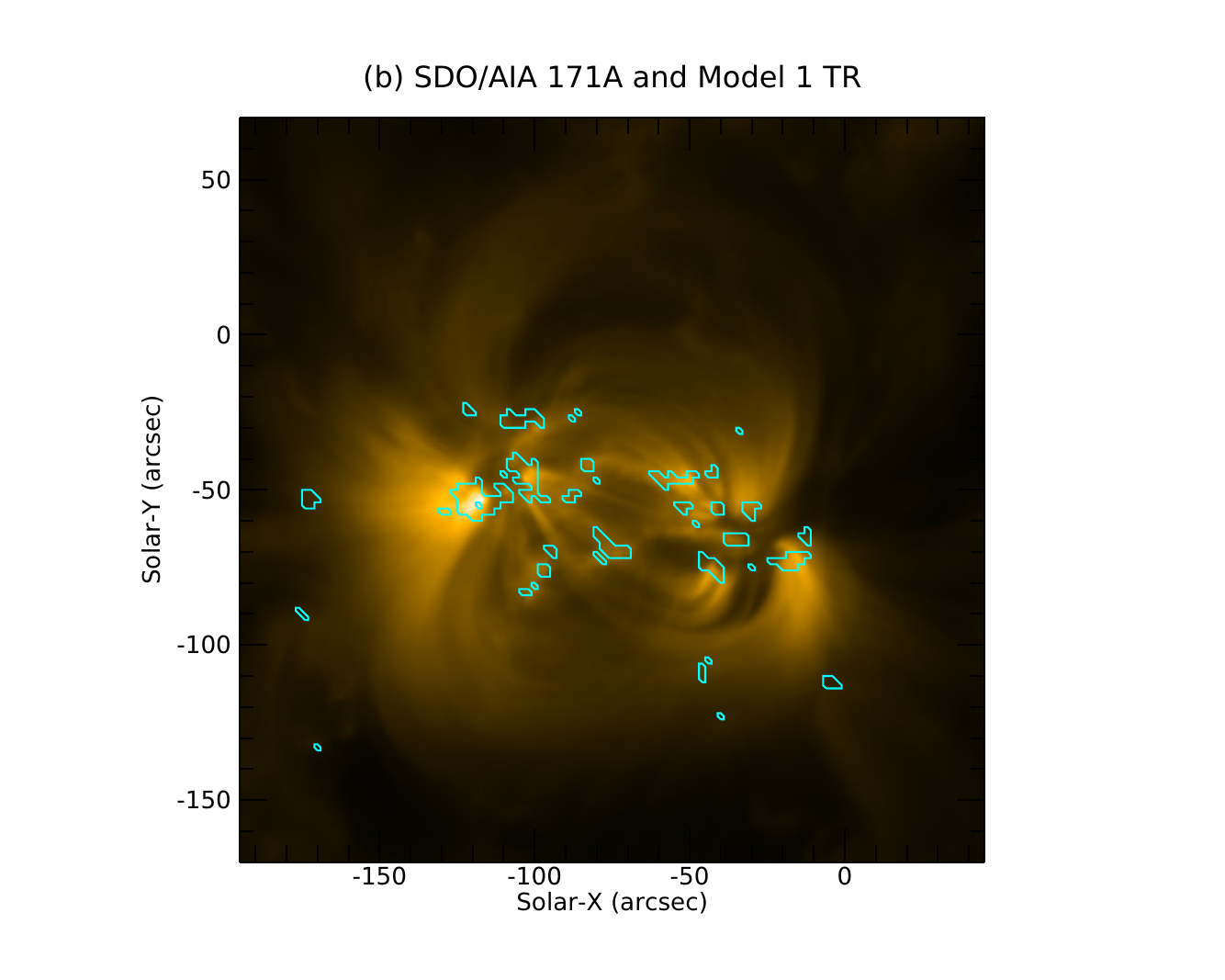}
\\
 \includegraphics[height=6cm, trim=40 10 40 10]{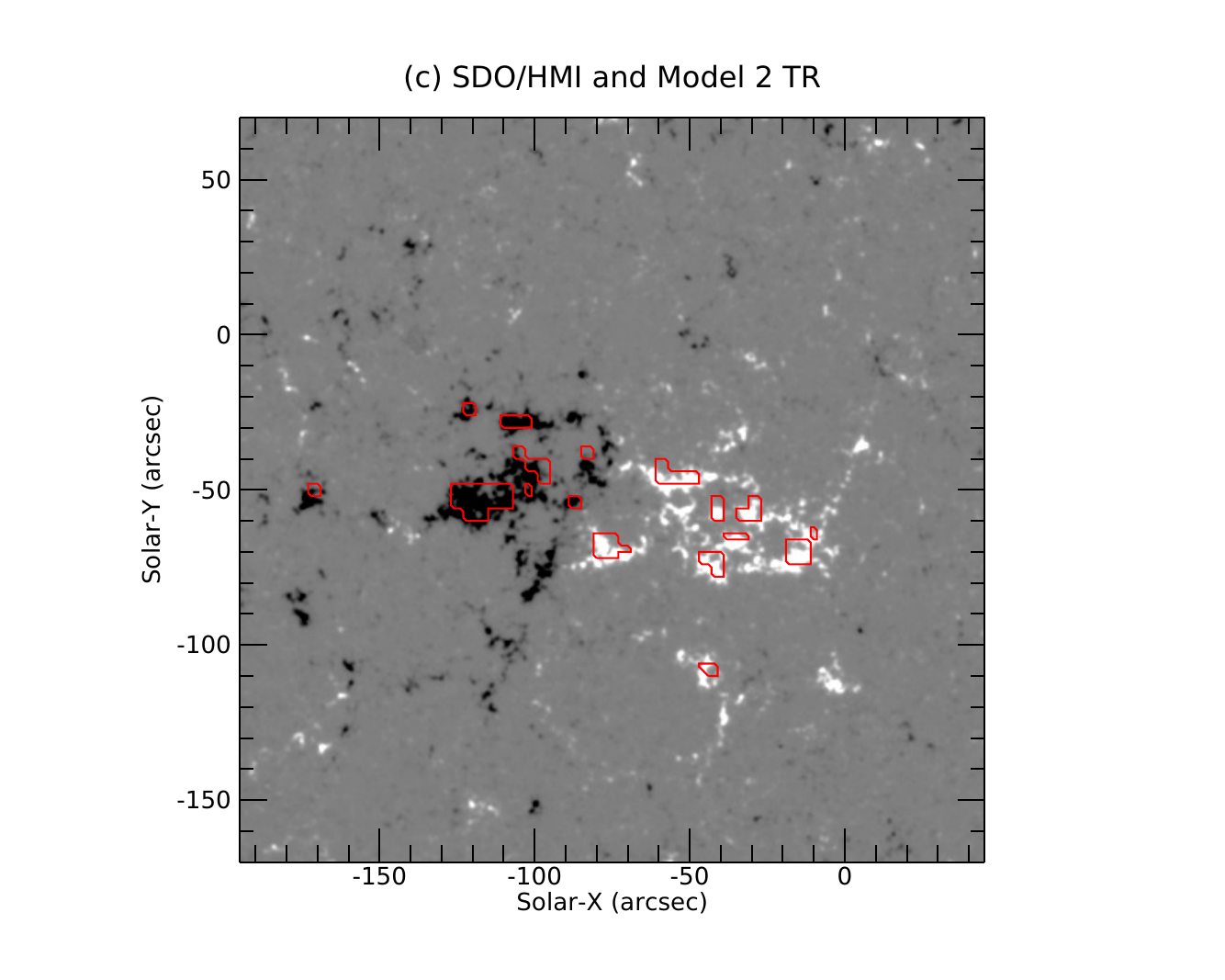}
 \includegraphics[height=6cm, trim=40 10 40 10]{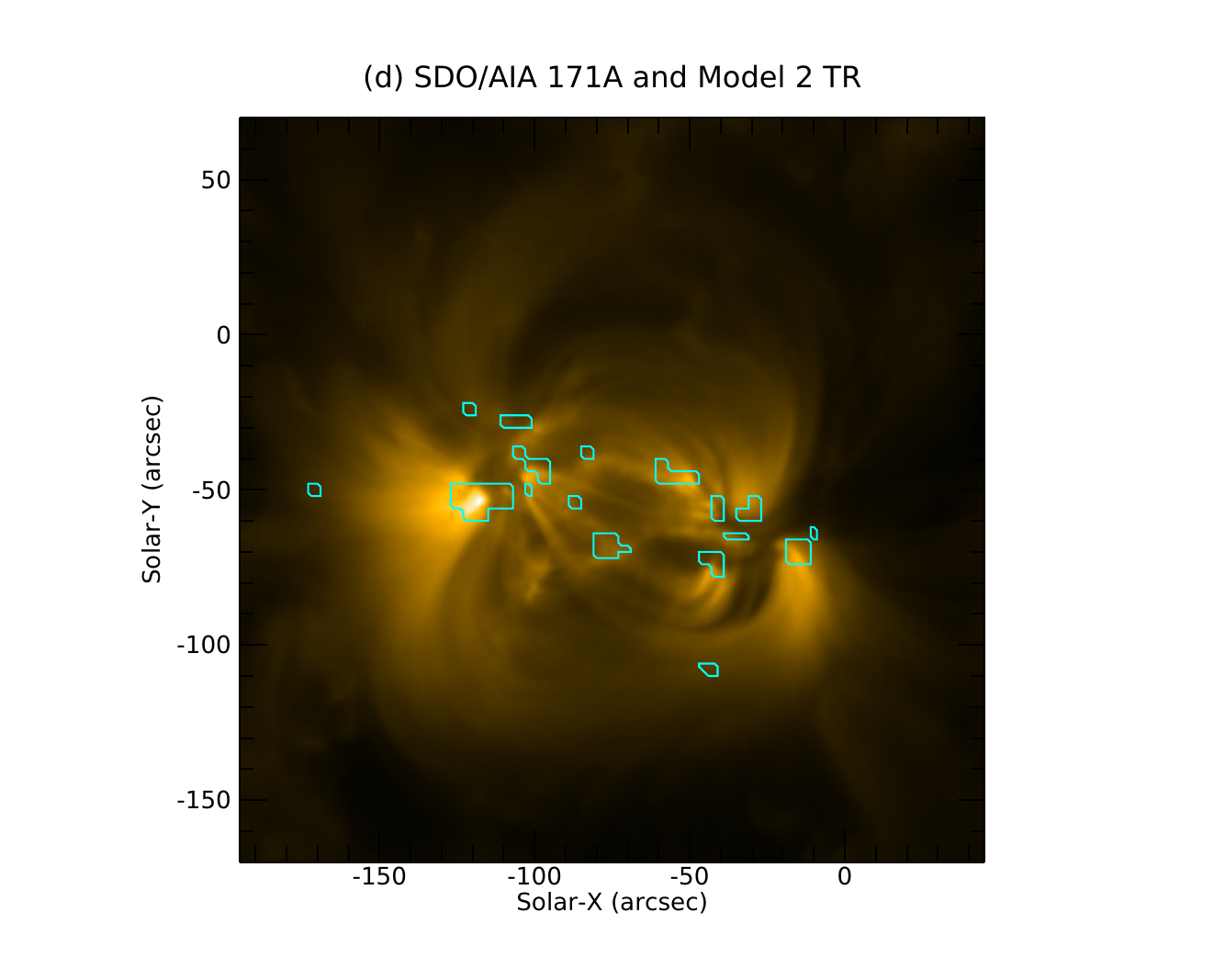}
\caption{Contours show regions in which it was assumed there was transition-region emission using Magnetic Model~1 (top row) and Magnetic Model~2 (bottom row) superimposed on a HMI longitudinal magnetogram (left) and AIA 171~\AA\  image (right).}
\label{f:TRLoc}
\end{figure}

In order to produce a lookup table of coronal and transition-region DEMs as a function of \Qavg\ and $L$, a large number of EBTEL runs are done. The $60 (Q) \times49 (L)$ grid provides logarithmically spaced values in the range $L/2 = 1$ to $6.3\times10^4$~Mm. For each $L/2$ value there is a different range of \Qavg\, with \Qavg\ ranging from 0 to 794 erg~cm$^{-3}$~s$^{-1}$ for $L/2=1$ and from  $1.5\times10^{-8}$ to 0.0126 for $L/2 = 6.3\times10^4$~Mm.  As discussed above, for each set of parameters a long run is averaged in order to represent a loop as a group of smaller strands heated randomly by an ensemble distribution of heating events.
The currently available EBTEL lookup tables in GX~Simulator are described in \citet{nita_23}. For this work we use ebtel\_scale=1\_alpha=-1.sav, which assumes a random distribution of nanoflares with an average time between nanoflares equal to the loop cooling time (scale = 1) and a power law distribution in the delay between events  with index $\alpha = -1$.  The energy  of the events is proportional to the time between them so that the energy also has a power-law distribution with the same index.

GX~Simulator calculates the transition region and corona emission independently and then sums them.
For each voxel the DEMs are determined from the EBTEL table using a bilinear interpolation. GX~Simulator places all the transition-region emission in a single layer at the bottom of the model cube.
We assume a loop with uniform cross-sectional area. This can affect the emission and the ratio between coronal and transition-region emission in particular \citep{cargill_22}. The transition region is relatively fainter in expanding loops, but how adding expansion would affect our model is a complex topic, as discussed in Section~\ref{s:Discuss_LoopExpansion}.

Currently the way GX~Simulator can model magnetic field lines that leave the model box is to either assume they contain plasma in hydrostatic equilibrium based on an assumed temperature and base pressure or to exclude them entirely. We have elected to leave out emission from these ``open'' field lines  (open in the sense that they leave the box, not necessarily open to the solar wind).
 In addition, the emission was also set to zero for voxels centered on a small number of very long field lines not within the boundaries of the \Qavg,$L$ lookup table and in cases in which the fitting algorithm was attempting to use unrealistically low values of $q_0$ in an effort to find a good fit.  The boundaries of the EBTEL table do not affect the results greatly in the parameter ranges with good fits to the data, although we discuss a possible issue related to open field lines in Sections~\ref{s:fit171} and \ref{s:Discuss_TR}.

To calculate the emission in each voxel the DEM was convolved with the temperature response function currently available in the routine provided by the AIA team, aia\_get\_response.pro, which uses the \citet{feldman_92a} coronal abundances.

\subsection{Fitting of the Model}
\label{s:Fitting}
The fit is done by comparing on a pixel by pixel basis the SDO/AIA image with the intensity image calculated by GX~Simulator.  We fit the $240\times240\arcsec$ region show in Figure~\ref{f:aia_img}. Most of the calculated model images had a resolution of 2\arcsec\ (resulting in a $120\times120$ pixel field of view), and were convolved with a 1.2\arcsec\ point spread function, but, as will be discussed in Section~\ref{s:fit211}, we also did fits at 8\arcsec\ resolution. The AIA data were rescaled to match, with the intensities rescaled to account for the change in the pixel area. 

GX~Simulator does any needed alignment of the AIA maps with the magnetic field model (see Section~\ref{s:BfieldModel}). Unlike the radio modeling done with GX~Simulator \citep[e.g.,][]{fleishman_21b}, the relative pointings of the data from the two SDO instruments is well understood, so no cross-correlation based alignment was done.

When calculating the model image we constrain the transition-region emission to areas where $|B_z|\ge 200$~G in the magnetogram, as in \citet{nita_18}. The resulting transition-region mask is shown in Figure~\ref{f:TRLoc}. As stated in the pervious section, all transition-region emission is restricted to the lowest layer of the model data cube. With this additional mask, we also confine the transition-region emission to only those areas where the photospheric field is strong and mostly vertical \citep{klimchuk_87b,nita_18}. The mask accounts approximately for the rapid expansion of the field in the upper photosphere and chromosphere, which occurs because of the transition from very large to very small plasma beta. This transition occurs over a thin but finite layer. In contrast, the magnetic field extrapolation assumes an infinitely sharp boundary, which leads to unrealistically high magnetic flux in areas with lower magnetic field, and thus strong transition-region emission everywhere, contrary to observations. Thus we suppress the model  transition-region emission in these low $B_z$ regions. The physical motivation for the mask is discussed further by \citet{nita_18} and in Appendix~\ref{a:TRmask}.

To emphasize the bright regions of the map, while still incorporating constrains from the the dark areas surrounding the active region, we assigned a constant uncertainty to all pixels in a given wave band, so that our model-to-data comparison metric is the sum of the squared residuals between the model and data, $\GoF=\sum(\Iobs-\Imod)^2$, where \Iobs\ and \Imod\ are the observed and modeled intensities respectively.  This means that the brighter areas of the image are weighted more heavily, but the emission is still constrained to low values outside the active region. This metric of success allows us to compare the quality of the fit within a given wave band, but it does not allow us to provide quantitative goodness-of-fit contours in our parameter space, or compare the goodness of fits between wave bands.   

For each ($a,b$) pair, we employed the Coronal Heating Modeling Pipeline utility \citep[CHMP,][]{nita_23}, which is part of the  GX~Simulator package, to compute maps for a range of values of $q_0$ converging to a minimum \GoF\ value, thus determining the best fit for that set of ($a,b$) values. For areas of the  ($a,b$) grid with the lower \GoF\ values the  $q_0$ vs. \GoF\ curves were smooth with clear minima. We considered a range of values for $a$ from $-1$ to 3 and $b$ from $-1$ to 2.5, in increments of 0.5.

We did two general types of data-model comparisons. Most included both the modeled corona and transition-region emission. As mentioned above, the modeled transition-region emission is limited to the bottom layer of the model cube in regions with photospheric $|B_z|\ge 200$~G as shown in Figure~\ref{f:TRLoc}.  All points in the array were used in the data-model comparison except for a one pixel wide band around the edge of the array to allow for shifts of the mask when interpolating to align with the AIA data.  We also did fits modeling coronal emission only.
GX~Simulator calculates the corona and transition region separately (as discussed in Section~\ref{s:Plasma}) and the transition-region emission can be turned off entirely. Thus in the corona-only fit the model transition-region emission was set to zero and the \GoF\ calculation excluded the points designated as transition region (Figure~\ref{f:TRLoc}) as well as the edge pixels. 

Below we focus on two AIA bands, 211~\AA, which shows chiefly coronal emission, and 171~\AA, which shows a combination of coronal and transition-region emission.

\section{Fit Results}
\label{s:FitResults}
\subsection{Fit to the 211~\AA\ band}
\label{s:fit211}

\begin{figure}
 \includegraphics[width=18cm, trim=0 0 0 0]{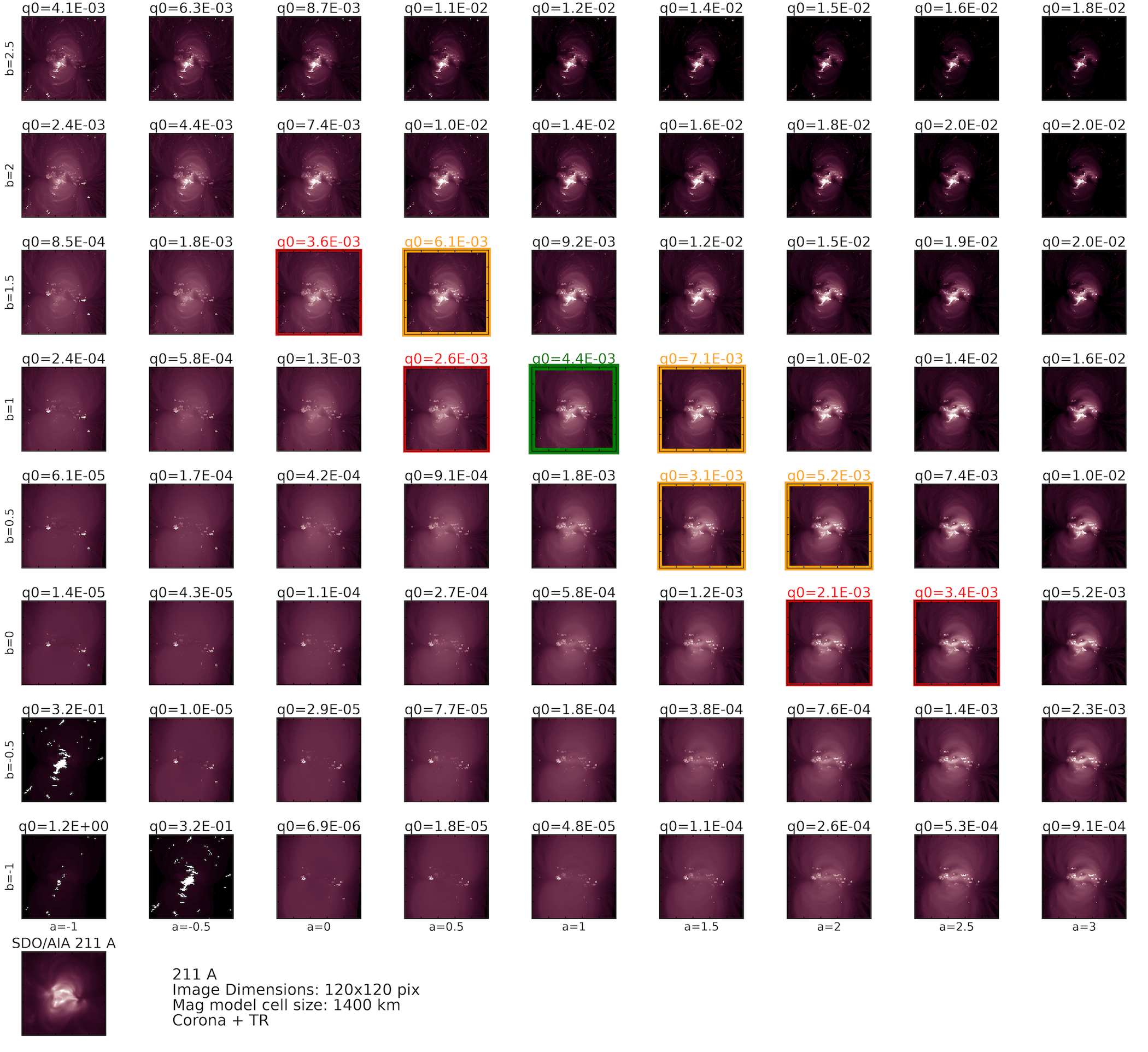}
\caption{Thumbnails of model 211~\AA\ images showing the best $q_0$ solutions for each of the ($a,b$) pairs considered for Magnetic Model~1, with a model/data comparison resolution of 2\arcsec, and including the model transition region. The actual 211~\AA\ image at the same resolution is at the bottom left. The $a=1$, $b=1$ image has the lowest \GoF\ value. Its title and border are in green. See Fig.~\ref{f:BestFit_211A_TRCorona}b for a larger version of the best model. Maps with \GoF\  within 10\% and 20\% of the minimum have titles and borders in orange and red, respectively. The images are all scaled linearly with the same intensity range.}
\label{f:Thumbnails_211A_YesTR}
\end{figure}

We first present the fits to the  211~\AA\ band, which is dominated by \ion{Fe}{14} with a peak emission temperature $\logT\approx 6.3$ \citep{{boerner_12}}.  In Figure~\ref{f:Thumbnails_211A_YesTR} we show a grid of thumbnails showing the best $q_0$ solutions for each of the ($a,b$) pairs considered for Model~1, including emission contributed by both the model corona and transition region.  The best fit corresponds to $a=1$, $b=1$, $q_0= 4.4\times10^{-3}$~erg~cm$^{-3}$~s$^{-1}$. This model image has a title and border in green.  Maps with \GoF\  within 10\% and 20\% of the minimum have titles and borders in orange and red, respectively. The result shows a linear area with similar fit quality.  We discuss this in more detail in Section~\ref{s:parameter_space}, which includes a plot of \GoF. 

However, the magnetic field in AR~12760 is weak, making it hard to model accurately because of observational uncertainties of the magnetic field inference. To try to compensate for that, we also compare the data and model maps at a lower resolution of 8\arcsec\ to reduce the effect of differences on a pixel-by-pixel basis (see Figure~\ref{f:Thumbnails_211A_nx30_YesTR} in the Appendix for the resulting thumbnail grid).  With the comparison at this resolution we find the best fit at $a=1.5$, $b=1.0$, $q_0= 7.1\times10^{-3}$~erg~cm$^{-3}$~s$^{-1}$.  We note that this best fit ($a,b$) pair corresponds to a pair that is within 10\% of the minimum \GoF\ for the higher-resolution fit (Figure~\ref{f:Thumbnails_211A_YesTR}). 
In general we would say that our uncertainty for $a$ and $b$ is around 0.5 because that is the step size in our ($a,b$) grid. However, the situation is rather more complicated due to the complex ($a$,$b$,$q_0$) parameter space, as will be discussed in Section~\ref{s:parameter_space}.

\begin{figure}
 \includegraphics[width=16cm, trim=0 0 0 0]{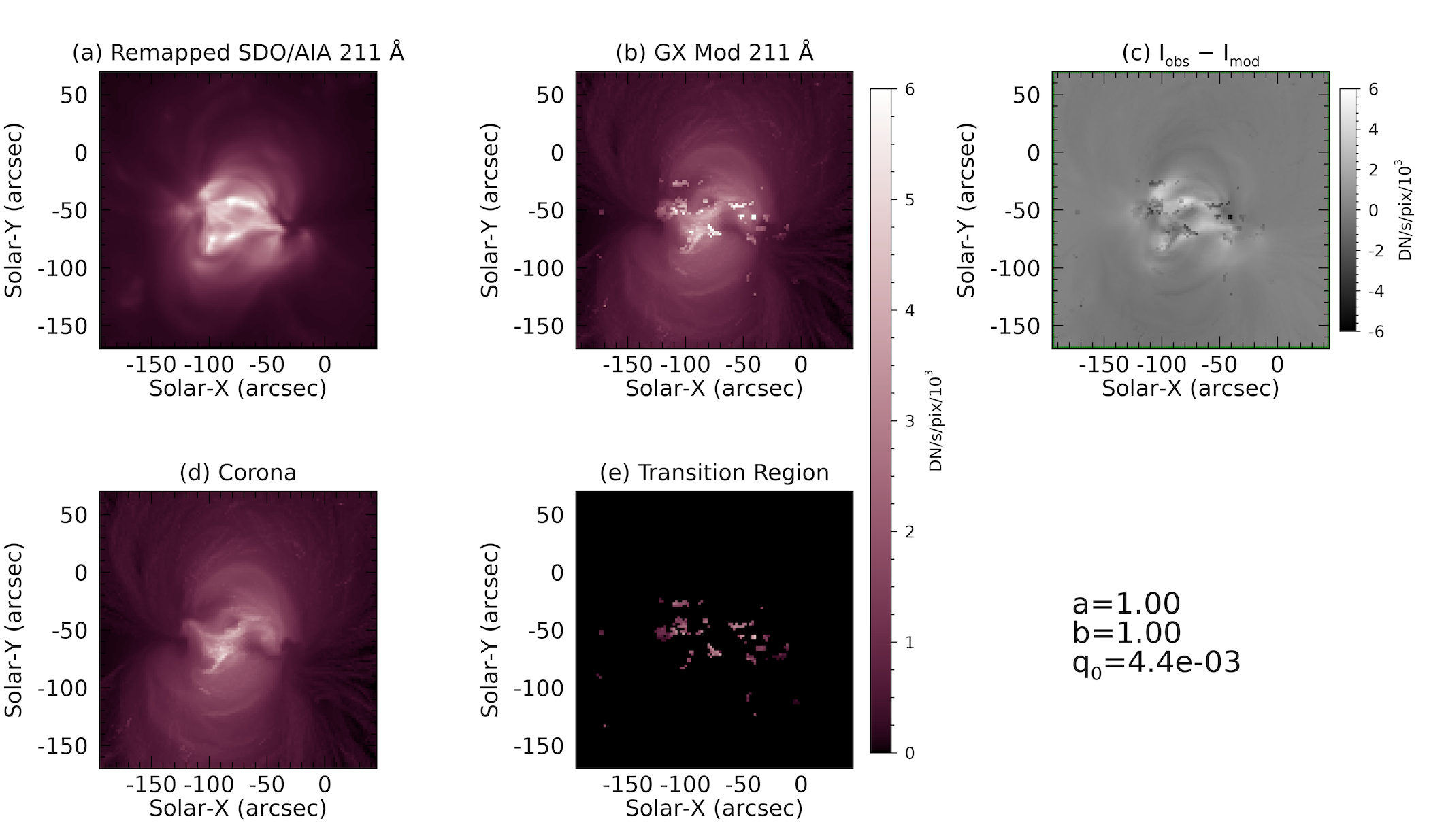}
\caption{Comparison of the best fit to 211~\AA\ data binned to 2\arcsec\ resolution using Magnetic Model 1 showing (a) the AIA data,  (b) the best-fit model map, (c) The residual, $\delta=I_{obs}-I_{mod}$, 
(d) the coronal component of the model, and (e) the transition-region component of the model. Panels (a),(b),(d), and (e) are scaled linearly from 0 to 95\% of the maximum data intensity. Panel (c) is scaled to $\pm 95\%$ of the maximum data intensity. See Fig.~\ref{f:Thumbnails_211A_YesTR} for grid of all ($a,b$) best matches.}
\label{f:BestFit_211A_TRCorona}
\end{figure}

\begin{figure}

  \includegraphics[width=16cm, trim=0 0 0 0]{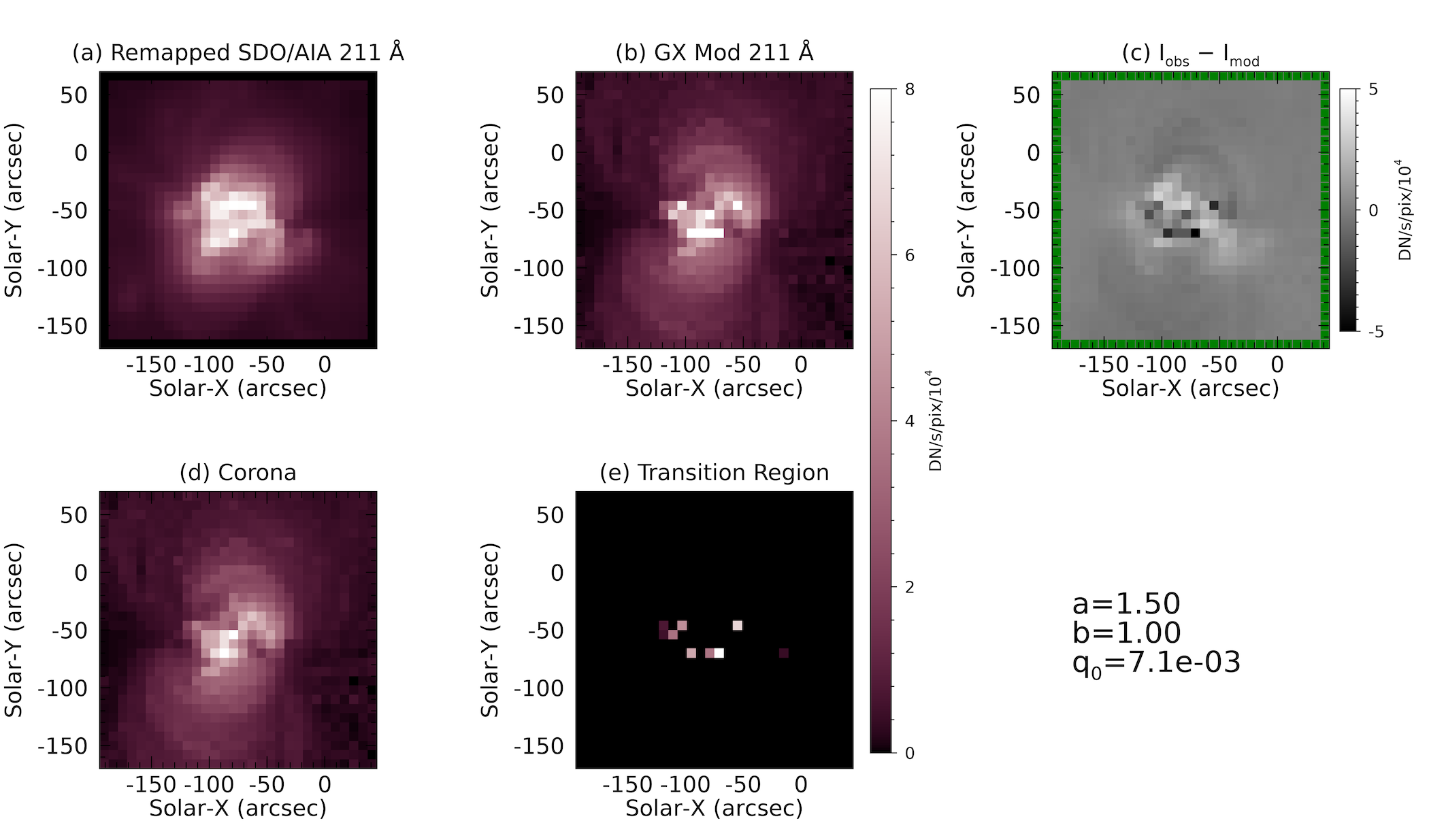}
\caption{Comparison of the best fit to 211~\AA\ data binned to 8\arcsec\ resolution using Magnetic Model 1 showing (a) the AIA data,  (b) the best-fit model map, (c) The residual, $\delta=I_{obs}-I_{mod}$, 
(d) the coronal component of the model, and (e) the transition-region component of the model. Panels (a),(b),(d), and (e) are scaled linearly from 0 to 95\% of the maximum data intensity. Panel (c) is scaled to $\pm 95\%$ of the maximum data intensity. The regions in green in the residual map were not used to calculate \GoF. See Fig.~\ref{f:Thumbnails_211A_nx30_YesTR} for grid of all ($a,b$) best matches.}
\label{f:BestFit_211A_TRCorona_8arcsec}
\end{figure}

The best fits for the 2\arcsec\ and 8\arcsec\ comparisons are shown in  Figure~\ref{f:BestFit_211A_TRCorona} and Figure~\ref{f:BestFit_211A_TRCorona_8arcsec} respectively,  along with the AIA data at the same resolution, and the residual, $\delta=I_{obs}-I_{mod}$. The plots also show the model coronal and transition-region contributions to the emission as separate maps.
The coronal and transition-region emissions are calculated separately and summed by GX~Simulator, and seperate transition-region and coronal maps can be synthesized in addition to the total emission.
The model maps look reasonable on a general level, with most of the emission coming from the active region core. However, the detailed correspondence of the bright features is not as good as we would like. A significant part of this can be attributed to inaccuracies in the magnetic field extrapolation \citep{fleishman_19}. Even if the hydrodynamics along field lines is modeled correctly, the spatial location of the emission will be off. We mitigate this in part by reducing the resolution to 8\arcsec, although differences are still apparent.  The transition-region emission is a significant contributor to the emission near loop footpoints, but it is not the main source of the 211~\AA\ emission.

\begin{figure}
 \includegraphics[width=16cm, trim=0 0 0 0]{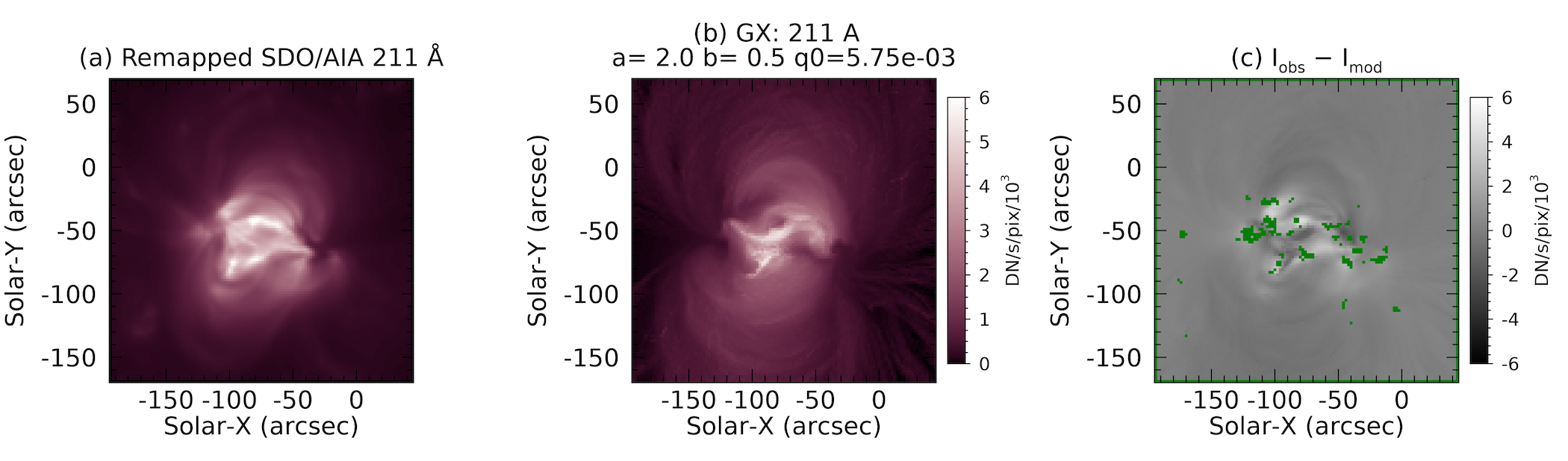}
\caption{AIA data (a) and best fit (b) to the corona only (leaving out transition-region emission) and the residual values ($\delta=\Iobs-\Imod$) for 211~\AA\ using Magnetic Model~1 and a fit comparison resolution of 2\arcsec. The regions in green in the residual map were not used to calculate \GoF. These include the areas we expected to contain transition-region emission (See Figure~\ref{f:TRLoc}). See Fig.~\ref{f:Thumbnails_211A_NoTR} for grid of all ($a,b$) best matches.}
\label{f:BestFit211A_noTR}
\end{figure}

The transition-region emission in the model seems especially bright and localized compared to the observations. We address the localization later. The brightness is a problem that has historically afflicted many models, including detailed 1D hydrodynamics models, i.e., ``loop'' models. As mentioned in Section~\ref{s:Plasma}, the discrepancy is reduced in expanding loops, but the excess brightness remains an unsolved problem. For this reason, we
also considered a fit to the corona only.  The resulting best-fit model is shown in Figure~\ref{f:BestFit211A_noTR}. For this fit we turned off the transition-region emission and used only pixels not designated as including transition region (Figure~\ref{f:TRLoc}) to calculate \GoF. In this case we found the best fit to be  $a=2$, $b=0.5$, $q_0= 5.8\times10^{-3}$~erg~cm$^{-3}$~s$^{-1}$. The intensity scale is the same as in Figure~\ref{f:BestFit_211A_TRCorona}. The model looks similar to the best-fit model for the combined corona and transition-region fit, although is a bit brighter than the corona alone in that fit. This ($a,b$) pair also corresponds to a pair that is within 10\% of the minimum \GoF\ for the corona and transition-region fit shown in Figure~\ref{f:Thumbnails_211A_YesTR}.

\subsection{Fit to the 171~\AA\ band}
\label{s:fit171}

\begin{figure}
 \includegraphics[width=18cm, trim=0 0 0 0]{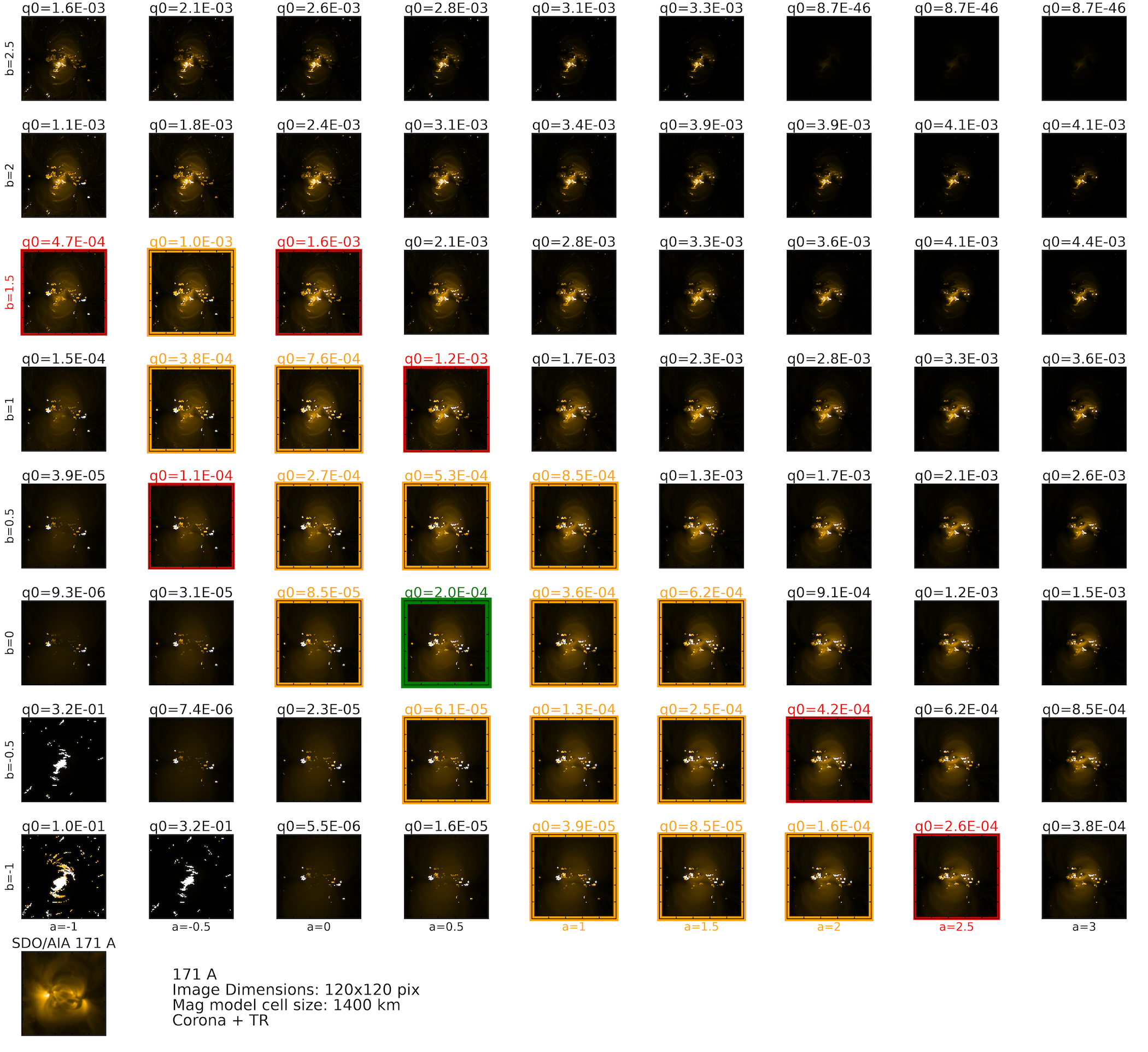}
\caption{Thumbnails of model 171~\AA\ images showing the best $q_0$ solutions for each of the ($a,b$) pairs considered for Model~1, a model/data comparison resolution of 2\arcsec, and  including both the corona and transition region. The actual 171~\AA\ image at the same resolution is at the bottom left. The $a=0.5$, $b=0.0$ image has the lowest \GoF\ value. Its title and border are in green. See Fig.~\ref{f:BestFit_171A_TRCorona}b for larger version of the best model.  Maps with \GoF\  within 10\% and 20\% of the minimum have titles and borders in orange and red, respectively. The images are all scaled linearly with the same intensity range.}
\label{f:Thumbnails_171A_YesTR}
\end{figure}

The fit results for the 171~\AA\ band, which has emission produced principally by \ion{Fe}{9} with a peak at \logT$\approx 5.8$, are rather different from those of the 211~\AA\ band.  Figure \ref{f:Thumbnails_171A_YesTR} shows the ($a,b$) grid of fits using Magnetic Model~1 with both the model corona and transition region included.  The best fit is $a=0.5$, $b=0.0$, $q_0=2.0\times10^{-4}$~erg~cm$^{-3}$~s$^{-1}$.
These are substantially different values for $a$ and $b$ than those found for 211~\AA. The best-fit 211~\AA\
model for these $a$ and $b$ values (Figure~\ref{f:Thumbnails_211A_YesTR}) looks like a terrible fit to 211~\AA, and, conversely, the best ($a,b$) values for 211~\AA\ do not appear to lead to good fits for 171~\AA, although there seem to be significant problems with all the 171~\AA\ fits.

\begin{figure}
 \includegraphics[width=16cm, trim=0 0 0 0]{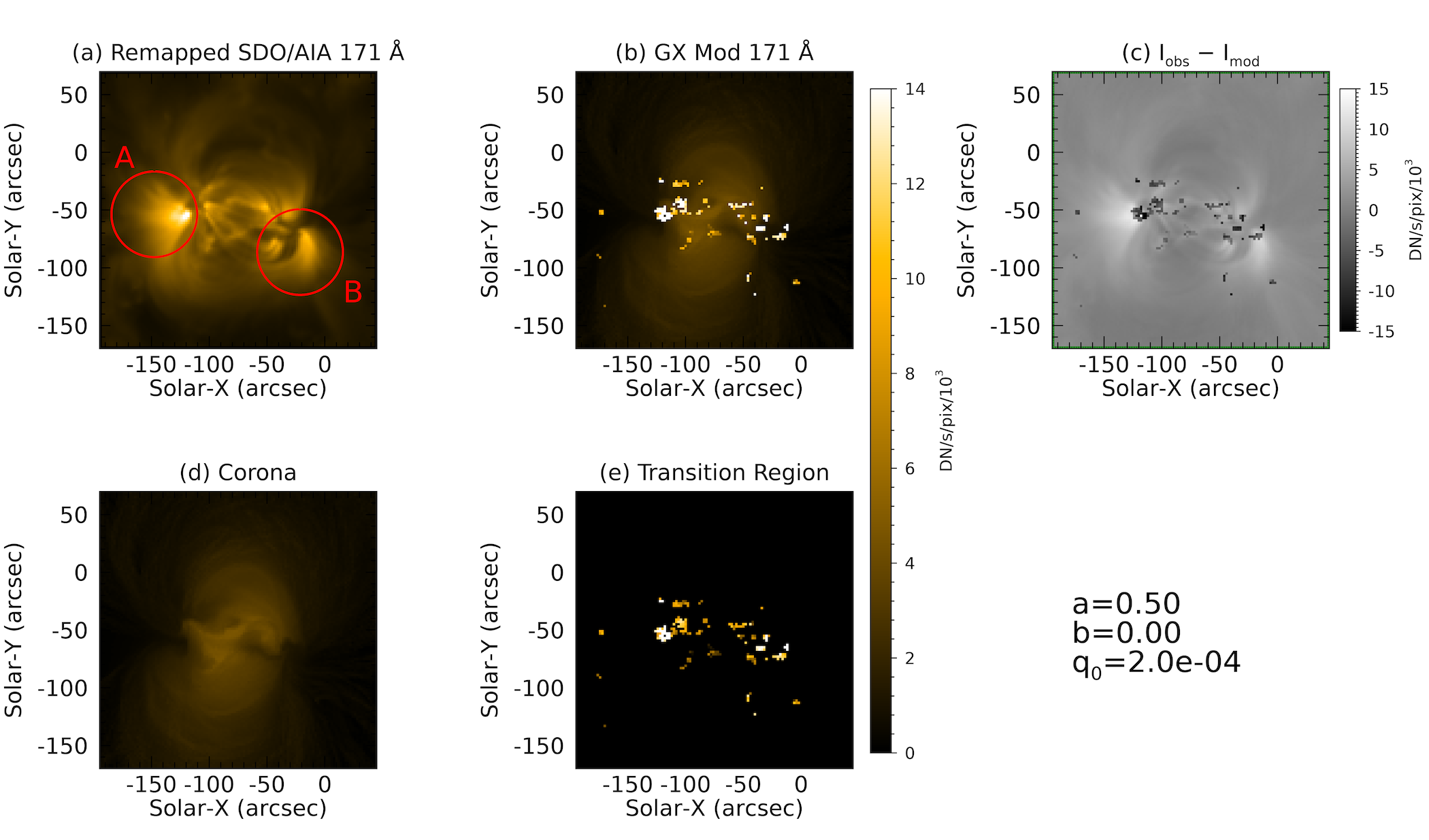}
\caption{Comparison of the best fit to 171~\AA\ data binned to 2\arcsec\ resolution and using Magnetic Model~1 showing (a) the AIA data,  (b) the best-fit model map, (c) The residual, $\delta=\Iobs-\Imod$, 
(d) the coronal component of the model, and (e) the transition-region component of the model. Panels (a),(b),(d), and (e) are scaled linearly from 0 to 95\% of the maximum data intensity. Panel (c) is scaled from $\pm 95\%$ of the maximum data intensity. See Fig.~\ref{f:Thumbnails_171A_YesTR} for grid of all ($a,b$) best matches.}
\label{f:BestFit_171A_TRCorona}
\end{figure}

The difference appears to be connected to longer loops on the periphery of the active region. Figure~\ref{f:BestFit_171A_TRCorona} shows the details of the best-fit model, again with separate images showing the corona and transition-region components.
As would be expected, the localized model transition-region emission is more evident in the 171~\AA\ band than in 211~\AA. However, in the ``best fit'' this seems to be exaggerated; almost all the emission is coming from the designated transition-region foot points with very little emission in the active region core. Also, although the 171~\AA\ image is dominated by the bright areas in the long, peripheral  loops (circled in red in Figure~\ref{f:BestFit_171A_TRCorona}a), these features are entirely missing from the model images. This is not just true of the best fit, but of all the fits shown in Figure~\ref{f:Thumbnails_171A_YesTR}. Feature A appears to be the base of a fan-loop, a large scale structure that we think closes in the corona. Feature B seems to form a more (although not entirely) complete arc, although its exact size and altitude are not clear.

\begin{figure}
 \includegraphics[height=8cm, trim=100 350 100 50]{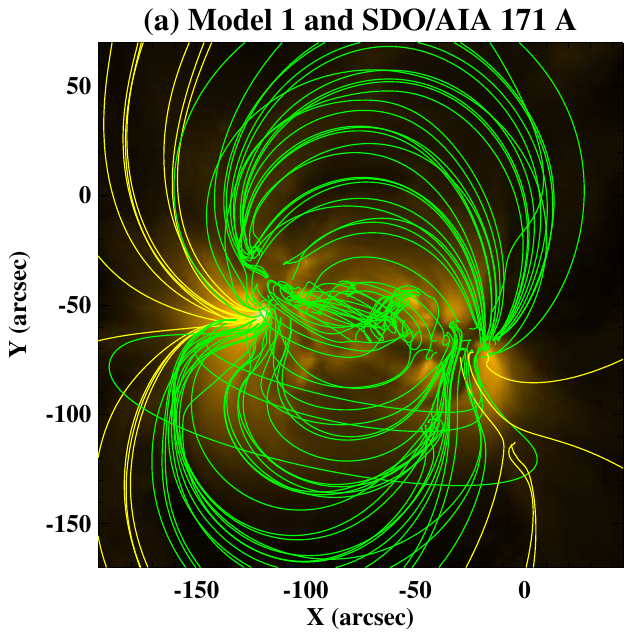}
 \includegraphics[height=8cm, trim=150 350 100 50]{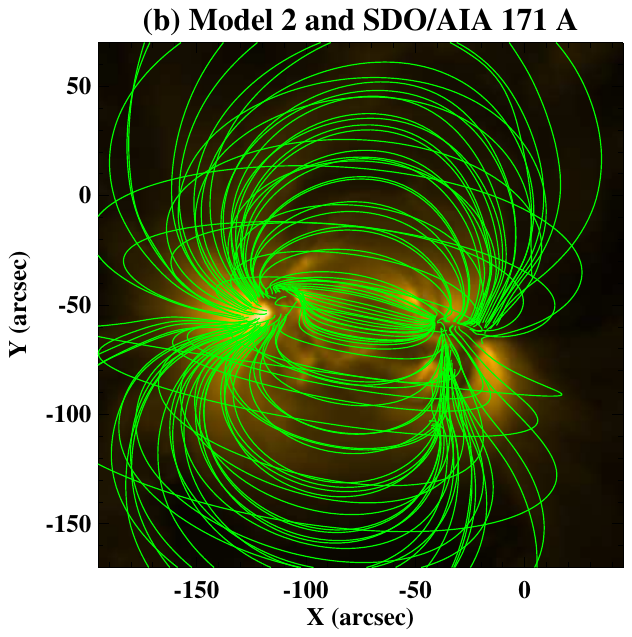}
 \caption{Magnetic field lines from (a) Magnetic Model~1 and (b) Magnetic Model~2 superimposed on the AIA 171 \AA\ band image. Field lines closing within the model box are shown in green while ones that do not are shown in yellow.}
 \label{f:Bfield_171A}
\end{figure}

One possible explanation for the total lack of emission in the peripheral loops is that the relevant loops did not close within the modeled cube, and thus were not included in the emission (see Section~\ref{s:BfieldModel}). This was the reason we developed Magnetic Model~2. Figure~\ref{f:Bfield_171A} shows magnetic field lines from both Magnetic Model~1 and Model~2 superimposed over the 171~\AA\ reference image. It can be seen that in Magnetic Model~1 some of the longer field lines extending from the periphery of the active region do not close down again inside of the box (these are shown in yellow), and so are being excluded from our model emission. In contrast, in Magnetic Model~2, the lines from these areas do close down  inside the NLFFF model cube, so we would expect them to be included in the emission.

\begin{figure}
  \includegraphics[width=16cm, trim=0 0 0 0]{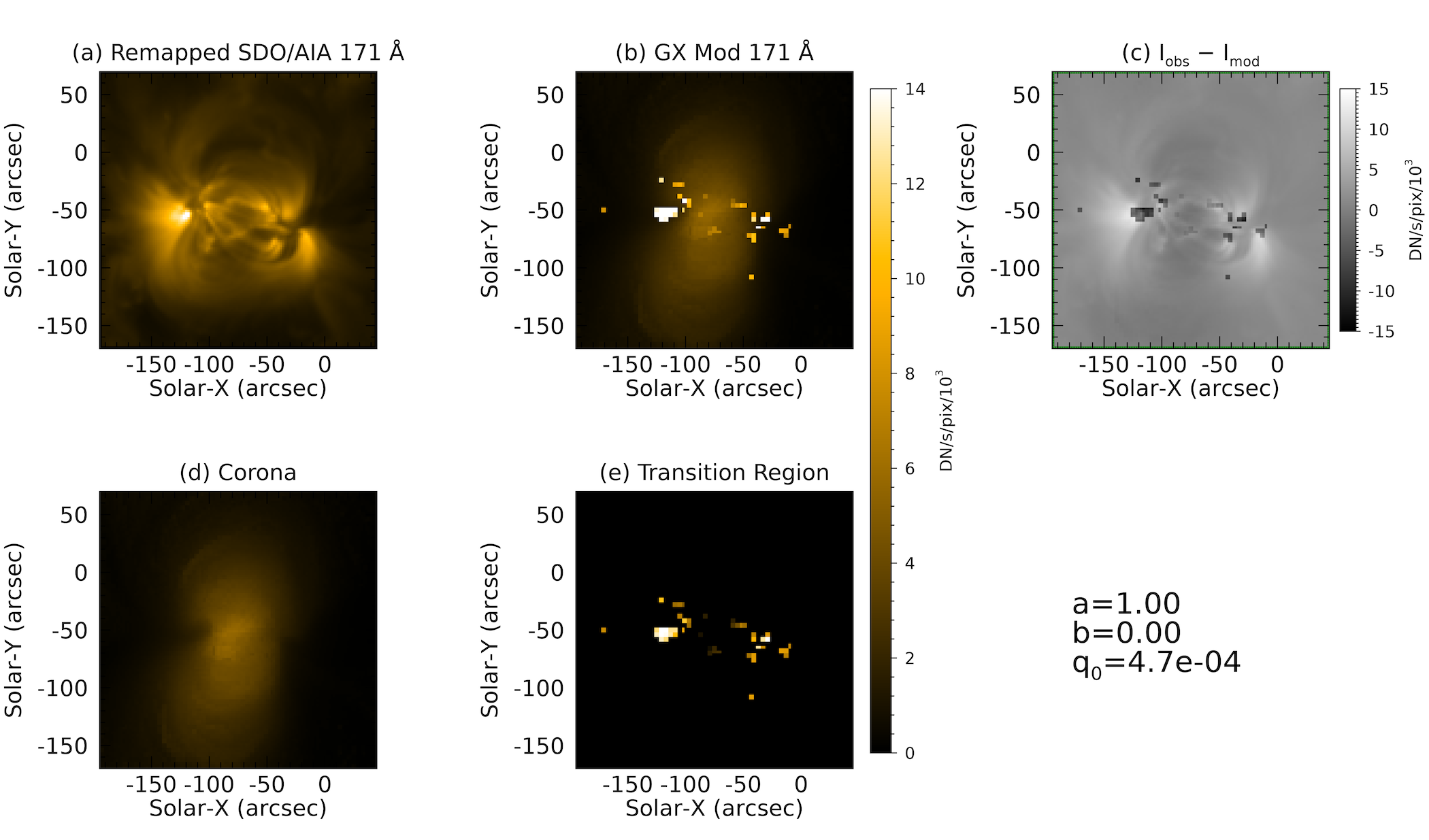}
\caption{Comparison of the best fit to 171~\AA\ data using Magnetic Model~2 and binned to 2\arcsec\ resolution showing (a) the AIA data  (b) the best-fit model map, (c) The residual, $\delta=\Iobs-\Imod$, 
(d) the coronal component of the model, and (e) the transition-region component of the model. Panels (a),(b),(d), and (e) are scaled linearly from 0 to 95\% of the maximum data intensity. Panel (c) is scaled to $\pm 95\%$ of the maximum data intensity.  See Fig.~\ref{f:Thumbnails_171A_dx2800_YesTR} for grid of all ($a,b$) best matches.}
\label{f:BestFitMod2_171A_TRCorona}
\end{figure}

However, this proposed explanation does not seem to be correct. In Figure~\ref{f:BestFitMod2_171A_TRCorona} we show the best fit result for model maps calculated with Magnetic Model~2. The best fit is $a=1.0$, $b=0.0$, $q_0=4.7\times10^{-4}$~rg~cm$^{-3}$~s$^{-1}$. The full ($a,b$) grid is shown in Appendix~\ref{a:fit_thumbnails}, in Figure \ref{f:Thumbnails_171A_dx2800_YesTR}. The longer field lines in the periphery of the active region are still not present in the emission, although the footpoints of the loops in the transition region are very bright.
 
 Thus, the use of the lower-resolution but larger-volume magnetic field model results in a less accurate distribution of emission in the core of the active region, but it does not result in the appearance of the longer loops, so the problem does \textit{not} seem to be that the loops are being left out of the magnetic field model. As we will discuss in Section~\ref{s:Discuss_TR}, we think the problem is in how the model transition-region emission is distributed along the loop.

\subsection{Fits to other AIA bands}
\label{s:fit_otherbands}
\begin{figure}
\includegraphics[width=14cm, trim=0 0 0 0]{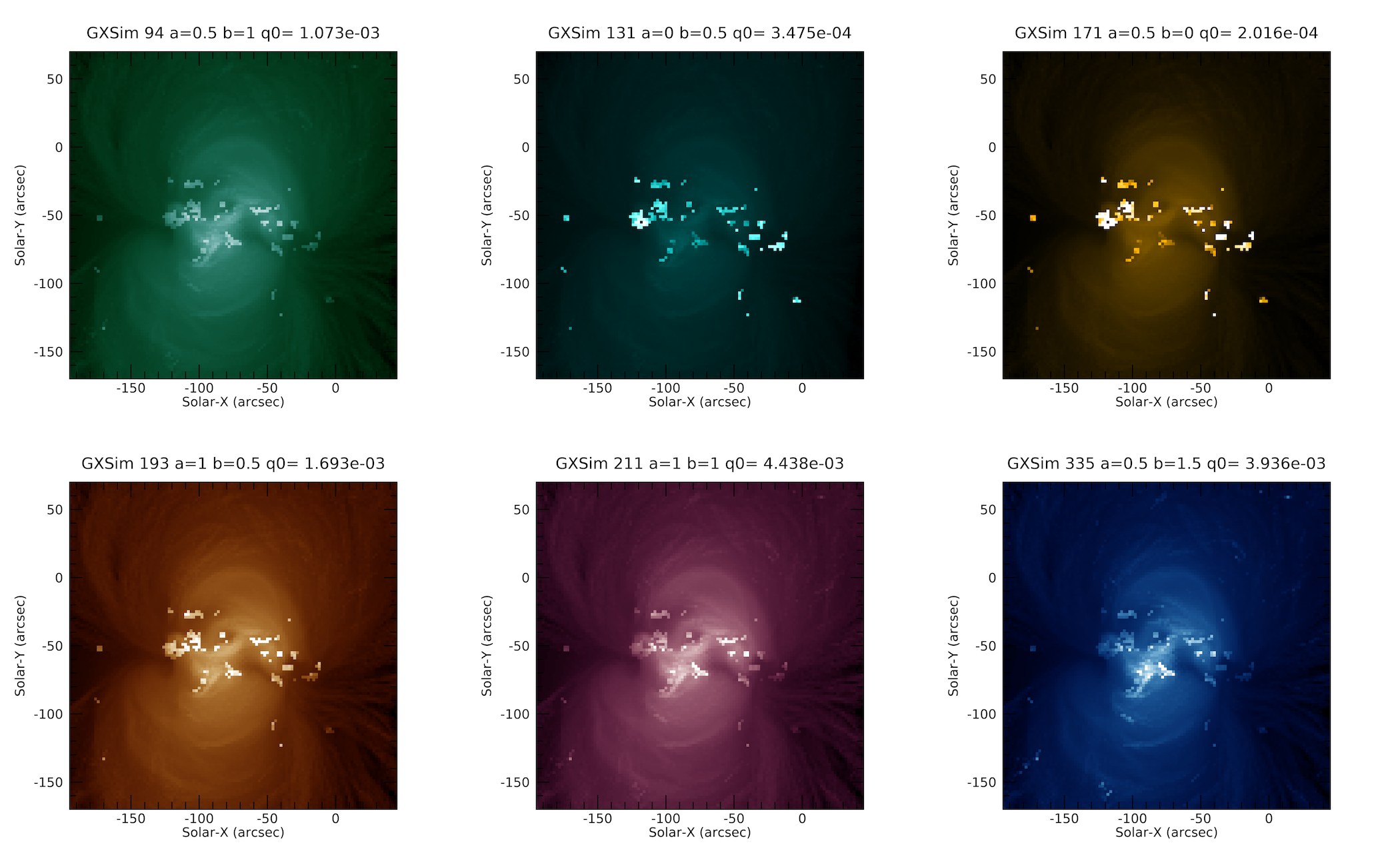}
\caption{The best fits for all AIA bands modeled, using Magnetic Model~1 and a fit comparison resolution of 2\arcsec. AIA images are shown in Figure~\ref{f:aia_img}.}
\label{f:BestFit_All_TRCorona}
\end{figure}

The other AIA bands we considered, 94, 131, 193, and 335~\AA, show a temperature dependent pattern similar to that seen in 171 and 211~\AA (see Figure~\ref{f:BestFit_All_TRCorona}). For the 211 and 335~\AA\  bands, which are dominated by emission at $\logT\ga 6.3$ \citep{boerner_12}, the fits look similar. Although they differ from the original AIA image in detail, they show similar amount of emission in the active region core, which dominates the emission in both data and model. The thumbnail plots for the comparisons of the model to the 335~\AA\ data binned to 8\arcsec\ are shown in Figure~\ref{f:Thumbnails_335A_nx30_YesTR}, and more discussion of the 211~\AA\ and 335~\AA\ results is in Section~\ref{s:parameter_space}.

The 131 and 171~\AA\ bands, on the other hand, which in the absence of flaring have peak emissions at $\logT\la 5.8$, both show bright peripheral loops that are missing from the models except at the loop bases. The 94~\AA\ band, in this case dominated by its response peak at $\logT\approx 6.05$,  and 193 \AA, with a response peaking at $\logT\approx 6.1$, are in between, showing the peripheral loops that are, again, missing from the model fits, but the peripheral loops are not as bright as for the bands dominated by cooler emission.

\section{Discussion}
\label{s:Discussion}

\subsection{Heating Parameters}
\label{s:parameter_space}

There are many measurements of $a$, $b$ and related parameters in the literature. There is a summary of measured and theoretical values \citep[the latter from][]{mandrini_00} presented by \citet{ishigami_24}, who include a summary plot of values that correspond to our $a$ and $1-b$. They vary quite a bit and our values of $b$ in particular are lower than those measured by others. It is important to note that these measured values - obtained by comparing models with observations as we do here - are for active regions in some studies and the global corona in other studies. Differences are not surprising because the dominant heating mechanism may be different in active regions and the quiet Sun.

\begin{figure}
 \includegraphics[width=9cm, trim=0 0 0 0]{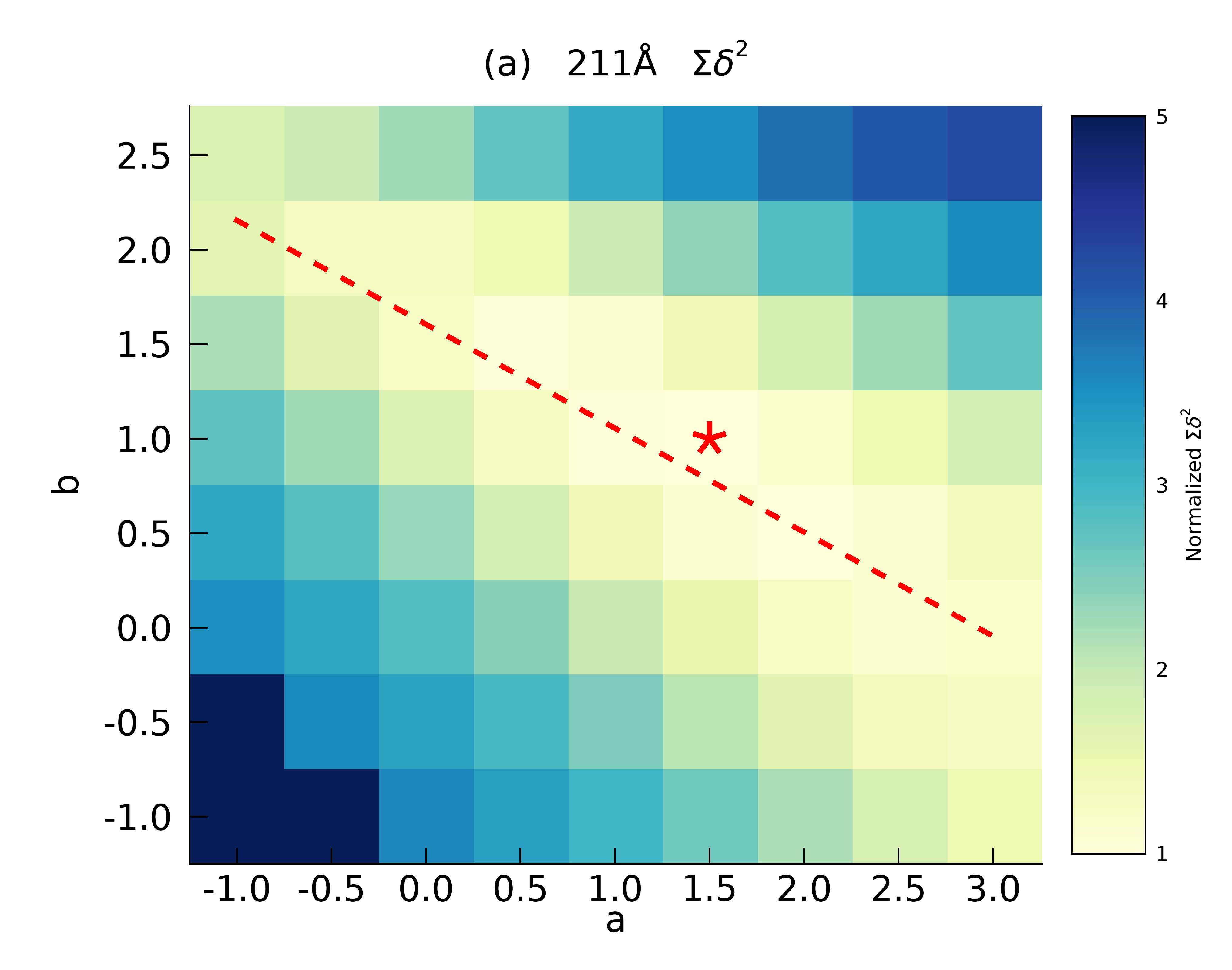} \includegraphics[width=9cm, trim=0 0 0 0]{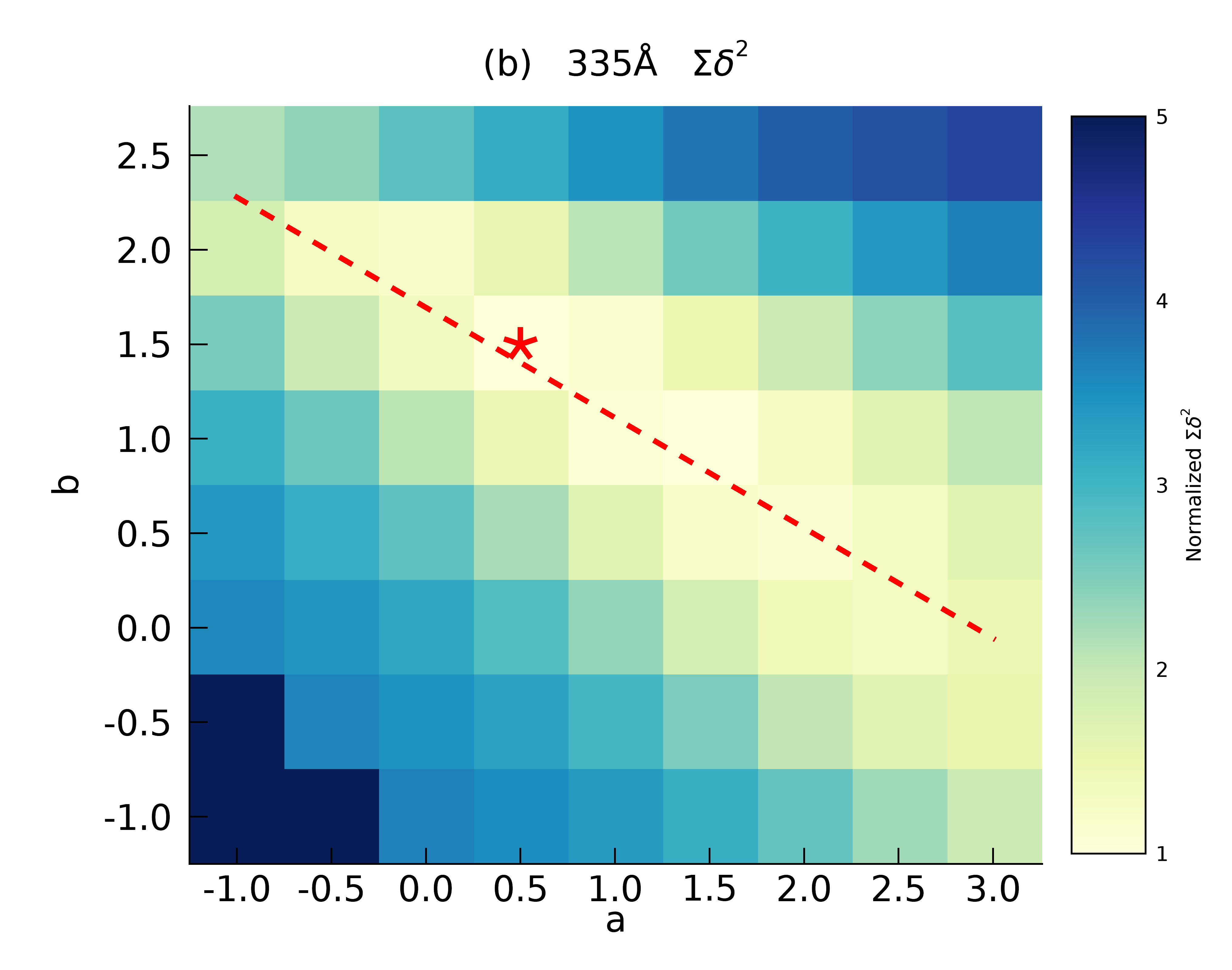}
\caption{\GoF\ values for the fit to the (a) 211~\AA\ data and (b) 335~\AA\ data using a 8\arcsec\ resolution map for the data to model comparison. The red asterisk shows the best-fit location, and the dashed line shows a fit to the minima in each column. The color bar shows the \GoF\ values normalized to the minimum \GoF.}
\label{f:chi2_plot_211A_nx30}
\end{figure}

Our most reliable fits are the low-resolution (8\arcsec) fits to the 211 and 335~\AA\ data, because the lower resolution smooths out some of the inaccuracies in the magnetic field model.
Figure~\ref{f:chi2_plot_211A_nx30} shows plots of our model-to-data comparison metric, \GoF, as a function of $a$ and $b$, where each fit has been optimized for $q_0$ for that fitted $a,b$ pair (see Figures~\ref{f:Thumbnails_211A_nx30_YesTR} and \ref{f:Thumbnails_335A_nx30_YesTR}). 
The best fit for 211~\AA, shown by the asterisk, is $a=1.5$, $b=1$, and $q_0= 7\times 10^{-3}$~erg~cm$^{-3}$~s$^{-1}$.  
If we consider the $q_0$ values for fits within 20\% of the minimum \GoF\ value then we get $q_0$ in the range $3 \times10^{-3}$ and $11\times10^{-3}$~erg~cm$^{-3}$~s$^{-1}$.
For 335~\AA\ it is $a=0.5$, $b=1.5$, and $q_0=4\times10^{-3}$~erg~cm$^{-3}$~s$^{-1}$.
The fits within 20\% of the minimum include $q_0$ values from $3\times10^{-3}$ to $7\times10^{-3}$~erg~cm$^{-3}$~s$^{-1}$.

\begin{table}
\begin{tabular}{|c c c c c c|}
\hline
$a$ & $b$ &  $q_0$ 211~\AA & $q_0$ 335~\AA\ & 211~\AA\ quality &   355~\AA\ quality \\
&&  \multicolumn{2}{c}{($10^{-3}$~erg~cm$^{-3}$~s$^{-1}$)} &  & \\
\hline
1.5 & 1  &7.1& 4.8 & Best fit & within 10\% of min \GoF  \\
0.5 & 1.5 & 3.9 & 6.1 & within 10\% of min \GoF & Best fit\\
\hline
\end{tabular}
\caption{211 and 355~\AA\ best-fit comparisons}
\label{t:comp211_355}
\end{table}

A recurring feature in these plots of model-to-data comparison metrics is the clear diagonal feature ranging from low $a$, high $b$ values on the top-left to high $a$, low $b$ values towards the bottom-right. This feature is consistent with other studies with EUV \citep{ugarte_19} and microwave data  \citep{nita_23,fleishman_25, kuznetsov_25}. For the 8\arcsec\ resolution 211~\AA\ fit we found that this feature roughly corresponds to $b\approx1.6-0.6a$ and for the 335~\AA\ fits we get $b\approx1.7-0.6a$. That our best fits for each wave band do not fall exactly on these lines is indicative of the coarseness of our ($a,b$) grid and uncertainties in our fitting of diagonal features. More specifically, if we assume a 0.25 uncertainty (half a ($a,b$) grid step) on both $a$ and $b$ we have $b$ intercepts of  $1.61\pm 0.12$ and $1.70\pm 0.13$ and slopes of $-0.56 \pm 0.07$ and $-0.59\pm 0.08$ for 211~\AA\ and 335~\AA, respectively, so the results for the two bands are the same within uncertainties.

We note that although the best $(a,b)$ values are different for the two wave bands, they both fall close to these best-fit lines. Table~\ref{t:comp211_355} compares the best model-to-data comparisons of the two wave bands. We see that for both wave bands the best ($a,b$) values fall within 10\%\ of the minimum \GoF\ for ($a,b$) values of the other wave band.  This gives an idea, albeit somewhat qualitative, of our confidence in our fit values. The model-to-data comparisons for the other bands are sufficiently poor (see Sections~\ref{s:fit171} and \ref{s:fit_otherbands}) that we do not include them in this analysis.

This linear feature in the model-to-data comparison metric parameter space indicates that $L$ and \Bavg\ are correlated. This is, of course, no surprise. It is well-known that the coronal magnetic field is on average stronger in the short loops of the active region core than in the long loops of the periphery. We can show that this feature is reasonably consistent with simple relationship between \Bavg\ and $L$.

As stated in Equation~\ref{e:Qavg}, we are assuming $\Qavg \propto \Bavg^a L^{-b}$.
So if 
\begin{equation}
\Bavg\propto L^{-\bl}    
\label{e:B_L_dependence}
\end{equation}
\citep{klimchuk_95} then we have $\Qavg\propto L^{-\bl a -b}$. Model fits should be similarly good when  $\bl a+b=$ constant; call it $b_0$. Good fits should therefore fall along a straight line in a plot of goodness of fit in the $(a,b)$ plane:
\begin{equation}
b=b_0 - \bl a.
\label{e:bprime}
\end{equation}

Thus, the slope of the best fit line in the goodness-of-fit parameter space gives us the relationship between $B$ and $L$ (Equation~\ref{e:B_L_dependence}), and the relationship for \Qavg\ and $L$ is dependent on the y-intercept of the line:
\begin{equation}
\Qavg\propto L^{-b_0}. 
\end{equation}

Our results shown in Figure~\ref{f:chi2_plot_211A_nx30} suggest that $b_0\approx 1.6$ or 1.7 and $\bl \approx 0.6$. As stated above, other analyses also show that similarly good fits fall along a straight line.
\citet{fleishman_25} using \gx\ to model optically thick gyroresonant microwave emission observed by the Siberian Solar Radio Telescope (SSRT) and Nobeyama Radioheliograph (NoRH) found $b=2.4-1.0a$ at 5.7~GHz and $b=2.2-0.7a$ at 17~GHz and a best solution at ($a,b$) = (0.6, 1.8). 
\citet{kuznetsov_25} analyzed eight active regions in 6, 8, and 10~GHz with the Siberian Radioheliograph and found that $b_0$ was mostly between 1.6 and 3.2, with the distribution peaking at 2.5, and  \bl\ ranged from 0.2 to 1.2, with a most probable value of 0.55. It is also interesting that \citet{porter_95} concluded that $\Qavg\propto L^{-2}$ from a study of soft X-ray loops observed by Yohkoh, which can be compared to the $b_0$ values derived from \gx.

\citet{mandrini_00} examined magnetic field extrapolation models and found $\bl = 0.88$ using Equation \ref{e:B_L_dependence} directly. \citet{fleishman_21a} also found a \Bavg-$L$ dependence based on a magnetic field model, although their fitting model was more sophisticated than a simple power law, i.e. $\log^2\Bavg = 48.61 - 1.84 \log L$. 

\begin{figure}
    \includegraphics[width=10cm]{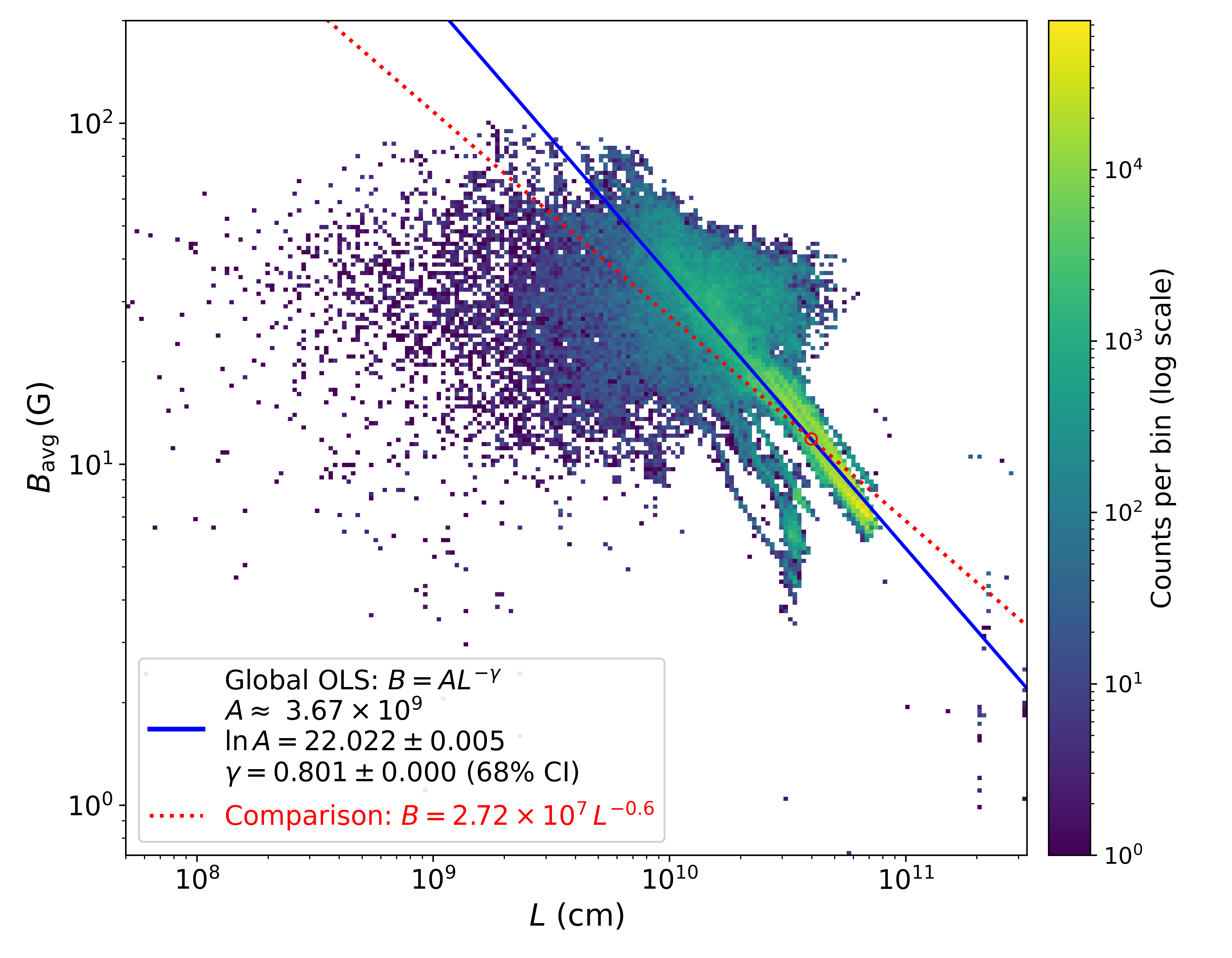}
    \caption{Density distribution in the $(B_{\mathrm{avg}},\,L)$ plane for Magnetic Model 1 (2D histogram, equal log--log bins). Lines shown are an ordinary least-squares (OLS) regression to the entire set of points (solid blue line) and a  line with a slope of $\bl=0.6$, as derived from the \gx\ fits (red dashed line). This line is forced to go through the center of mass of the magnetic model points (red circle).}
    \label{f:BavgLfit}
\end{figure}

With this in mind, we fit the \Bavg\ vs $L$ dependence in our Magnetic Model 1, as shown in Figure~\ref{f:BavgLfit}, using a least-squares fit in log-log space. We derived $\bl=0.8$. The plot also shows a line with a slope corresponding to $\bl=0.6$ as suggested by the \gx\ modeling. The value derived directly from the magnetic model is slightly higher than the value derived from \GoF\ plots, but it is clear from the distribution of points in Fig.~\ref{f:BavgLfit} that the distribution of the field lines in the model is more complex that a simple power-law, with noticeable substructure. We expect Equation~\ref{e:B_L_dependence} to be a general trend rather than an equation applying precisely to all parts of the active region. These results are, however, consistent with the idea that the diagonal feature in  \gx\ model-to-data comparison metrics is due to a dependence of \Bavg\ on $L$.

We should note that \citet{fleishman_25} and \citet{kuznetsov_25} are looking at different active regions, which in addition to a difference in wave bands and a different model-to-data comparison metric could lead to differences in the properties of the zones of best fit.  \citet{kuznetsov_25} in particular find a range of values over the many active regions they modeled.

However, the broad agreement of these results is very encouraging. This validates the power-law dependencies assumed in Equations \ref{e:Qavg} and \ref{e:B_L_dependence}.  Had similarly good fits not fallen along a straight line, one or more of the dependencies would be invalidated. 

\subsection{Transition-Region Emission}
\label{s:Discuss_TR}
\begin{figure}
\includegraphics[width=13cm]{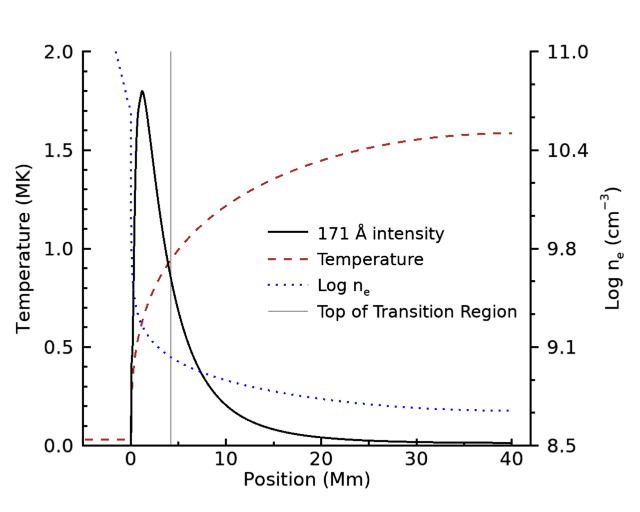}
\caption{The temperature, electron density, and normalized 171~\AA\ intensity of half of an equilibrium loop with uniform heating as modeled by ARGOS. The vertical line shows the location of the top of the transition region/base of the corona. It is about 5.4\% of the full loop length, $L$, from the top of the chromosphere}
\label{f:LoopModel}
\end{figure}

An important issue in our modeling effort is that the emission from long features at the periphery of the active region is missing from the models. For this active region these are dominant features in the cooler bands, 131, and 171~\AA, and significant contributors to other bands with peak contributions below about 1.6~MK (94 and 193~\AA).  We think this is due to an unrealistic distribution of the transition-region emission in these long loops. In this work and other modeling studies, the transition region is defined to be the locations in the loops where thermal conduction serves as a heating term rather than a cooling term \citep{vesecky_79}. As discussed in \citet{klimchuk_08}, this is a precise, physically motivated definition that avoids ambiguities that arise from vague observational definitions based on temperature or altitude. As physically defined, the transition region extends from the top of the chromosphere to about 10\% of a loop half-length ($0.05L$) \citep[see][]{klimchuk_08, cargill_22}. An example of this is shown in Figure~\ref{f:LoopModel}, which depicts the temperature, electron density, and normalized 171~\AA\ intensity as a function of position of a 1D hydrodynamic model of a loop in equilibrium with steady uniform heating. The model was computed with the ARGOS code \citep{antiochos_99} and is broadly representative of static equilibrium loops with a wide range of lengths and heating rates. The vertical line shows the top of the transition region at $0.054L$ from the loop footpoints. This same spatial fraction applies approximately to other loops, but the temperature of the top of the transition region varies, always about 60\% of the apex temperature.

For relatively short loops it is sufficient to assume that the transition-region emission would be confined to a limited range approximated by a single voxel along the loop in our model cube. In our higher-resolution Model 1 each voxel is 1.4~Mm on a side. That would be 5\% of a loop with $L=28$~Mm, within a factor of two or so of loop length for the core of an active region. However, based on Magnetic Model~2, the loops at the periphery of the active region likely have $L= 200-700$ Mm,
implying transition regions on the order of 10-50\arcsec\ in the AIA images, much larger than the pixels immediately around the loop base and roughly consistent with the large cool loops we see in the data.  
Thus, in order to get a better fit to EUV images showing extended emission in loop legs, it will be necessary to account for the finite spatial extent of the transition region. Note that it is only the upper transition region, the upper third as determined by temperature, where this is important. The temperature scale lengths in the lower and even middle transition region are extremely short.
The amount of transition-region emission predicted by EBTEL is reasonable (although see caveats about general transition-region modeling below), but because all of it is assigned to the footpoints by GX~Simulator, the footpoints are too bright and the extended legs do not appear in model images.

This explanation would explain Feature~A, the fan-loop feature in Figure~\ref{f:BestFit_171A_TRCorona}. It is not as clear if it accounts for Feature~B in that plot, in which the emission appears to extend farther than $0.05L$ along a magnetic tube. We think it unlikely Feature A extends into the solar wind because we would expect the emission along open magnetic field lines to be fainter, as they are in coronal holes. It is possible the actual feature is on closed magnetic field lines that extend outside the volume we are modeling. However, that would still leave the question of why we would see emission on such a structure but not along the model field lines that \textit{are} in Model~2.

Our group is working on multi-strand MHD simulations that will make it possible to determine DEMs for both corona and transition region as a function of location along the loop, which, if implemented in GX Simulator, should take care of this problem \citep{johnston_25,sow_mondal_25a}.

As discussed in \citet{fleishman_25}, one possible way to obtain more information about the transition region is to combine analysis of microwave emission, which is sensitive to only the corona, with EUV emission, which includes both transition region and coronal components. This will help to separate out the emission from the two sources. Another approach would be to consider active regions from multiple points of view using  AIA combined with either STEREO/EUVI \citep[Solar TErrestrial RElations Observatory/Extreme UltraViolet Imager;][]{howard_08} or Solar Orbiter/EUI \citep[Extreme Ultraviolet Imager;][]{rochus_20} near quadrature,  so that lower-altitude transition-region emission could be separated out. However, for both of these approaches it would be important to consider the extended transition-region emission from the long loops discussed in this paper.

\subsection{Loop Expansion}
\label{s:Discuss_LoopExpansion}

We note an inconsistency in our models. Loops expand with height in the magnetic field extrapolation of GX~Simulator, whereas the EBTEL simulations assume a constant loop cross section. Often, multiple voxels in a cut across the magnetic field map magnetically to a single transition-region voxel. Each of the coronal voxels is assigned a coronal DEM that corresponds to a companion transition-region DEM from the EBTEL simulation. The transition-region DEM appears one time in the active region model, but the coronal DEM appears multiple times. Consequently, these geometrical considerations suggest that the volume integrated coronal emission in the model is overestimated relative to the transition-region emission by a factor approximately equal to the area expansion factor of the loop.
 
However, this geometric effect is at least partially offset by the different thermodynamic properties of expanding and non-expanding loops. \citet[][Fig.~2]{cargill_22} found that the ratio of coronal to transition-region emission \textit{per unit area} is approximately independent of the expansion factor for a fixed heating rate. Therefore, for a given amount of transition-region emission, the volume integrated coronal emission should scale roughly with the expansion factor. This means that the thermodynamic effects missing from the EBTEL models (because they assume a constant cross section) fortuitously offset the geometric effects not accounted for in GX Simulator. In this sense, the current modeling is approximately correct. These effects must, of course, be investigated and verified more rigorously in future work. We note that EBTEL has the capability of simulating expanding loops, and this feature may need to be implemented in next-generation models. This is a significant undertaking as it adds a third dimension to the model lookup table.

We note also that \citet{cargill_22} find that expansion can affect the ratio of the length of the transition region to $L$, so incorporating loop expansion may also affect the treatment of the location of the transition region in the model (see discussion in Section~\ref{s:Discuss_TR} above.)

\subsection{Further Considerations}
\label{s:DiscussOther}
In addition to the treatment of the transition region and loop expansion there are other areas in which our modeling could be improved.  As stated above, the details of the core magnetic field lines do not reflect the data well. This is likely because this active region had a rather weak magnetic field. It was selected because we wished to have a very quiet active region so as to avoid solar flares in the time-average images. Also, we wished to avoid sunspots because it is highly unlikely that sunspots obey the same heating scaling law that applies throughout most of the active region. For one thing, loops rooted in sunspot umbrae have much different emission properties from other active region loops; they are cooler and fainter \citep{tiwari_17}. For another, many coronal heating models are based on the small-scale stressing of the field by photospheric convection, and convection is suppressed in sunspots. We therefore preferred to avoid this complication in our initial study, but of course sunspots must be included in an eventual complete model. A more typical active region, with stronger fields, may be easier to model magnetically. It would optimally make sense to have an iterative fit in which both the plasma and magnetic field parameters could be varied (see \citet{fleishman_21a} for a case where fits to radio data were done using multiple magnetic models), but that is beyond the scope of this paper and our readily available computing power.

Another area to consider is the assumptions going into the heating calculations in the lookup tables.
\citet{fleishman_25} and \citet{kuznetsov_25} considered the various lookup tables currently available in GX Simulator and, in analyses utilizing microwaves, concluded that the best fits for the diffuse coronal component come from assuming heating parameters based on higher frequency nanoflares than we used for this study or even constant heating.  So, ultimately, heating with various nanoflare distributions should be considered for the EUV as well. Furthermore, the heating frequency may vary across the active region \citep{mondal_25}, which should be incorporated into future models.
Power-law forms for the nanoflare energy distribution  are commonly assumed in the modeling community \citep[e.g.,][]{pauluhn_07,hood_16,knizhnik_18}  and are motivated by observed distributions \citep[see review by][]{reale_14}. However, the observed distributions are typically based on microflares and full-sized flares, and it is not obvious that the distributions can be extrapolated to smaller energies. Furthermore, the measured events are volume-integrated, whereas the nanoflares inputs to the EBTEL models are energy per unit cross-sectional area. Our multi-strand MHD simulations indicate that the energy per unit area actually obeys a log-normal distribution \citep{sow_mondal_25a}. The MHD simulations also indicate that the delay and energy of the nanoflares are uncorrelated \citep{sow_mondal_26}, contrary to the assumption in the current GX~Simulator lookup tables. These properties must be included  when new lookup tables are computed.

Although we concluded that it would not have helped in our case, it would be good to develop a way to estimate the loop lengths of field lines that exit the model box. At the moment the options are to either ignore the plasma in such lines or assume it is hydrostatic, neither of which is consistent with the modeling of the loops that close in the box. The calculation and manipulation of large volume NLFFF models needed to include more loops within the modeled box uses a lot of computer resources, so a way of estimating $L$ and \Bavg\ for large loops exiting smaller boxes would be helpful.


\section{Summary and Conclusions}
\label{s:Summary}

We used GX~Simulator to model the EUV flux from a small, quiescent active region averaged over a 4 hour period in order to determine the dependence of the volumetric heating, \Qavg, on the average magnetic field strength along the loop, \Bavg, and loop length, $L$, assuming $\Qavg\propto\Bavg^aL^{-b}$.  We found that for the AIA bands dominated by emission with $\logT\ga  6.3$ (211 and 335~\AA) the larger spatial scale intensity distribution was reasonably well fit, although the detailed structure  was different due to difficulties in the magnetic modeling of an active region with relatively weak field. Using a relatively low-resolution version of the model (8\arcsec) we found the best fit for 211~\AA\ has $\Qavg\approx7\times 10^{-3} (\Bavg/{100\mbox{G}})^{1.5} (L/{10^9\mbox{cm}})^{-1}$~erg~cm$^{-3}$~s$^{-1}$, and for 335~\AA\ the best fit is $\Qavg\approx4\times 10^{-3} (\Bavg/{100\mbox{G}})^{0.5} (L/{10^9\mbox{cm}})^{-1.5}$~erg~cm$^{-3}$~s$^{-1}$.
However, for all fits there is a range of parameter values that give qualitatively reasonable fits lying roughly in a line in ($a,b$) space, indicative of a dependence between $L$ and \Bavg. This line was approximately defined by $b\approx 1.6-0.6a$. This relationship is consistent with a \Bavg\ dependence on $L$.

We also found that for EUV lines showing a cooler (but still optically thin) plasma the models did not replicate the emission from loops along the edge of the active region, which dominates the emission in the 171 and 131~\AA\ bands and also contributes significantly to the 94 and 193~\AA\ bands.  We conclude that this emission is due to relatively high-altitude transition-region emission in these long loops. EBTEL defines the transition region to be the section of the loop in which thermal conduction serves to heat the plasma rather than cool it (as occurs in the coronal section), and as such the transition region generally fills a length about 5\% of the full loop length from the footpoint. GX~Simulator and the EBTEL model informing it assume that all transition-region emission comes from one voxel at the footpoint, which is not sufficient to model the spatial distribution of the emission in these AIA bands. Thus we conclude that it is important in modeling active regions to consider the spatial extent of the upper transition region as a function of temperature. The cooler part of the transition region is confined to low altitudes close to the footpoints.

\begin{acknowledgments}

T.A.K.\ and J.A.K.\ were supported by the GSFC Heliophysics Internal Scientist Funding Model (competitive work package program). This work is also part of the effort of the Committee on Space Research International Space Weather Action Team (COSPAR-ISWAT) S2-06: Origins of the Spectral Irradiance and its Intermediate Timescale Variability. GX~Simulator is a package available as part of Solar Soft \citep{freeland_98} and also on github at \url{https://github.com/Gelu-Nita/GX_SIMULATOR.} G.D.F. and G.M.N. were partly supported by the NSF grants RISE-2324724 and AST-2206424, NASA grant 80NSSC23K0090 to the New Jersey Institute of Technology and by CSTR/NJIT. The authors gratefully acknowledge the use of data from SDO/AIA and HMI. Thanks to the referee for thoughtful suggestions that have significantly enhanced the paper.

\software{SolarSoft}
\facility{SDO (AIA, HMI)}

\end{acknowledgments}

\bibliographystyle{apj}
\bibliography{bibliography}

@preamble{" \newcommand{\noop}[1]{} "}

@string{apj = {ApJ}}

@string{apjl = {Astrophys. Journ. Lett.}}

@string{apjs = {Astrophys. Journ. Supp.}}

@string{ssr = {Space Science Reviews}}

@article{antiochos_99,
	adsurl = {http://adsabs.harvard.edu/abs/1999ApJ...512..985A},
	author = {{Antiochos}, S.~K. and {MacNeice}, P.~J. and {Spicer}, D.~S. and {Klimchuk}, J.~A.},
	doi = {10.1086/306804},
	eprint = {astro-ph/9808199},
	journal = {\apj},
	month = feb,
	pages = {985-991},
	title = {{The Dynamic Formation of Prominence Condensations}},
	volume = 512,
	year = 1999}

@ARTICLE{barnes_19,
       author = {{Barnes}, W.~T. and {Bradshaw}, S.~J. and {Viall}, N.~M.},
        title = "{Understanding Heating in Active Region Cores through Machine Learning. I. Numerical Modeling and Predicted Observables}",
      journal = {\apj},
         year = 2019,
        month = jul,
       volume = {880},
       number = {1},
          eid = {56},
        pages = {56},
          doi = {10.3847/1538-4357/ab290c},
archivePrefix = {arXiv},
       eprint = {1906.03350},
 primaryClass = {astro-ph.SR},
       adsurl = {https://ui.adsabs.harvard.edu/abs/2019ApJ...880...56B}
}

@article{boerner_12,
	adsurl = {http://adsabs.harvard.edu/abs/2012SoPh..275...41B},
	author = {{Boerner}, P. and {Edwards}, C. and {Lemen}, J. and {Rausch}, A. and {Schrijver}, C. and {Shine}, R. and {Shing}, L. and {Stern}, R. and {Tarbell}, T. and {Title}, A. and {Wolfson}, C.~J. and {Soufli}, R. and {Spiller}, E. and {Gullikson}, E. and {McKenzie}, D. and {Windt}, D. and {Golub}, L. and {Podgorski}, W. and {Testa}, P. and {Weber}, M.},
	doi = {10.1007/s11207-011-9804-8},
	journal = {\solphys},
	month = jan,
	pages = {41-66},
	title = {{Initial Calibration of the Atmospheric Imaging Assembly (AIA) on the Solar Dynamics Observatory (SDO)}},
	volume = 275,
	year = 2012}

@ARTICLE{bourdin_13,
       author = {{Bourdin}, Ph. -A. and {Bingert}, S. and {Peter}, H.},
        title = "{Observationally driven 3D magnetohydrodynamics model of the solar corona above an active region}",
      journal = {\aap},
         year = 2013,
        month = jul,
       volume = {555},
          eid = {A123},
        pages = {A123},
          doi = {10.1051/0004-6361/201321185},
archivePrefix = {arXiv},
       eprint = {1305.5693},
 primaryClass = {astro-ph.SR},
       adsurl = {https://ui.adsabs.harvard.edu/abs/2013A&A...555A.123B}
}

@ARTICLE{cargill_12a,
       author = {{Cargill}, P.~J. and {Bradshaw}, S.~J. and {Klimchuk}, J.~A.},
        title = "{Enthalpy-based Thermal Evolution of Loops. II. Improvements to the Model}",
      journal = {\apj},
         year = 2012,
        month = jun,
       volume = {752},
       number = {2},
          eid = {161},
        pages = {161},
          doi = {10.1088/0004-637X/752/2/161},
archivePrefix = {arXiv},
       eprint = {1204.5960},
 primaryClass = {astro-ph.SR},
       adsurl = {https://ui.adsabs.harvard.edu/abs/2012ApJ...752..161C}
}

@ARTICLE{cargill_12b,
       author = {{Cargill}, P.~J. and {Bradshaw}, S.~J. and {Klimchuk}, J.~A.},
        title = "{Enthalpy-based Thermal Evolution of Loops. III. Comparison of Zero-dimensional Models}",
      journal = {\apj},
         year = 2012,
        month = oct,
       volume = {758},
       number = {1},
          eid = {5},
        pages = {5},
          doi = {10.1088/0004-637X/758/1/5},
       adsurl = {https://ui.adsabs.harvard.edu/abs/2012ApJ...758....5C}
}

@ARTICLE{cargill_22,
       author = {{Cargill}, P.~J. and {Bradshaw}, S.~J. and {Klimchuk}, J.~A. and {Barnes}, W.~T.},
        title = "{Static and dynamic solar coronal loops with cross-sectional area variations}",
      journal = {\mnras},
         year = 2022,
        month = jan,
       volume = {509},
       number = {3},
        pages = {4420-4429},
          doi = {10.1093/mnras/stab3163},
archivePrefix = {arXiv},
       eprint = {2111.09339},
 primaryClass = {astro-ph.SR},
       adsurl = {https://ui.adsabs.harvard.edu/abs/2022MNRAS.509.4420C}
}

@ARTICLE{cranmer_19,
       author = {{Cranmer}, Steven R. and {Winebarger}, Amy R.},
        title = "{The Properties of the Solar Corona and Its Connection to the Solar Wind}",
      journal = {\araa},
         year = 2019,
        month = aug,
       volume = {57},
        pages = {157-187},
          doi = {10.1146/annurev-astro-091918-104416},
archivePrefix = {arXiv},
       eprint = {1811.00461},
 primaryClass = {astro-ph.SR},
       adsurl = {https://ui.adsabs.harvard.edu/abs/2019ARA&A..57..157C}
}

@article{feldman_92a,
	adsurl = {http://adsabs.harvard.edu/abs/1992ApJS...81..387F},
	author = {{Feldman}, U. and {Mandelbaum}, P. and {Seely}, J.~F. and {Doschek}, G.~A. and {Gursky}, H.},
	doi = {10.1086/191698},
	journal = {\apjs},
	month = jul,
	pages = {387-408},
	title = {{The potential for plasma diagnostics from stellar extreme-ultraviolet observations}},
	volume = 81,
	year = 1992}

@ARTICLE{fleishman_17,
       author = {{Fleishman}, Gregory D. and {Anfinogentov}, Sergey and {Loukitcheva}, Maria and {Mysh'yakov}, Ivan and {Stupishin}, Alexey},
        title = "{Casting the Coronal Magnetic Field Reconstruction Tools in 3D Using the MHD Bifrost Model}",
      journal = {\apj},
         year = 2017,
        month = apr,
       volume = {839},
       number = {1},
          eid = {30},
        pages = {30},
          doi = {10.3847/1538-4357/aa6840},
archivePrefix = {arXiv},
       eprint = {1703.06360},
 primaryClass = {astro-ph.SR},
       adsurl = {https://ui.adsabs.harvard.edu/abs/2017ApJ...839...30F}
}

@ARTICLE{fleishman_19,
       author = {{Fleishman}, Gregory and {Mysh'yakov}, Ivan and {Stupishin}, Alexey and {Loukitcheva}, Maria and {Anfinogentov}, Sergey},
        title = "{Force-free Field Reconstructions Enhanced by Chromospheric Magnetic Field Data}",
      journal = {\apj},
         year = 2019,
        month = jan,
       volume = {870},
       number = {2},
          eid = {101},
        pages = {101},
          doi = {10.3847/1538-4357/aaf384},
archivePrefix = {arXiv},
       eprint = {1811.02093},
 primaryClass = {astro-ph.SR},
       adsurl = {https://ui.adsabs.harvard.edu/abs/2019ApJ...870..101F}
}

@ARTICLE{fleishman_21a,
       author = {{Fleishman}, Gregory D. and {Anfinogentov}, Sergey A. and {Stupishin}, Alexey G. and {Kuznetsov}, Alexey A. and {Nita}, Gelu M.},
        title = "{Coronal Heating Law Constrained by Microwave Gyroresonant Emission}",
      journal = {\apj},
         year = 2021,
        month = mar,
       volume = {909},
       number = {1},
          eid = {89},
        pages = {89},
          doi = {10.3847/1538-4357/abdab1},
archivePrefix = {arXiv},
       eprint = {2101.03651},
 primaryClass = {astro-ph.SR},
       adsurl = {https://ui.adsabs.harvard.edu/abs/2021ApJ...909...89F}
}

@ARTICLE{fleishman_21b,
       author = {{Fleishman}, Gregory D. and {Kuznetsov}, Alexey A. and {Landi}, Enrico},
        title = "{Gyroresonance and Free-Free Radio Emissions from Multithermal Multicomponent Plasma}",
      journal = {\apj},
         year = 2021,
        month = jun,
       volume = {914},
       number = {1},
          eid = {52},
        pages = {52},
          doi = {10.3847/1538-4357/abf92c},
archivePrefix = {arXiv},
       eprint = {2104.07655},
 primaryClass = {astro-ph.SR},
       adsurl = {https://ui.adsabs.harvard.edu/abs/2021ApJ...914...52F}
}

@ARTICLE{fleishman_25,
       author = {{Fleishman}, Gregory D. and {Kuznetsov}, Alexey A. and {Nita}, Gelu M.},
        title = "{Steady-State Heating of Diffuse Coronal Plasma in a Solar Active Region}",
      journal = {\apj\ in press},
     pubstate = {in press},
         year = 2025,
        month = jun,
          eid = {arXiv:2506.18723},
        pages = {arXiv:2506.18723},
archivePrefix = {arXiv},
       eprint = {2506.18723},
 primaryClass = {astro-ph.SR},
       adsurl = {https://ui.adsabs.harvard.edu/abs/2025arXiv250618723F}
}

@article{Fleishman_2025,
doi = {10.3847/1538-4357/ade3dd},
url = {https://dx.doi.org/10.3847/1538-4357/ade3dd},
year = {2025},
month = {jul},
publisher = {The American Astronomical Society},
volume = {988},
number = {1},
pages = {100},
author = {Fleishman, Gregory D. and Kuznetsov, Alexey A. and Nita, Gelu M.},
title = {Steady-state Heating of Diffuse Coronal Plasma in a Solar Active Region},
journal = {The Astrophysical Journal}
}

@article{freeland_98,
	adsurl = {http://adsabs.harvard.edu/abs/1998SoPh..182..497F},
	author = {{Freeland}, S.~L. and {Handy}, B.~N.},
	doi = {10.1023/A:1005038224881},
	journal = {\solphys},
	month = oct,
	pages = {497-500},
	title = {{Data Analysis with the SolarSoft System}},
	volume = 182,
	year = 1998}

@ARTICLE{gudiksen_02,
       author = {{Gudiksen}, Boris Vilhelm and {Nordlund}, {\r{A}}ke},
        title = "{Bulk Heating and Slender Magnetic Loops in the Solar Corona}",
      journal = {\apjl},
         year = 2002,
        month = jun,
       volume = {572},
       number = {1},
        pages = {L113-L116},
          doi = {10.1086/341600},
       adsurl = {https://ui.adsabs.harvard.edu/abs/2002ApJ...572L.113G}
}

@ARTICLE{gudiksen_05a,
       author = {{Gudiksen}, Boris Vilhelm and {Nordlund}, {\r{A}}ke},
        title = "{An Ab Initio Approach to the Solar Coronal Heating Problem}",
      journal = {\apj},
         year = 2005,
        month = jan,
       volume = {618},
       number = {2},
        pages = {1020-1030},
          doi = {10.1086/426063},
archivePrefix = {arXiv},
       eprint = {astro-ph/0407266},
 primaryClass = {astro-ph},
       adsurl = {https://ui.adsabs.harvard.edu/abs/2005ApJ...618.1020G}
}

@ARTICLE{hood_16,
       author = {{Hood}, A.~W. and {Cargill}, P.~J. and {Browning}, P.~K. and {Tam}, K.~V.},
        title = "{An MHD Avalanche in a Multi-threaded Coronal Loop.}",
      journal = {\apj},
         year = 2016,
        month = jan,
       volume = {817},
       number = {1},
          eid = {5},
        pages = {5},
          doi = {10.3847/0004-637X/817/1/5},
archivePrefix = {arXiv},
       eprint = {1512.00628},
 primaryClass = {astro-ph.SR},
       adsurl = {https://ui.adsabs.harvard.edu/abs/2016ApJ...817....5H}
}

@article{howard_08,
	adsurl = {http://adsabs.harvard.edu/abs/2008SSRv..136...67H},
	author = {{Howard}, R.~A. and {Moses}, J.~D. and {Vourlidas}, A. and {Newmark}, J.~S. and {Socker}, D.~G. and {Plunkett}, S.~P. and {Korendyke}, C.~M. and {Cook}, J.~W. and {Hurley}, A. and {Davila}, J.~M. and {Thompson}, W.~T. and {St Cyr}, O.~C. and {Mentzell}, E. and {Mehalick}, K. and {Lemen}, J.~R. and {Wuelser}, J.~P. and {Duncan}, D.~W. and {Tarbell}, T.~D. and {Wolfson}, C.~J. and {Moore}, A. and {Harrison}, R.~A. and {Waltham}, N.~R. and {Lang}, J. and {Davis}, C.~J. and {Eyles}, C.~J. and {Mapson-Menard}, H. and {Simnett}, G.~M. and {Halain}, J.~P. and {Defise}, J.~M. and {Mazy}, E. and {Rochus}, P. and {Mercier}, R. and {Ravet}, M.~F. and {Delmotte}, F. and {Auchere}, F. and {Delaboudiniere}, J.~P. and {Bothmer}, V. and {Deutsch}, W. and {Wang}, D. and {Rich}, N. and {Cooper}, S. and {Stephens}, V. and {Maahs}, G. and {Baugh}, R. and {McMullin}, D. and {Carter}, T.},
	doi = {10.1007/s11214-008-9341-4},
	journal = {\ssr},
	month = apr,
	pages = {67-115},
	title = {{Sun Earth Connection Coronal and Heliospheric Investigation (SECCHI)}},
	volume = 136,
	year = 2008}

@ARTICLE{ishigami_24,
       author = {{Ishigami}, Shun and {Hara}, Hirohisa and {Oba}, Takayoshi},
        title = "{Spectroscopic Study of Heating Distributions and Mechanisms Using Hinode/EIS}",
      journal = {\apj},
         year = 2024,
        month = nov,
       volume = {975},
       number = {2},
          eid = {289},
        pages = {289},
          doi = {10.3847/1538-4357/ad7def},
       adsurl = {https://ui.adsabs.harvard.edu/abs/2024ApJ...975..289I}
}

@ARTICLE{johnston_25,
       author = {{Johnston}, Craig D. and {Daldorff}, Lars K.~S. and {Klimchuk}, James A. and {Mondal}, Shanwlee Sow and {Barnes}, Will T. and {Leake}, James E. and {Reid}, Jack and {Parker}, Jacob D.},
        title = "{Self-Consistent Heating of the Magnetically Closed Solar Corona: Generation of Nanoflares, Thermodynamic Response of the Plasma and Observational Signatures}",
      journal = {\apj},
         year = 2025,
        month = dec,
       volume = {994},
       number = {2},
          eid = {139},
        pages = {139},
          doi = {10.3847/1538-4357/ae08a2},
archivePrefix = {arXiv},
       eprint = {2508.12952},
 primaryClass = {astro-ph.SR},
       adsurl = {https://ui.adsabs.harvard.edu/abs/2025ApJ...994..139J}
}

@ARTICLE{klimchuk_87b,
       author = {{Klimchuk}, James A.},
        title = "{On the Large-Scale Dynamics and Magnetic Structure of Solar Active Regions}",
      journal = {\apj},
         year = 1987,
        month = dec,
       volume = {323},
        pages = {368},
          doi = {10.1086/165834},
       adsurl = {https://ui.adsabs.harvard.edu/abs/1987ApJ...323..368K}
}

@ARTICLE{klimchuk_95,
       author = {{Klimchuk}, James A. and {Porter}, Lisa J.},
        title = "{Scaling of heating rates in solar coronal loops}",
      journal = {\nat},
         year = 1995,
        month = sep,
       volume = {377},
       number = {6545},
        pages = {131-133},
          doi = {10.1038/377131a0},
       adsurl = {https://ui.adsabs.harvard.edu/abs/1995Natur.377..131K}
}

@ARTICLE{klimchuk_08,
       author = {{Klimchuk}, J.~A. and {Patsourakos}, S. and {Cargill}, P.~J.},
        title = "{Highly Efficient Modeling of Dynamic Coronal Loops}",
      journal = {\apj},
         year = 2008,
        month = aug,
       volume = {682},
       number = {2},
        pages = {1351-1362},
          doi = {10.1086/589426},
archivePrefix = {arXiv},
       eprint = {0710.0185},
 primaryClass = {astro-ph},
       adsurl = {https://ui.adsabs.harvard.edu/abs/2008ApJ...682.1351K}
}

@ARTICLE{klimchuk_20,
       author = {{Klimchuk}, James A. and {DeForest}, Craig E.},
        title = "{Cross Sections of Coronal Loop Flux Tubes}",
      journal = {\apj},
         year = 2020,
        month = sep,
       volume = {900},
       number = {2},
          eid = {167},
        pages = {167},
          doi = {10.3847/1538-4357/abab09},
archivePrefix = {arXiv},
       eprint = {2007.15085},
 primaryClass = {astro-ph.SR},
       adsurl = {https://ui.adsabs.harvard.edu/abs/2020ApJ...900..167K}
}

@ARTICLE{klimchuk_23a,
       author = {{Klimchuk}, James A. and {Knizhnik}, Kalman J. and {Uritsky}, Vadim M.},
        title = "{Observational Signatures of Coronal Heating in Magnetohydrodynamic Simulations without Radiation or a Lower Atmosphere}",
      journal = {\apj},
         year = 2023,
        month = jan,
       volume = {942},
       number = {1},
          eid = {10},
        pages = {10},
          doi = {10.3847/1538-4357/ac9f41},
       adsurl = {https://ui.adsabs.harvard.edu/abs/2023ApJ...942...10K}
}

@ARTICLE{knizhnik_18,
       author = {{Knizhnik}, K.~J. and {Uritsky}, V.~M. and {Klimchuk}, J.~A. and {DeVore}, C.~R.},
        title = "{Power-law Statistics of Driven Reconnection in the Magnetically Closed Corona}",
      journal = {\apj},
         year = 2018,
        month = jan,
       volume = {853},
       number = {1},
          eid = {82},
        pages = {82},
          doi = {10.3847/1538-4357/aaa0d9},
archivePrefix = {arXiv},
       eprint = {1801.05245},
 primaryClass = {astro-ph.SR},
       adsurl = {https://ui.adsabs.harvard.edu/abs/2018ApJ...853...82K}
}

@ARTICLE{kuznetsov_25,
       author = {{Kuznetsov}, Alexey A. and {Fleishman}, Gregory D. and {Nita}, Gelu M. and {Anfinogentov}, Sergey A.},
        title = "{Magneto-thermal Coupling and Coronal Heating in Solar Active Regions Inferred from Microwave Observations}",
      journal = {\apj},
         year = 2025,
        month = oct,
       volume = {991},
       number = {2},
          eid = {186},
        pages = {186},
          doi = {10.3847/1538-4357/adfc4b},
archivePrefix = {arXiv},
       eprint = {2508.18647},
 primaryClass = {astro-ph.SR},
       adsurl = {https://ui.adsabs.harvard.edu/abs/2025ApJ...991..186K}
}

@article{lemen_12,
	adsurl = {http://adsabs.harvard.edu/abs/2012SoPh..275...17L},
	author = {{Lemen}, J.~R. and {Title}, A.~M. and {Akin}, D.~J. and {Boerner}, P.~F. and {Chou}, C. and {Drake}, J.~F. and {Duncan}, D.~W. and {Edwards}, C.~G. and {Friedlaender}, F.~M. and {Heyman}, G.~F. and {Hurlburt}, N.~E. and {Katz}, N.~L. and {Kushner}, G.~D. and {Levay}, M. and {Lindgren}, R.~W. and {Mathur}, D.~P. and {McFeaters}, E.~L. and {Mitchell}, S. and {Rehse}, R.~A. and {Schrijver}, C.~J. and {Springer}, L.~A. and {Stern}, R.~A. and {Tarbell}, T.~D. and {Wuelser}, J.-P. and {Wolfson}, C.~J. and {Yanari}, C. and {Bookbinder}, J.~A. and {Cheimets}, P.~N. and {Caldwell}, D. and {Deluca}, E.~E. and {Gates}, R. and {Golub}, L. and {Park}, S. and {Podgorski}, W.~A. and {Bush}, R.~I. and {Scherrer}, P.~H. and {Gummin}, M.~A. and {Smith}, P. and {Auker}, G. and {Jerram}, P. and {Pool}, P. and {Soufli}, R. and {Windt}, D.~L. and {Beardsley}, S. and {Clapp}, M. and {Lang}, J. and {Waltham}, N.},
	doi = {10.1007/s11207-011-9776-8},
	journal = {\solphys},
	month = jan,
	pages = {17-40},
	title = {{The Atmospheric Imaging Assembly (AIA) on the Solar Dynamics Observatory (SDO)}},
	volume = 275,
	year = 2012}

@ARTICLE{lundquist_08a,
       author = {{Lundquist}, L.~L. and {Fisher}, G.~H. and {McTiernan}, J.~M.},
        title = "{Forward Modeling of Active Region Coronal Emissions. I. Methods and Testing}",
      journal = {\apjs},
         year = 2008,
        month = dec,
       volume = {179},
       number = {2},
        pages = {509-533},
          doi = {10.1086/592775},
       adsurl = {https://ui.adsabs.harvard.edu/abs/2008ApJS..179..509L}
}

@ARTICLE{lundquist_08b,
       author = {{Lundquist}, L.~L. and {Fisher}, G.~H. and {Metcalf}, T.~R. and {Leka}, K.~D. and {McTiernan}, J.~M.},
        title = "{Forward Modeling of Active Region Coronal Emissions. II. Implications for Coronal Heating}",
      journal = {\apj},
         year = 2008,
        month = dec,
       volume = {689},
       number = {2},
        pages = {1388-1405},
          doi = {10.1086/592760},
       adsurl = {https://ui.adsabs.harvard.edu/abs/2008ApJ...689.1388L}
}

@ARTICLE{mandrini_00,
       author = {{Mandrini}, C.~H. and {D{\'e}moulin}, P. and {Klimchuk}, J.~A.},
        title = "{Magnetic Field and Plasma Scaling Laws: Their Implications for Coronal Heating Models}",
      journal = {\apj},
         year = 2000,
        month = feb,
       volume = {530},
       number = {2},
        pages = {999-1015},
          doi = {10.1086/308398},
       adsurl = {https://ui.adsabs.harvard.edu/abs/2000ApJ...530..999M}
}

@ARTICLE{mok_05,
       author = {{Mok}, Yung and {Miki{\'c}}, Zoran and {Lionello}, Roberto and {Linker}, Jon A.},
        title = "{Calculating the Thermal Structure of Solar Active Regions in Three Dimensions}",
      journal = {\apj},
         year = 2005,
        month = mar,
       volume = {621},
       number = {2},
        pages = {1098-1108},
          doi = {10.1086/427739},
       adsurl = {https://ui.adsabs.harvard.edu/abs/2005ApJ...621.1098M}
}

@ARTICLE{mok_08,
       author = {{Mok}, Yung and {Miki{\'c}}, Zoran and {Lionello}, Roberto and {Linker}, Jon A.},
        title = "{The Formation of Coronal Loops by Thermal Instability in Three Dimensions}",
      journal = {\apjl},
         year = 2008,
        month = jun,
       volume = {679},
       number = {2},
        pages = {L161},
          doi = {10.1086/589440},
       adsurl = {https://ui.adsabs.harvard.edu/abs/2008ApJ...679L.161M}
}

@ARTICLE{mok_16,
       author = {{Mok}, Yung and {Miki{\'c}}, Zoran and {Lionello}, Roberto and {Downs}, Cooper and {Linker}, Jon A.},
        title = "{A Three-dimensional Model of Active Region 7986: Comparison of Simulations with Observations}",
      journal = {\apj},
         year = 2016,
        month = jan,
       volume = {817},
       number = {1},
          eid = {15},
        pages = {15},
          doi = {10.3847/0004-637X/817/1/15},
       adsurl = {https://ui.adsabs.harvard.edu/abs/2016ApJ...817...15M}
}

@ARTICLE{mondal_25,
       author = {{Mondal}, Biswajit and {Klimchuk}, James A. and {Winebarger}, Amy R. and {Athiray}, P.~S. and {Liu}, Jiayi},
        title = "{Spatial and Temporal Distribution of Nanoflare Heating during Active Region Evolution}",
      journal = {\apj},
         year = 2025,
        month = feb,
       volume = {980},
       number = {1},
          eid = {75},
        pages = {75},
          doi = {10.3847/1538-4357/ada3d6},
archivePrefix = {arXiv},
       eprint = {2412.20348},
 primaryClass = {astro-ph.SR},
       adsurl = {https://ui.adsabs.harvard.edu/abs/2025ApJ...980...75M}
}

@ARTICLE{nita_15,
       author = {{Nita}, Gelu M. and {Fleishman}, Gregory D. and {Kuznetsov}, Alexey A. and {Kontar}, Eduard P. and {Gary}, Dale E.},
        title = "{Three-dimensional Radio and X-Ray Modeling and Data Analysis Software: Revealing Flare Complexity}",
      journal = {\apj},
         year = 2015,
        month = feb,
       volume = {799},
       number = {2},
          eid = {236},
        pages = {236},
          doi = {10.1088/0004-637X/799/2/236},
archivePrefix = {arXiv},
       eprint = {1409.0896},
 primaryClass = {astro-ph.SR},
       adsurl = {https://ui.adsabs.harvard.edu/abs/2015ApJ...799..236N}
}

@ARTICLE{nita_18,
       author = {{Nita}, Gelu M. and {Viall}, Nicholeen M. and {Klimchuk}, James A. and {Loukitcheva}, Maria A. and {Gary}, Dale E. and {Kuznetsov}, Alexey A. and {Fleishman}, Gregory D.},
        title = "{Dressing the Coronal Magnetic Extrapolations of Active Regions with a Parameterized Thermal Structure}",
      journal = {\apj},
         year = 2018,
        month = jan,
       volume = {853},
       number = {1},
          eid = {66},
        pages = {66},
          doi = {10.3847/1538-4357/aaa4bf},
       adsurl = {https://ui.adsabs.harvard.edu/abs/2018ApJ...853...66N}
}

@ARTICLE{nita_23,
       author = {{Nita}, Gelu M. and {Fleishman}, Gregory D. and {Kuznetsov}, Alexey A. and {Anfinogentov}, Sergey A. and {Stupishin}, Alexey G. and {Kontar}, Eduard P. and {Schonfeld}, Samuel J. and {Klimchuk}, James A. and {Gary}, Dale E.},
        title = "{Data-constrained Solar Modeling with GX Simulator}",
      journal = {\apjs},
         year = 2023,
        month = jul,
       volume = {267},
       number = {1},
          eid = {6},
        pages = {6},
          doi = {10.3847/1538-4365/acd343},
archivePrefix = {arXiv},
       eprint = {2301.00795},
 primaryClass = {astro-ph.SR},
       adsurl = {https://ui.adsabs.harvard.edu/abs/2023ApJS..267....6N}
}

@ARTICLE{pauluhn_07,
       author = {{Pauluhn}, A. and {Solanki}, S.~K.},
        title = "{A nanoflare model of quiet Sun EUV emission}",
      journal = {\aap},
         year = 2007,
        month = jan,
       volume = {462},
       number = {1},
        pages = {311-322},
          doi = {10.1051/0004-6361:20065152},
archivePrefix = {arXiv},
       eprint = {astro-ph/0612585},
 primaryClass = {astro-ph},
       adsurl = {https://ui.adsabs.harvard.edu/abs/2007A&A...462..311P}
}

@article{pesnell_12,
	adsurl = {http://adsabs.harvard.edu/abs/2012SoPh..275....3P},
	author = {{Pesnell}, W.~D. and {Thompson}, B.~J. and {Chamberlin}, P.~C.},
	doi = {10.1007/s11207-011-9841-3},
	journal = {\solphys},
	month = jan,
	pages = {3-15},
	title = {{The Solar Dynamics Observatory (SDO)}},
	volume = 275,
	year = 2012}

@ARTICLE{plowman_21,
       author = {{Plowman}, Joseph},
        title = "{Three-dimensional Reconstruction of Coronal Plasma Properties from a Single Perspective}",
      journal = {\apj},
         year = 2021,
        month = dec,
       volume = {922},
       number = {2},
          eid = {109},
        pages = {109},
          doi = {10.3847/1538-4357/ac2664},
archivePrefix = {arXiv},
       eprint = {2103.02028},
 primaryClass = {astro-ph.SR},
       adsurl = {https://ui.adsabs.harvard.edu/abs/2021ApJ...922..109P}
}

@ARTICLE{plowman_23,
       author = {{Plowman}, Joseph},
        title = "{Validation and Testing of the CROBAR 3D Coronal Reconstruction Method with a MURaM Simulation}",
      journal = {\apj},
         year = 2023,
        month = apr,
       volume = {947},
       number = {1},
          eid = {5},
        pages = {5},
          doi = {10.3847/1538-4357/acbc71},
archivePrefix = {arXiv},
       eprint = {2209.01753},
 primaryClass = {astro-ph.SR},
       adsurl = {https://ui.adsabs.harvard.edu/abs/2023ApJ...947....5P}
}

@ARTICLE{porter_95,
       author = {{Porter}, Lisa J. and {Klimchuk}, James A.},
        title = "{Soft X-Ray Loops and Coronal Heating}",
      journal = {\apj},
         year = 1995,
        month = nov,
       volume = {454},
        pages = {499},
          doi = {10.1086/176501},
       adsurl = {https://ui.adsabs.harvard.edu/abs/1995ApJ...454..499P}
}

@ARTICLE{reale_14,
       author = {{Reale}, Fabio},
        title = "{Coronal Loops: Observations and Modeling of Confined Plasma}",
      journal = {Living Reviews in Solar Physics},
         year = 2014,
        month = dec,
       volume = {11},
       number = {1},
          eid = {4},
        pages = {4},
          doi = {10.12942/lrsp-2014-4},
       adsurl = {https://ui.adsabs.harvard.edu/abs/2014LRSP...11....4R}
}

@ARTICLE{rempel_17,
       author = {{Rempel}, M.},
        title = "{Extension of the MURaM Radiative MHD Code for Coronal Simulations}",
      journal = {\apj},
         year = 2017,
        month = jan,
       volume = {834},
       number = {1},
          eid = {10},
        pages = {10},
          doi = {10.3847/1538-4357/834/1/10},
archivePrefix = {arXiv},
       eprint = {1609.09818},
 primaryClass = {astro-ph.SR},
       adsurl = {https://ui.adsabs.harvard.edu/abs/2017ApJ...834...10R}
}

@ARTICLE{rochus_20,
       author = {{Rochus}, P. and {Auch{\`e}re}, F. and {Berghmans}, D. and {Harra}, L. and {Schmutz}, W. and {Sch{\"u}hle}, U. and {Addison}, P. and {Appourchaux}, T. and {Aznar Cuadrado}, R. and {Baker}, D. and {Barbay}, J. and {Bates}, D. and {BenMoussa}, A. and {Bergmann}, M. and {Beurthe}, C. and {Borgo}, B. and {Bonte}, K. and {Bouzit}, M. and {Bradley}, L. and {B{\"u}chel}, V. and {Buchlin}, E. and {B{\"u}chner}, J. and {Cab{\'e}}, F. and {Cadiergues}, L. and {Chaigneau}, M. and {Chares}, B. and {Choque Cortez}, C. and {Coker}, P. and {Condamin}, M. and {Coumar}, S. and {Curdt}, W. and {Cutler}, J. and {Davies}, D. and {Davison}, G. and {Defise}, J. -M. and {Del Zanna}, G. and {Delmotte}, F. and {Delouille}, V. and {Dolla}, L. and {Dumesnil}, C. and {D{\"u}rig}, F. and {Enge}, R. and {Fran{\c{c}}ois}, S. and {Fourmond}, J. -J. and {Gillis}, J. -M. and {Giordanengo}, B. and {Gissot}, S. and {Green}, L.~M. and {Guerreiro}, N. and {Guilbaud}, A. and {Gyo}, M. and {Haberreiter}, M. and {Hafiz}, A. and {Hailey}, M. and {Halain}, J. -P. and {Hansotte}, J. and {Hecquet}, C. and {Heerlein}, K. and {Hellin}, M. -L. and {Hemsley}, S. and {Hermans}, A. and {Hervier}, V. and {Hochedez}, J. -F. and {Houbrechts}, Y. and {Ihsan}, K. and {Jacques}, L. and {J{\'e}r{\^o}me}, A. and {Jones}, J. and {Kahle}, M. and {Kennedy}, T. and {Klaproth}, M. and {Kolleck}, M. and {Koller}, S. and {Kotsialos}, E. and {Kraaikamp}, E. and {Langer}, P. and {Lawrenson}, A. and {Le Clech'}, J. -C. and {Lenaerts}, C. and {Liebecq}, S. and {Linder}, D. and {Long}, D.~M. and {Mampaey}, B. and {Markiewicz-Innes}, D. and {Marquet}, B. and {Marsch}, E. and {Matthews}, S. and {Mazy}, E. and {Mazzoli}, A. and {Meining}, S. and {Meltchakov}, E. and {Mercier}, R. and {Meyer}, S. and {Monecke}, M. and {Monfort}, F. and {Morinaud}, G. and {Moron}, F. and {Mountney}, L. and {M{\"u}ller}, R. and {Nicula}, B. and {Parenti}, S. and {Peter}, H. and {Pfiffner}, D. and {Philippon}, A. and {Phillips}, I. and {Plesseria}, J. -Y. and {Pylyser}, E. and {Rabecki}, F. and {Ravet-Krill}, M. -F. and {Rebellato}, J. and {Renotte}, E. and {Rodriguez}, L. and {Roose}, S. and {Rosin}, J. and {Rossi}, L. and {Roth}, P. and {Rouesnel}, F. and {Roulliay}, M. and {Rousseau}, A. and {Ruane}, K. and {Scanlan}, J. and {Schlatter}, P. and {Seaton}, D.~B. and {Silliman}, K. and {Smit}, S. and {Smith}, P.~J. and {Solanki}, S.~K. and {Spescha}, M. and {Spencer}, A. and {Stegen}, K. and {Stockman}, Y. and {Szwec}, N. and {Tamiatto}, C. and {Tandy}, J. and {Teriaca}, L. and {Theobald}, C. and {Tychon}, I. and {van Driel-Gesztelyi}, L. and {Verbeeck}, C. and {Vial}, J. -C. and {Werner}, S. and {West}, M.~J. and {Westwood}, D. and {Wiegelmann}, T. and {Willis}, G. and {Winter}, B. and {Zerr}, A. and {Zhang}, X. and {Zhukov}, A.~N.},
        title = "{The Solar Orbiter EUI instrument: The Extreme Ultraviolet Imager}",
      journal = {\aap},
         year = 2020,
        month = oct,
       volume = {642},
          eid = {A8},
        pages = {A8},
          doi = {10.1051/0004-6361/201936663},
       adsurl = {https://ui.adsabs.harvard.edu/abs/2020A&A...642A...8R}
}

@article{scherrer_12,
	adsurl = {https://ui.adsabs.harvard.edu/abs/2012SoPh..275..207S},
	author = {{Scherrer}, P.~H. and {Schou}, J. and {Bush}, R.~I. and {Kosovichev}, A.~G. and {Bogart}, R.~S. and {Hoeksema}, J.~T. and {Liu}, Y. and {Duvall}, T.~L. and {Zhao}, J. and {Title}, A.~M. and {Schrijver}, C.~J. and {Tarbell}, T.~D. and {Tomczyk}, S.},
	doi = {10.1007/s11207-011-9834-2},
	journal = {\solphys},
	month = {Jan},
	number = {1-2},
	pages = {207-227},
	title = {{The Helioseismic and Magnetic Imager (HMI) Investigation for the Solar Dynamics Observatory (SDO)}},
	volume = {275},
	year = {2012}}

@article{schrijver_04,
	adsurl = {http://adsabs.harvard.edu/abs/2004ApJ...615..512S},
	author = {{Schrijver}, C.~J. and {Sandman}, A.~W. and {Aschwanden}, M.~J. and {De Rosa}, M.~L.},
	doi = {10.1086/424028},
	journal = {\apj},
	month = nov,
	pages = {512-525},
	title = {{The Coronal Heating Mechanism as Identified by Full-Sun Visualizations}},
	volume = 615,
	year = 2004}

@ARTICLE{sow_mondal_25a,
       author = {{Sow Mondal}, Shanwlee and {Daldorff}, Lars K.~S. and {Klimchuk}, James A. and {Johnston}, Craig. D.},
        title = "{Characterizing Nanoflare Energy and Frequency through Field Line Analysis}",
      journal = {\apj},
         year = 2025,
        month = nov,
       volume = {994},
       number = {1},
          eid = {71},
        pages = {71},
          doi = {10.3847/1538-4357/ae0cac},
       adsurl = {https://ui.adsabs.harvard.edu/abs/2025ApJ...994...71S}
}

@ARTICLE{sow_mondal_26,
       author = {{Sow Mondal}, Shanwlee and {Klimchuk}, James A. and {Johnston}, Craig D. and {Daldorff}, Lars K.~S.},
        title = "{On the Relationship between Nanoflare Energy and Delay in the Closed Solar Corona}",
      journal = {\apj},
         year = 2026,
        month = feb,
       volume = {998},
       number = {1},
          eid = {185},
        pages = {185},
          doi = {10.3847/1538-4357/ae3158},
archivePrefix = {arXiv},
       eprint = {2512.20875},
 primaryClass = {astro-ph.SR},
       adsurl = {https://ui.adsabs.harvard.edu/abs/2026ApJ...998..185S}
}

@ARTICLE{tiwari_17,
       author = {{Tiwari}, Sanjiv K. and {Thalmann}, Julia K. and {Panesar}, Navdeep K. and {Moore}, Ronald L. and {Winebarger}, Amy R.},
        title = "{New Evidence that Magnetoconvection Drives Solar-Stellar Coronal Heating}",
      journal = {\apjl},
         year = 2017,
        month = jul,
       volume = {843},
       number = {2},
          eid = {L20},
        pages = {L20},
          doi = {10.3847/2041-8213/aa794c},
archivePrefix = {arXiv},
       eprint = {1706.08035},
 primaryClass = {astro-ph.SR},
       adsurl = {https://ui.adsabs.harvard.edu/abs/2017ApJ...843L..20T}
}

@ARTICLE{ugarte_17,
       author = {{Ugarte-Urra}, Ignacio and {Warren}, Harry P. and {Upton}, Lisa A. and {Young}, Peter R.},
        title = "{Modeling Coronal Response in Decaying Active Regions with Magnetic Flux Transport and Steady Heating}",
      journal = {\apj},
         year = 2017,
        month = sep,
       volume = {846},
       number = {2},
          eid = {165},
        pages = {165},
          doi = {10.3847/1538-4357/aa8597},
archivePrefix = {arXiv},
       eprint = {1708.04324},
 primaryClass = {astro-ph.SR},
       adsurl = {https://ui.adsabs.harvard.edu/abs/2017ApJ...846..165U}
}

@ARTICLE{ugarte_19,
       author = {{Ugarte-Urra}, Ignacio and {Crump}, Nicholas A. and {Warren}, Harry P. and {Wiegelmann}, Thomas},
        title = "{The Magnetic Properties of Heating Events on High-temperature Active-region Loops}",
      journal = {\apj},
         year = 2019,
        month = jun,
       volume = {877},
       number = {2},
          eid = {129},
        pages = {129},
          doi = {10.3847/1538-4357/ab1d4d},
archivePrefix = {arXiv},
       eprint = {1904.11976},
 primaryClass = {astro-ph.SR},
       adsurl = {https://ui.adsabs.harvard.edu/abs/2019ApJ...877..129U}
}

@ARTICLE{vesecky_79,
       author = {{Vesecky}, J.~F. and {Antiochos}, S.~K. and {Underwood}, J.~H.},
        title = "{Numerical modeling of quasi-static coronal loops. I. Uniform energy input.}",
      journal = {\apj},
         year = 1979,
        month = nov,
       volume = {233},
       number = {3},
        pages = {987-997},
          doi = {10.1086/157462},
       adsurl = {https://ui.adsabs.harvard.edu/abs/1979ApJ...233..987V}
}

@ARTICLE{warnecke_19,
       author = {{Warnecke}, J. and {Peter}, H.},
        title = "{Data-driven model of the solar corona above an active region}",
      journal = {\aap},
         year = 2019,
        month = apr,
       volume = {624},
          eid = {L12},
        pages = {L12},
          doi = {10.1051/0004-6361/201935385},
archivePrefix = {arXiv},
       eprint = {1903.00455},
 primaryClass = {astro-ph.SR},
       adsurl = {https://ui.adsabs.harvard.edu/abs/2019A&A...624L..12W}
}

@ARTICLE{warren_06,
       author = {{Warren}, Harry P. and {Winebarger}, Amy R.},
        title = "{Hydrostatic Modeling of the Integrated Soft X-Ray and Extreme Ultraviolet Emission in Solar Active Regions}",
      journal = {\apj},
         year = 2006,
        month = jul,
       volume = {645},
       number = {1},
        pages = {711-719},
          doi = {10.1086/504075},
archivePrefix = {arXiv},
       eprint = {astro-ph/0602052},
 primaryClass = {astro-ph},
       adsurl = {https://ui.adsabs.harvard.edu/abs/2006ApJ...645..711W}
}

@ARTICLE{wiegelmann_04,
       author = {{Wiegelmann}, T.},
        title = "{Optimization code with weighting function for the reconstruction of coronal magnetic fields}",
      journal = {\solphys},
         year = 2004,
        month = jan,
       volume = {219},
       number = {1},
        pages = {87-108},
          doi = {10.1023/B:SOLA.0000021799.39465.36},
archivePrefix = {arXiv},
       eprint = {0802.0124},
 primaryClass = {astro-ph},
       adsurl = {https://ui.adsabs.harvard.edu/abs/2004SoPh..219...87W}
}

@misc{stupishin_20,
  author    = {{Stupishin}, A.},
title = "{Magnetic-Field\_Library: NLFFF and magnetic lines}",
 year = 2023,
month = dec,
  eid = {10.5281/zenodo.3896222},
  doi = {10.5281/zenodo.3896222},
version = {v3.4.23.1203},
publisher = {Zenodo},
adsurl = {https://ui.adsabs.harvard.edu/abs/2022zndo...3896222A}
}

@misc{yu_25,
  author       = {Sijie Yu and
                  Fedenyov, Viktor and
                  Stupishin, Alexey and
                  Nita, Gelu},
  title        = {Python Automatic Model Production Pipeline for solar coronal modeling (pyAMPP)
                  },
  month        = jun,
  year         = 2025,
  publisher    = {Zenodo},
  version      = {v0.1.0-alpha},
  doi          = {10.5281/zenodo.15660388},
  url          = {https://doi.org/10.5281/zenodo.15660388},
  swhid        = {swh:1:dir:5332ce7c72bc1b156d14514d0d7e6d80a33f72fd
                   ;origin=https://doi.org/10.5281/zenodo.15660387;vi
                   sit=swh:1:snp:b3130bee7c55dcd59fd9bda6db8fac8f936e
                   116b;anchor=swh:1:rel:4115edfaba0e4ecffe7d23081632
                   d92e9436a698;path=suncast-org-pyAMPP-70316f2
                  },
}

\appendix

\section{Transition-Region Mask}
\label{a:TRmask}


We present here a short physical justification for the transition-region mask (see Section~\ref{s:Fitting}), which is expanded upon in more detail by \citet{nita_18} in their Section 4.1 and Appendix A.

For a given loop length and heating rate, the brightness density of the transition-region emission at the solar surface decreases as the loop leg becomes more inclined to vertical, as discussed after Equation~5 of Nita et al. (2018). One way to understand this is as follows. The radiative energy losses from the transition region per unit cross-sectional area depend on the loop length and heating rate and are independent of the base inclination. By definition, the cross section is perpendicular to the loop axis. It is different from the footprint of the loop on the solar surface, which is larger than the cross section by an amount that increases with the tilt from vertical. Because the fixed radiative losses are spread out over a larger area, the brightness density of the footprint decreases with increasing tilt.
 
Photospheric magnetograms of active regions reveal regions of strong vertical magnetic field surrounded by regions of much weaker vertical field \citep[e.g.,][]{klimchuk_87b}. This gives rise to a magnetic canopy, where the strong vertical fields bend over sharply to become nearly horizontal in the chromosphere and transition region above the weak fields. From the discussion above, the transition region is observed to be much fainter above weak fields than strong fields.
 
Canopies do not arise in extrapolated magnetic field models because the transition from large to small plasma beta happens discontinuously at the lower boundary of the model. The field is much less inclined above regions of photospheric weak field than on the real Sun and therefore the predicted transition-region emission is much brighter. To correct this, we impose a transition-region mask. We set the transition-region emission to zero wherever the photospheric field is weaker than a critical value, which we take to be 200~G.

\section{AIA Model Grids}
\label{a:fit_thumbnails}
\begin{figure}
 \includegraphics[width=18cm, trim=0 0 0 0]{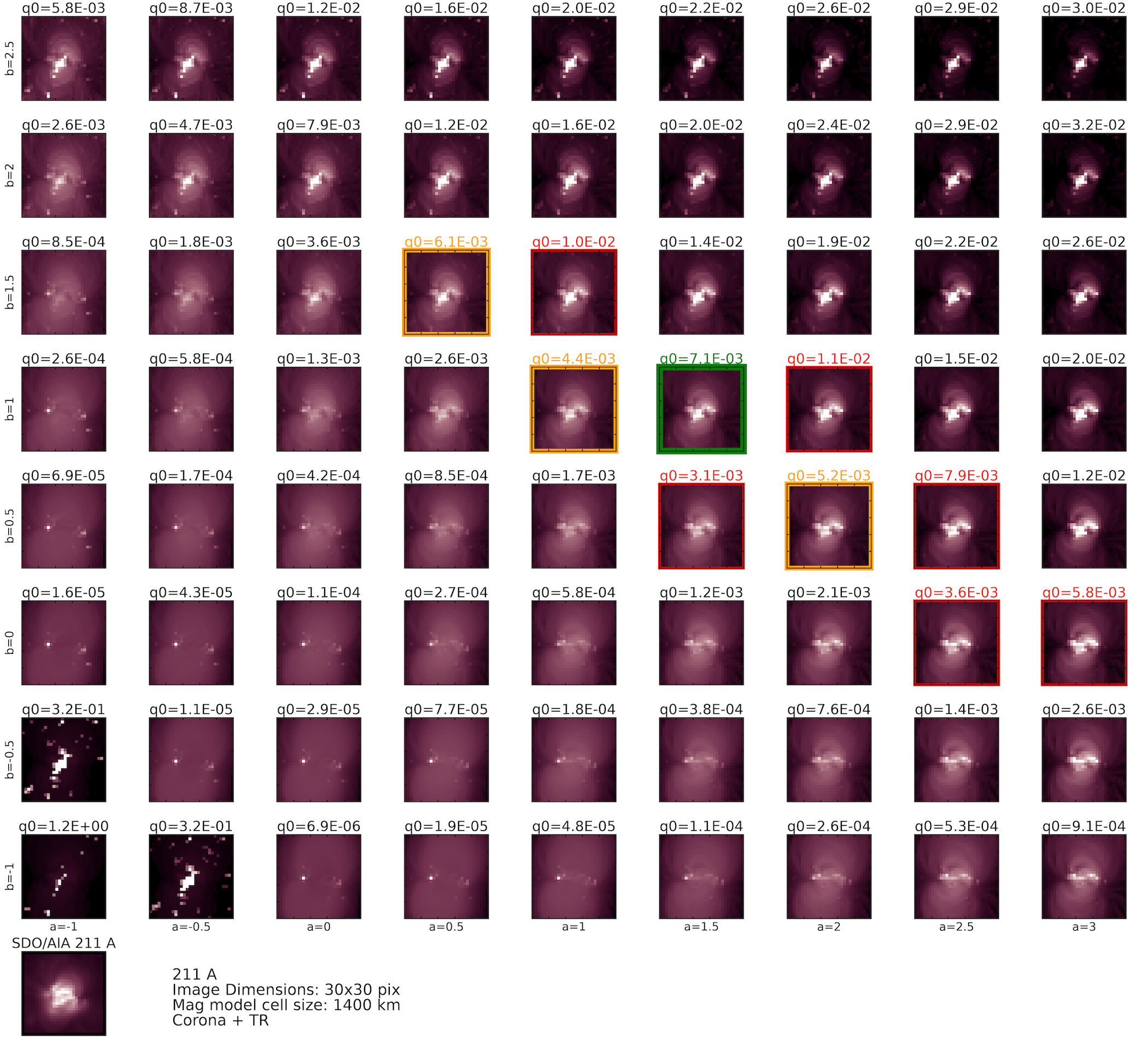}
\caption{Thumbnails of model 211~\AA\ images showing the best $q_0$ solutions for each of the ($a,b$) pairs considered for calculations using Magnetic Model~1, a model/data comparison resolution of 8\arcsec, and including both the corona and transitions region. The actual 211~\AA\ image at the same resolution is on the bottom left. The $a=1.5$, $b=1$ image has the lowest \GoF\ value. Its title and border are in green. See Fig.~\ref{f:BestFit_211A_TRCorona_8arcsec}b for larger version of the best model. Maps with \GoF\  within 10\% and 20\% of the minimum have titles and borders in orange and red, respectively. The images are all scaled linearly with the same intensity range. }
\label{f:Thumbnails_211A_nx30_YesTR}
\end{figure}

\begin{figure}
 \includegraphics[width=18cm, trim=0 0 0 0]{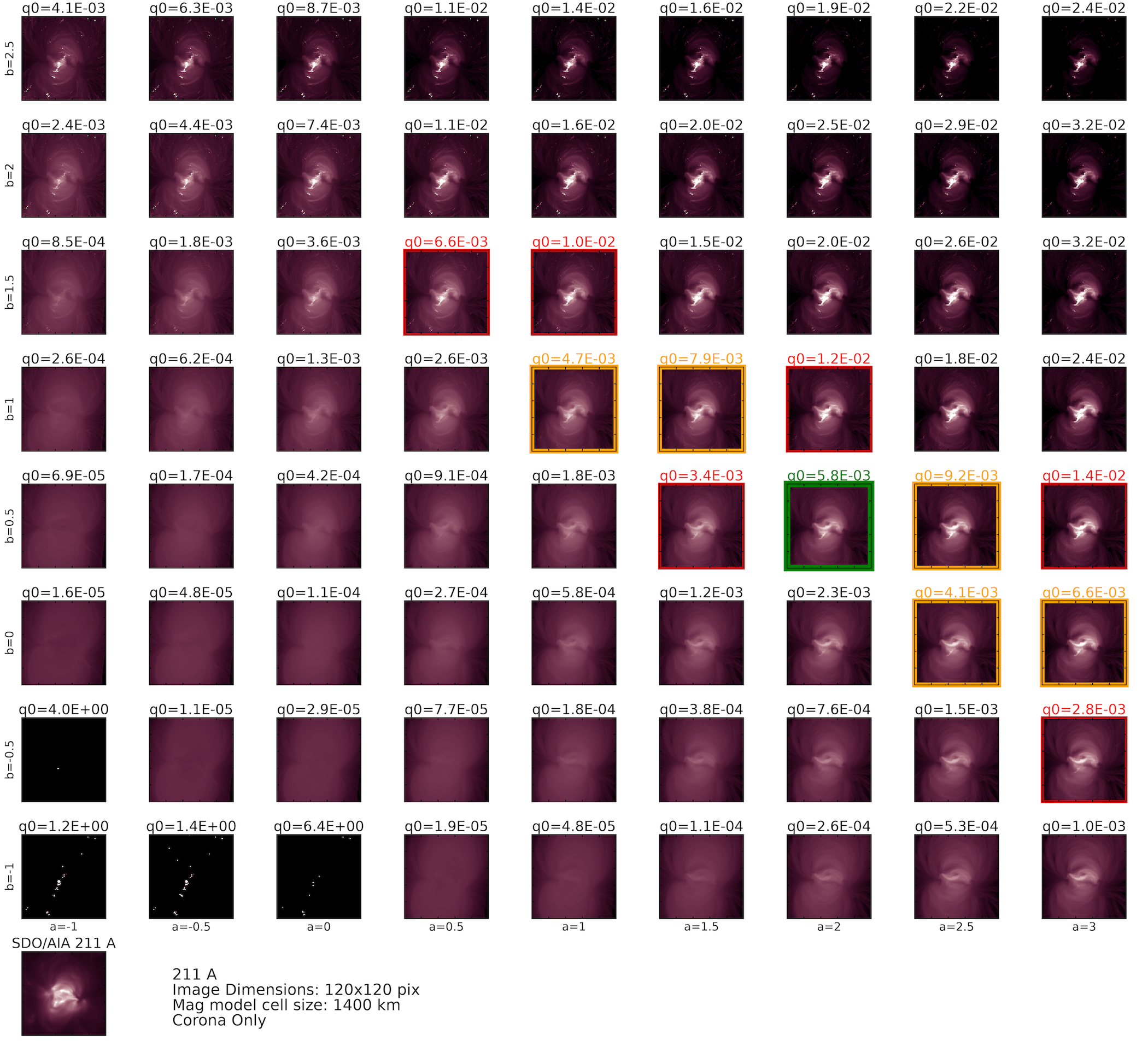}
\caption{Thumbnails of model 211~\AA\ images showing the best $q_0$ solutions for each of the ($a,b$) pairs considered for calculations using Magnetic Model~1, a model/data comparison resolution of 2\arcsec, and including the corona only. The actual 211~\AA\ image at the same resolution is at the bottom left. The $a=2$, $b=0.5$ image has the lowest \GoF\ value. Its title and border are in green. See Fig.~\ref{f:BestFit211A_noTR}b for larger version of the best model. Maps with \GoF\  within 10\% and 20\% of the minimum have titles and borders in orange and red, respectively. The images are all scaled linearly with the same intensity range. }
\label{f:Thumbnails_211A_NoTR}
\end{figure}

\begin{figure}
\includegraphics[width=18cm, trim=0 0 0 0]{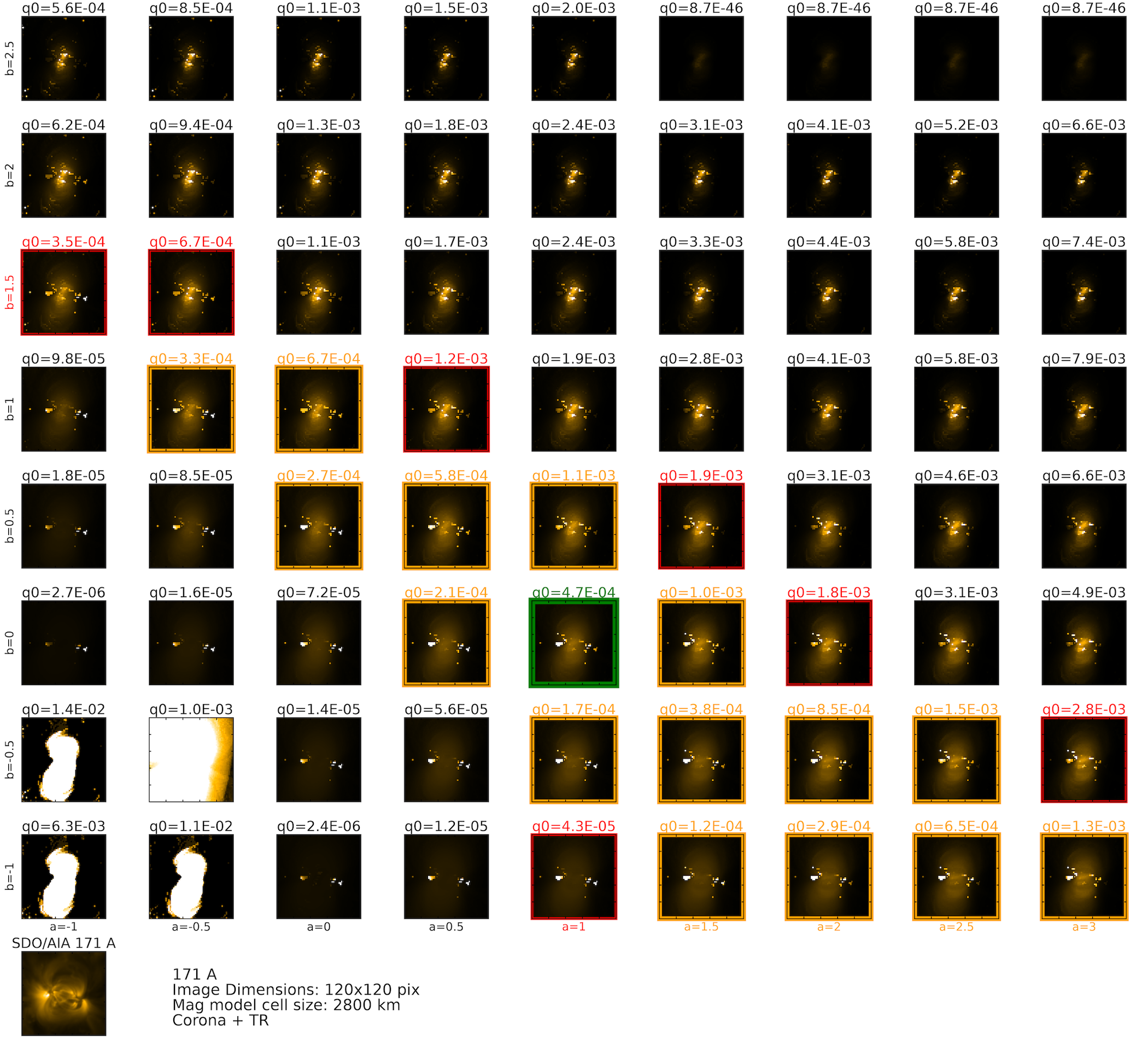}
\caption{Thumbnails of model 171~\AA\ images showing the best $q_0$ solutions for each of the ($a,b$) pairs considered for calculations using Magnetic Model~2, a model/data comparison resolution of 2\arcsec, and including the corona and transition region. The actual 171~\AA\ image at the same resolution is at the bottom left. The $a=1.0$, $b=0$ image has the lowest \GoF\ value. Its title and border are in green. See Fig.~\ref{f:BestFitMod2_171A_TRCorona}b for larger version of the best model. Maps with \GoF\  within 10\% and 20\% of the minimum have titles and borders in orange and red, respectively. The images are all scaled linearly with the same intensity range.}
\label{f:Thumbnails_171A_dx2800_YesTR}
\end{figure}

\begin{figure}
 \includegraphics[width=18cm, trim=0 0 0 0]{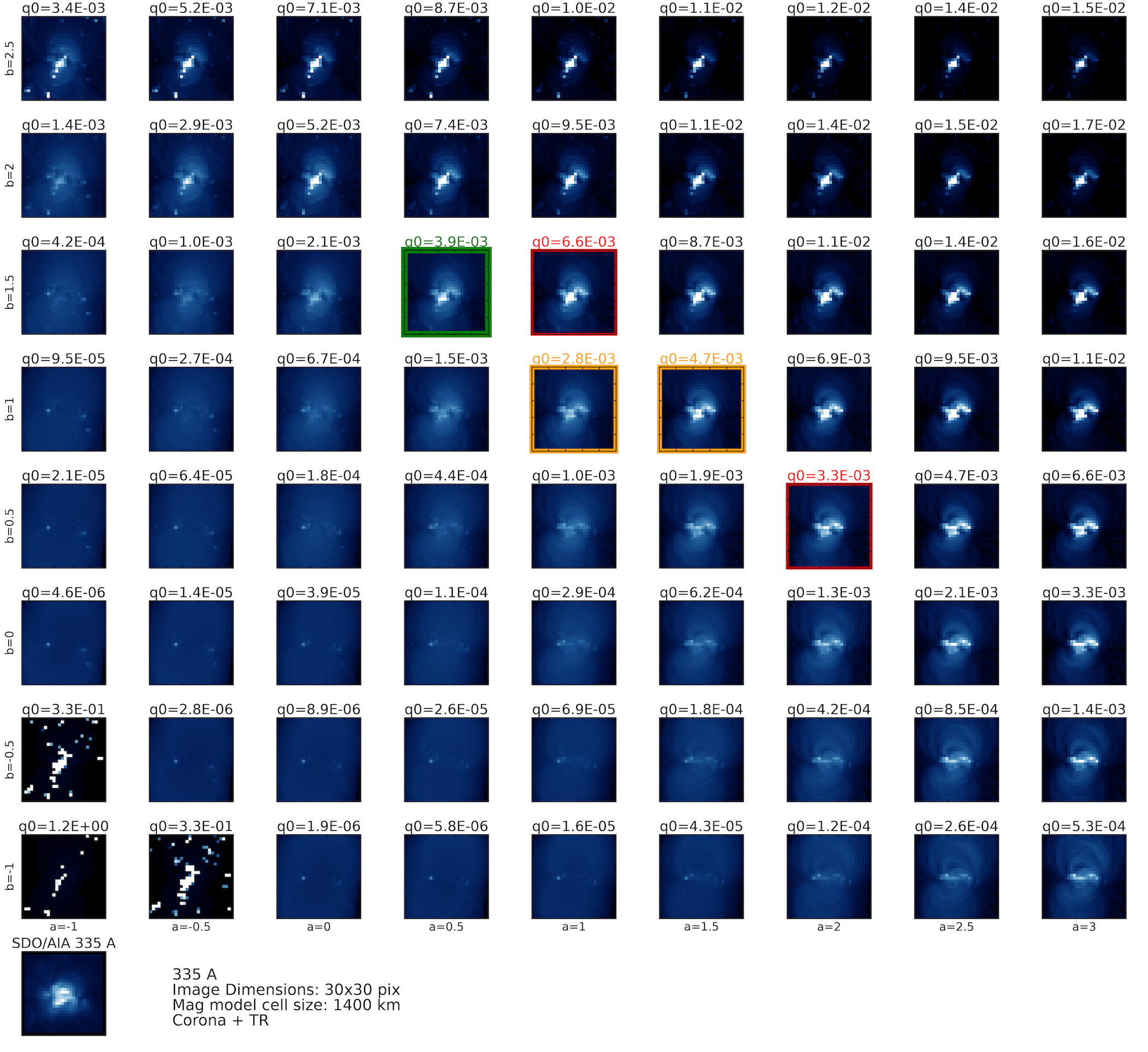}
\caption{Thumbnails of model 335~\AA\ images showing the best $q_0$ solutions for each of the ($a,b$) pairs considered for calculations using Magnetic Model~1, a model/data comparison resolution of 8\arcsec, and including both the corona and transitions region. The actual 335~\AA\ image at the same resolution is on the bottom left. The $a=0.5$, $b=1.5$ image has the lowest \GoF\ value. Its title and border are in green. See Fig.~\ref{f:BestFit_All_TRCorona} bottom right panel for larger version of the best model. Maps with \GoF\  within 10\% and 20\% of the minimum have titles and borders in orange and red, respectively. The images are all scaled linearly with the same intensity range. }
\label{f:Thumbnails_335A_nx30_YesTR}
\end{figure}

This appendix contains more examples of calculated models that give the lowest \GoF\ values as a function of $q_0$ for each of a grid of ($a,b$) values, Figures~\ref{f:Thumbnails_211A_nx30_YesTR}-\ref{f:Thumbnails_335A_nx30_YesTR}.

\end{document}